\documentclass[runningheads,a4paper]{llncs}

\usepackage{amssymb}
\usepackage{graphicx}
\usepackage{caption, booktabs}
\usepackage{subcaption}
\usepackage{listings}
\usepackage{xcolor}
\usepackage{algorithm2e}
\usepackage{hyperref}

\usepackage{footnote}

\usepackage{url}

\urldef{\mailsa}\path|{alfred.hofmann, ursula.barth, ingrid.haas, frank.holzwarth,|
\urldef{\mailsb}\path|anna.kramer, leonie.kunz, christine.reiss, nicole.sator,|
\urldef{\mailsc}\path|erika.siebert-cole, peter.strasser, lncs}@springer.com|    
\newcommand{\keywords}[1]{\par\addvspace\baselineskip
\noindent\keywordname\enspace\ignorespaces#1}

\begin{document}


\definecolor{forestgreen}{RGB}{34,139,34}
\definecolor{orangered}{RGB}{239,134,64}
\definecolor{darkblue}{rgb}{0.0,0.0,0.6}
\definecolor{gray}{rgb}{0.4,0.4,0.4}
\definecolor{codegreen}{rgb}{0,0.6,0}
\definecolor{codegray}{rgb}{0.5,0.5,0.5}
\definecolor{codepurple}{rgb}{0.58,0,0.82}
\definecolor{backcolour}{rgb}{0.95,0.95,0.92}

\lstdefinestyle{XML} {
    language=XML,
    extendedchars=true, 
    breaklines=true,
    breakatwhitespace=true,
    emph={},
    emphstyle=\color{red},
    basicstyle=\ttfamily,
    columns=fullflexible,
    commentstyle=\color{gray}\upshape,
    morestring=[b]",
    morecomment=[s]{<?}{?>},
    morecomment=[s][\color{forestgreen}]{<!--}{-->},
    keywordstyle=\color{orangered},
    stringstyle=\ttfamily\color{black}\normalfont,
    tagstyle=\color{darkblue}\bf,
    morekeywords={attribute,xmlns,version,type,release},
    otherkeywords={attribute=, xmlns=},
}

\lstdefinelanguage{PDDL}
{
  sensitive=false,    
  morecomment=[l]{;}, 
  alsoletter={:,-},   
  morekeywords={
    define,domain,problem,not,and,or,when,forall,exists,either,
    :domain,:requirements,:types,:objects,:constants,
    :predicates,:action,:parameters,:precondition,:effect,
    :fluents,:primary-effect,:side-effect,:init,:goal,
    :strips,:adl,:equality,:typing,:conditional-effects,
    :negative-preconditions,:disjunctive-preconditions,
    :existential-preconditions,:universal-preconditions,:quantified-preconditions,
    :functions,assign,increase,decrease,scale-up,scale-down,
    :metric,minimize,maximize,
    :durative-actions,:duration-inequalities,:continuous-effects,
    :durative-action,:duration,:condition
  }
}

\lstdefinelanguage{Srv}{
keywords = [1]{bool, uint8, int32, uint64, float32, float64, string, Header, Point, Quaternion, time},
comment=[l]{\#}
}

\lstdefinestyle{mystyle}{
  backgroundcolor=\color{backcolour},   commentstyle=\color{codegray},
  keywordstyle=\color{codegreen},
  numberstyle=\tiny\color{codegray},
  stringstyle=\color{codepurple},
  basicstyle=\ttfamily\footnotesize,
  breakatwhitespace=false,         
  breaklines=true,                 
  captionpos=b,                    
  keepspaces=true,                 
  numbers=left,                    
  numbersep=5pt,                  
  showspaces=false,                
  showstringspaces=false,
  showtabs=false,                  
  tabsize=2
}
\lstset{style=mystyle}

\mainmatter  

\title{Technical Report: Comparative Evaluation of AR-based, VR-based, and Traditional Basic Life Support Training}

\titlerunning{Technical Report: AR- and VR-based Basic Life Support Training}

%
%
\author{Enes Yigitbas \and Sebastian Krois \and Timo Renzelmann \and Gregor Engels}

\authorrunning{E. Yigitbas et al.}

\institute{Paderborn University\\ Zukunftsmeile 2, 33102 Paderborn, Germany\\
\email{enes@mail.upb.de, skrois@mail.upb.de, tire@mail.upb.de, engels@upb.de}, 
}
%




%
%

\toctitle{Technical Report: Comparative Evaluation of AR-based, VR-based, and Traditional Basic Life Support Training}
\tocauthor{Authors' Instructions}
\maketitle

\begin{abstract}
Basic life support (BLS) is crucial in the emergency response system as sudden cardiac arrest is still a major cause of death worldwide. In the majority of cases, cardiac arrest is witnessed out-of-hospital where execution of BLS including resuscitation through by-standers gets indispensable. However, survival rates of cardiac arrest victims could majorly increase if BLS skills would be trained regularly. In this context, technology-enhanced BLS training approaches utilizing augmented (AR) and virtual reality (VR) have been proposed in recent works. However, these approaches are not compliant with the medical BLS guidelines or focus only on specific steps of BLS training such as resuscitation. Furthermore, most of the existing training approaches do not focus on automated assessment to enhance efficiency and effectiveness through fine-grained real-time feedback. To overcome these issues, we present a novel AR- and VR-based training environment which supports a comprehensive BLS training compliant with the medical guidelines. Our training environment combines AR-/VR-based BLS training with an interactive haptic manikin that supports automated assessment, real-time feedback, and debriefing in an integrated environment. We have conducted a usability evaluation where we analyze the efficiency, effectiveness, and user satisfaction of BLS training based on our AR and VR environment against traditional BLS training. Results of the evaluation indicate that AR and VR technology have the potential to increase engagement in BLS training, improve high-quality resuscitation, and reduce the cognitive workload compared to traditional training.
\keywords{basic life support, resuscitation, augmented reality, virtual reality, usability evaluation}
\end{abstract}

\section{Introduction}
\label{section:intro}

Basic Life Support, or BLS, generally refers to the type of care that bystanders, first-responders, healthcare providers, and public safety professionals provide to anyone who is experiencing cardiac arrest, respiratory distress, or an obstructed airway \cite{olasveengen2021european}. BLS mainly requires knowledge and skills in cardiopulmonary resuscitation (CPR), using automated external defibrillators (AED), and relieving airway obstructions in patients of every age. 

According to the American Heart Association’s Heart and Stroke Statistics from the year 2020 \cite{virani2020heart}, there are more than 356,000 out-of-hospital cardiac arrests (OHCAs) annually in the U.S., nearly 90 percent of them fatal. Similar results can be observed across Europe according to the EuReCa 2 report \cite{grasner2020survival} from the year 2020, where a total of 37,054 OHCAs were confirmed where out of 8 percent of patients were discharged from hospital alive. 

In the majority of cases, cardiac arrest is witnessed out-of-hospital, but emergency medical services may not arrive at the emergency location in a short time interval. Thus, execution of BLS and including steps such as resuscitation through bystanders becomes indispensable. However, survival rates of cardiac arrest victims could majorly increase if lay resuscitation rates would be higher. Despite the importance of becoming active in BLS and starting the foreseen steps, BLS skills are usually not trained regularly among people. Regular training, for instance in schools or at regular training courses, is missing in most cases. Thus, many people are not skilled in BLS procedures. The vital links for successful BLS are
commonly summarized with the chain of survival metaphor. It describes
four links for successful BLS: early recognition and call for help,
early bystander cardiopulmonary resuscitation (CPR), early defibrillation, and post-resuscitation care. As the chain of survival indicates, the need for bystanders to become active as early as possible is critical for increased survival chances. CPR, which describes the act of performing chest compressions and ventilation, and defibrillation with the aid of an Automated External Defibrillator (AED), are simple procedures that every adult person can execute.

However, the question remains on how to teach people BLS skills and how to motivate them to refresh their skills regularly. Even though there is a large selection of training formats available, such as local Basic Life Support (BLS) training courses, online courses, or training sets including video tutorials on DVD and other materials, people may not be motivated or willing to pay for these courses or materials. Especially for younger people, who are accustomed to using modern technology gadgets, traditional
teaching methods may feel outdated and may create a barrier concerning their self-motivation for training. In addition, these training formats are usually limited with regard to their capability to combine theoretical teaching of knowledge with practical skill training and assessment, such that they are suitable for independent self-training only to a limited extent.

Motivated by the need to provide access to such BLS self-training approaches for everyone, modern training formats that utilize Augmented Reality (AR) or Virtual Reality (VR) technology were proposed in recent years. While VR and AR technologies have shown great potential to enhance BLS training in general and especially cardio-pulmonary resuscitation (CPR) performance, most of the existing approaches are in a rather prototypical state of implementation or do not leverage the full potential of an immersive learning environment for basic life support training. For instance, most of the existing approaches are not compliant with standardized medical guidelines for BLS or only focus on specific sub-steps of the BLS training such as CPR. Also, most of the existing approaches are experimenting by implementing different AR- or VR-based solutions, but still, the question remains which technology is most suitable for supporting BLS training in practice. To the best of our knowledge, there are no approaches that compare AR- and VR-based solutions with traditional BLS training approaches in terms of efficiency, effectiveness, and user satisfaction. Therefore, the following research questions can be raised:

\textbf{RQ1}: How can we exploit AR/VR technology to support comprehensive BLS training in an efficient, effective, and user-friendly way?

\textbf{RQ2}: How does AR-/VR-based BLS self-training perform in comparison to traditional training in terms of efficiency, effectiveness, and user satisfaction?

To answer the above-mentioned research questions, based on a collaboration with medical experts, we have identified the following requirements for the development of a novel VR- and AR-based BLS training environment to overcome the issues of the existing training approaches (see also Section 2 for a detailed analysis).

\begin{itemize}
\item \textbf{R1 - Compliance with BLS Guidelines}: In 2021, the European Resuscitation Council (ERC) has published the latest basic life support guidelines which are based on the 2020 International Consensus on Cardiopulmonary Resuscitation Science with Treatment Recommendations \cite{olasveengen2021european}. Procedures and techniques taught by the AR- and VR-based BLS training solution should be compliant with these ERC guidelines for BLS. All steps of the BLS procedure from "Ensuring Scene Safety" until "Usage of AED" should be covered to provide a comprehensive training approach.
\item \textbf{R2 - Automated Assessment} 
The AR- and VR-based training solution should be able to automatically assess user actions. For this purpose, the solution must be able to recognize user actions, match these actions to the predefined tasks of the BLS sequence, and finally evaluate the task execution. The assessment should cover the following aspects: correctness of task execution for each task, the correctness of task execution order for the whole BLS sequence, execution time, and CPR quality including compression rate and depth.  
\item \textbf{R3 - Real-Time Feedback and Debriefing}
Implementation guidelines on BLS training emphasize the importance of debriefing and real-time feedback during training in order to increase learning effectiveness. Users should receive real-time feedback on their actions in visual or audible form
when an action has been recognized or a task has been completed. In addition
to these general feedback mechanisms that are valid for all steps, task-specific
feedback should be provided by the system during CPR by showing the current compression
rate and depth. When a training session has finished, the system should present an overview of the training results (Debriefing). Results should show the list of tasks in combination with their individual completion result, i.e. whether the task has been completed correctly or not. Further, the system should provide hints for improvement where appropriate, such for instance when a task has been completed successfully, but a time constraint associated with the corresponding task has been exceeded.
\item \textbf{R4 - Learning and Training Mode} There are two major purposes of the training environment: acquisition of knowledge and skills, and practical training of the acquired skills. To fulfill both, the training environment should be divided into two modes of operation. In the learning mode, each step of the training procedure, which by default is the BLS sequence, is taught to the user in a detailed manner by providing step-by-step instructions. In the training mode, users can test their BLS skills and knowledge in practice. While using this mode users can find out whether they are able to successfully complete all steps of a BLS scenario on their own and without external help. The steps and actions necessary to successfully pass the training mode are equivalent to the ones presented in the learning mode. However, in the training mode, the user will not get any explanations or additional help from the system. Instead, the system is built to automatically recognize and assess all the actions that the user performs.
\item \textbf{R5 - Haptic Reality} To support an interactive and haptic learning environment, the AR- and VR-based training environment should be integrated with a haptic manikin to represent the victim who is in charge of BLS. The haptic manikin should deliver a hands-on learning experience for BLS and support the training of high-quality CPR especially with regard to compression rate and depth. 
\end{itemize}

Our contributions in this technical report are as follows: Firstly, we introduce a novel AR- and VR-based BLS training environment which tackles all the above-mentioned requirements. Our training environment consists of a generic training application kernel that can be instantiated for AR and VR usage scenarios. Furthermore, it integrates a custom build haptic manikin which supports an interactive learning mode where AR/VR technology is combined with haptics. Apart from the constructive approach, we have conducted a usability evaluation where we analyze the efficiency, effectiveness, and user satisfaction of BLS training based on our AR and VR environment against traditional BLS training.

The rest of the technical report is structured as follows. In Section 2, we present and discuss the related work. In Section 3, we describe the conceptual solution of our AR- and VR-based BLS training environment. In Section 4, we show the details of its implementation. In Section 5, we present and discuss the main results of the usability evaluation. In Section 6, we conclude the technical report and give an outlook for future work.  

\section{Related Work}
\label{section:relWork}

Augmented Reality (AR) and Virtual Reality (VR) have been a topic of intense research in the last decades. In the past few years, massive advances in affordable consumer hardware and accessible software frameworks are now bringing AR and VR to the masses. While VR interfaces support the interaction in an immersive computer generated 3D world and have been used in different application domains such as training~\cite{DBLP:conf/vrst/YigitbasJSE20}, robotics \cite{DBLP:conf/seams/YigitbasKJE21}, education (e.g.,\cite{DBLP:conf/mc/YigitbasTE20}, \cite{DBLP:conf/models/YigitbasGWE21}), prototyping \cite{DBLP:conf/vl/YigitbasKGE21}, or healthcare~\cite{DBLP:conf/mc/YigitbasHE19}, AR enables the augmentation of real-world physical objects with virtual elements. In previous works, AR has been already applied for different aspects such as product configuration (e.g., \cite{DBLP:conf/hcse/GottschalkYSE20}, \cite{DBLP:conf/hcse/GottschalkYSE20a}), prototyping \cite{DBLP:conf/hcse/JovanovikjY0E20}, planning and measurements (e.g., \cite{DBLP:conf/eics/EnesScaffolding}, \cite{DBLP:conf/eics/KringsYBE22}), robot programming \cite{DBLP:conf/interact/YigitbasJE21}, or for realizing smart interfaces (e.g., \cite{DBLP:conf/eics/KringsYJ0E20}, \cite{DBLP:conf/interact/YigitbasJ0E19}). Besides this broad view of application domains, several approaches for AR- and VR-based training in the medical domain have been proposed in recent years \cite{hsieh2018preliminary}. Improvements in modern AR and VR technology make researchers gain increasing interest in using this technology for medical education, including resuscitation training (\cite{everson2021virtual}, \cite{kuyt2021use}). Following the problem statement, research questions, and requirements overview as outlined in Section 1, we present and describe related approaches for AR- and VR-based BLS training. 

\subsection{AR-based approaches}

Strada et al.\cite{strada2019holo} propose an AR-based self-training and self-evaluation tool for Basic Life Support and Defibrillation which is called Holo-BLSD. It utilizes Microsoft HoloLens to create a first aid emergency scenario where the user is confronted with a recumbent cardiac arrest victim lying on the floor. Several objects that are involved in a typical BLS scenario are modeled via holograms, such as a public phone station used for emergency calls, an AED, or other non-player characters (NPCs) that can be given instructions to. Holo-BLSD offers three modes of operation. In the learning mode, detailed instructions on the BLS procedure are given step by step via voice output from a virtual trainer. In the rehearsal mode, trainees are asked to execute the procedure by themselves, while still getting feedback on their actions from the application. Finally, in the evaluation mode, the trainees do not get any feedback, but their actions are assessed by the application. The automated assessment evaluates all steps except CPR quality. When performing CPR, only the compression frequency is automatically assessed, an automatic evaluation of compression depth and chest recoil is not included. The suitability of the proposed tool is analyzed by conducting a user study. However, there is no direct comparison with VR-based approaches or traditional training which can be used as a basis to address our stated research questions. 

HoloCPR by Johnson et al. \cite{johnson2018holocpr} is a further AR-based approach that targets helping uneducated bystanders with performing timely resuscitation of a cardiac arrest victim. This approach makes use of simple visualizations, such as arrows, circles, texts, and simple animations, in order to give instructions to the user and guide the user’s attention towards points of interest. There is no virtual emergency scenario created, and there are no virtual environmental objects or persons modeled that the user could interact with. To demonstrate the effectiveness of their proposed application, the authors conduct a between-subjects experiment with several participants. Experiment results indicate that, while using HoloCPR, participants show reduced initial reaction times and are faster in interpreting CPR instructions given by the AR system as compared to traditional 2D instructions, resulting in faster action. Further, HoloCPR facilitates procedural adherence, i.e. it is easier for HoloCPR users to follow individual steps of the BLS procedure. However, HoloCPR is not specifically designed as a BLS training application as it does not fully conform to the BLS guidelines. Also, it enables no further insides about a comparison to traditional training methods.

Javaheri et al. \cite{javaheri2018stayin} combine a traditional training manikin and Microsoft HoloLens with a dedicated client-server application running on a third device. The training manikin is equipped with several pressure sensors, which are able to measure data on CPR compression depth and frequency. Via the client-server application, direct data communication between the manikin’s sensors and the HoloLens is possible, such that direct feedback on current CPR quality can be given to the user in real-time. The sensor-based approach allows for high-quality automated assessment of CPR performance.
Similar to the previous approaches, the application comprises different modes of operation. In the first phase, a virtual trainer explains the basics of how to perform CPR. In the second phase, the learned techniques have to be practically applied to the physical training manikin, and the trainee’s CPR performance is automatically assessed. In both phases, the application can be controlled via voice commands. While this approach supports a detailed CPR tutorial it is not comprehensive enough to cover a whole BLS training process. Furthermore, a usability evaluation and comparison against traditional approaches are missing such that our research questions are not fully tackled by this approach. 

Similar to the previous approach, Kwon et al. \cite{kwon2014heartisense} developed a CPR training system employing a dedicated sensor kit that is inserted into a traditional training manikin. Via these sensors, the CPR quality can be measured in terms of compression location, depth, and frequency. Further, sensors can measure the strength and volume of artificial respiration, and they can detect whether the victim’s airways have been cleared. The concept has been implemented in two ways: As an AR training simulator using the projection method, and as a mobile application. In both cases, visual and auditory feedback is given, for instance when the user carries out correct or incorrect activities, respectively. An automated assessment of the user’s CPR performance is given, and hints for possible improvements are provided. By conducting a user study, the authors analyze the effectiveness of their proposed system. They conclude that their system is able to provide accurate training assessment and that by using their system the retention of training content can be enhanced. However, the proposed system focuses on CPR training and assessment, thus it is not suitable for comprehensive BLS training. A dedicated learning mode in which the required methods and techniques are taught is not part of the work.

In the ARLIST project conducted by Pretto et al. \cite{pretto2009augmented}, an AR environment for general life support training has
been developed and evaluated. The main goal of the project is to enhance traditional training by augmenting it with auditory feedback and visualizations in the form of AR projections onto the training manikin. For instance, facial expressions and skin injuries projected onto the manikin reflect the physical state of the virtual patient, and cardiac or pulmonary sounds played via speakers installed in the back of the manikin’s neck enable the trainee to autonomously assess the patient’s consciousness. Additionally, a waistcoat and an adapted stethoscope have been created that allow the trainees to execute cardiac and pulmonary auscultation during training. Actions carried out by the trainees are automatically recorded
during training, and the created logs can later be viewed by instructors via a companion tool that has been developed along with the AR environment. In a user study, the authors evaluate whether their proposed system is able to increase the realism of life support training. They conclude that their system improves accuracy and objectivity as well as the trainees’ autonomy during training.

Balian et al. \cite{balian2019feasibility} conducted a feasibility study of an AR CPR training approach that seeks to supplement traditional training with the potential advantages AR technology offers. The system uses an AR headset (HoloLens) to show a model of the human circulatory system to the user next to a Laerdal recording CPR manikin. To provide feedback to the user the model changes the rate of blood flow displayed proportional to the rate of chest compressions applied. Further, to train consistent CPR the system uses an audible heartbeat of 110 bpm when applied compression rate is not within the tolerance of ±10 bpm to remind and correct the user. The authors conducted a study with 51 health care providers using their AR system and collected performance data and quantitative survey data. The survey responses showed promising results concerning the usefulness of the blood flow
visualization and the willingness to use the system again [34]. The authors aggregated similar responses to open-ended questions. The most liked
feature was the real-time audio-visual feedback. However, the two most common suggested improvements were placing the blood flow simulation above the manikin (not to the side), and providing users with real-time performance
statistics. While this approach addresses many requirements, it is specialized in CPR and does not cover the whole BLS training. Also, real-time performance statistics are missing and comparison against traditional training methods has not been conducted. 

Similarly, Ingrassia et al. \cite{ingrassia2020augmented} conducted a usability study of an AR system for basic life support training. The system uses the HoloLens and overlays an AR scenario onto a CPR training
manikin equipped with computer vision markers. This system does not only focus on CPR but demonstrates the breadth of potential for AR
systems. The methods of interaction enabled by the HoloLens
are introduced individually to lead the participants through a
series of AR simulated basic life support actions. These include
safety, checking responsiveness, contacting, and interacting with
emergency services, and AED retrieval and application. A total
of 26 participants were guided through the multistep AR simulated
resuscitation procedure and recorded survey responses to
identify the system usability and ISO 9241-400 ergonomics on
a Likert scale. Results show that the system was easy to use, comfortable,
and required minimal cognitive effort. While this approach supports the wider application of AR training for basic life support, it is not leveraging insides about the efficiency, effectiveness, and user satisfaction of such a solution compared to traditional BLS training. 

\subsection{VR-based approaches}

VReanimate II \cite{bucher2019vreanimate} (successor of VReanimate\cite{blome2017vreanimate}) is a VR application intended for non-verbal guidance in virtual first aid scenarios. In VReanimate, no text or speech is used. Giving instructions via graphical visualizations only avoids the need for the user to read and interpret texts during training, thus allowing the creation of fast-paced, uninterrupted real-world
scenarios. The application is based on the ERC 2015 life support guidelines \cite{perkins2015european} and comprises the basic steps of the BLS algorithm. However, a fully automated assessment of the trainee’s performance is not part of the application. The authors evaluate the usability and effectiveness of their approach by conducting interviews and an experiment.

EMERGENZA \cite{ferracani2015natural} is a serious game designed for the training of emergency medical personnel. Different virtual emergency scenarios are created, including NPCs that the user can interact with. The scenarios are designed to train the basic life support procedure. However, since this approach is completely virtual and does not make use of a dedicated training manikin, high-quality CPR training cannot be provided. The application is controlled via gestures. The authors utilize the Microsoft Kinect in order to translate the user’s gestures into the virtual environment. All actions performed during training are logged and can later be used for debriefing. A usability evaluation of the proposed system is conducted in the form of a small user study. Training effectiveness is not evaluated.

While the above-described approaches have the BLS training in scope, there are also specialized VR-based approaches focusing on CPR training. Semeraro et al. \cite{semeraro2009virtual}, for instance, demonstrated traditional CPR training enhanced with VR in 2009 with a system using a VR headset, a CPR training manikin (Laerdal HeartSim 4000), data gloves, and tracking devices. The VR headset displays a first-person perspective cardiac arrest simulation to the user, and the data gloves track the user's hand position for representation within the virtual environment. The researchers use audio and visual detail with the intent to make the training scenario more realistic. A refined prototype VR CPR standard manikin training system with hand tracking and handsfree Kinect motion detection was presented in their follow-up work \cite{semeraro2019virtual}. This system was trialed with 43 medical students. The test collected compression rate and compression depth simultaneously
with Resusci-Anne and the developed motion tracking tool. The results showed that the data collected by both systems were comparable demonstrating that it is possible to collect CPR metrics externally from the manikin and use them in a VR scenario. Furthermore, Vaughan et al. \cite{vaughan2019cpr} demonstrated a VR CPR training system specifically for schools. Similarly, it uses a commercially available VR headset (Oculus Rift) and a hand tracking solution that does not require gloves (Leap Motion). The simulation presents
common incident scenarios and utilizes a 3-D scan of the training
manikin used to ensure correct visual-physical alignment. The
authors stress the importance of CPR training at an early age and
point to the growing trend in medical simulation to underpin the
decision to develop this system. Finally, Almousa et al. \cite{almousa2019virtual} developed a VR-CPR system that focuses on the gamification of CPR training. This approach also uses a first-person perspective simulation of a cardiac arrest patient and uses hand
tracking for hand position measurement relative to the manikin,
determining compression rate and compression depth. The novelty of this system is the exploration of gamification features, real-time feedback, and a progressive difficulty system. As the difficulty is advanced, the user experiences reduced feedback and guidance and increased distractions. Competition with other trainees is incentivized with a leader board,
to encourage repeated training and prevent degradation of CPR
skills over time. Evaluation results of their VR-CPR solution
compared with traditional CPR training methods show that the testers liked the idea of gamification and found it exciting and engaging. While the above-described approaches combine innovative ideas such as VR and gamification to enhance CPR training, their scope is restricted to CPR and comprehensive BLS training is not covered. Furthermore, a detailed comparison between VR-, AR-based training versus traditional training is not covered in a comprehensive usability evaluation such that strengths and weaknesses of each technology can be identified. 

\subsection{Discussion}

\begin{figure}
    \centering
    \includegraphics[width=1.0\linewidth]{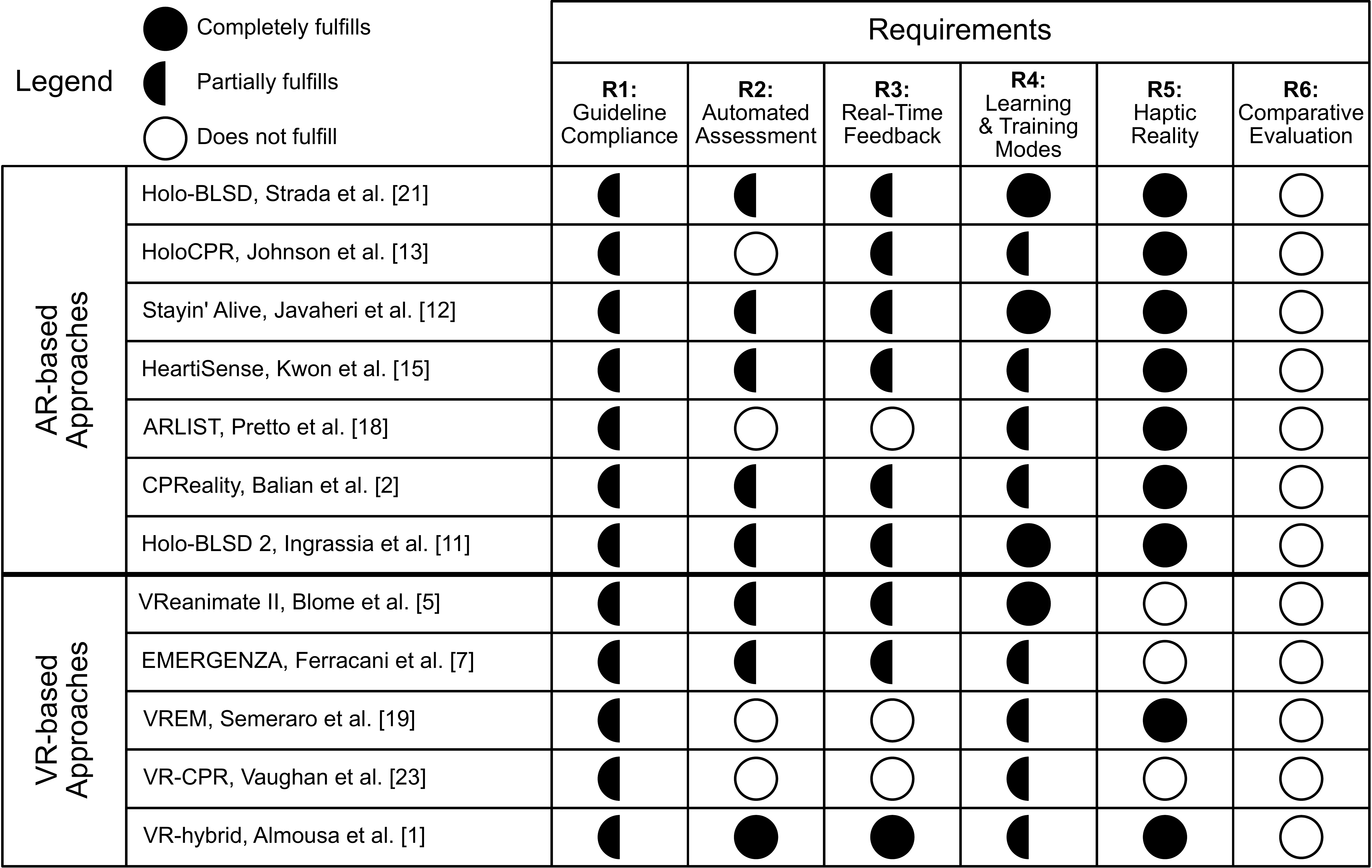}
    \caption{Related Work Overview}
    \label{fig:relWOrk}
\end{figure}

A summary overview of the presented related work approaches is depicted in Figure \ref{fig:relWOrk}. As it can be seen in the overview, none of the proposed systems are fully compliant with the new ERC basic life support guidelines \cite{olasveengen2021european} in terms of the taught methods, techniques, and procedures. While some of the existing approaches focus on specific steps of the BLS training, the necessary BLS steps are usually not included in their entirety. For instance, assessment of respiration checks and airway clearance is only included in one of the systems \cite{kwon2014heartisense}. Since properly executed CPR is one of the most important steps in life
support, the majority of the works focus on this topic. High-quality automatic
assessment of CPR performance is non-trivial to achieve, as it usually requires the
employment of dedicated sensors that are attached to or inserted into the training
manikin. Thus, automated assessment of CPR quality is only included in those
approaches that put their entire focus on CPR assessment, but do not create a
training environment that can be used for independent self-training which covers the whole BLS process. An important feature for virtual training systems is a dedicated learning mode that teaches necessary skills. Without teaching skills, trainees are not able to learn how to perform actions properly, and they may not be able to recognize their mistakes.
Thus, training becomes ineffective, and independent self-training without
a human instructor is not possible. Those systems that are meant to be used for independent self-training usually include such a learning mode such as \cite{strada2019holo} or \cite{semeraro2009virtual}. However, most of the existing approaches do not separate between a training and exam mode to systematically enhance the learning outcomes of the training session. 
The comparison shows that a complete and automatic assessment of the trainees’
performance is difficult. None of the considered approaches, except the VR-based approach by Almousa et al. \cite{almousa2019virtual}, includes a fully automated assessment for all steps of the BLS procedure. Another important limitation of the considered approaches is the lack of an included debriefing feature. Some approaches state that data acquired during training are logged and those data can later be used for debriefing. However, debriefing is never part of the existing AR and VR application as such but is always carried out via separate companion tools, such as a usual desktop application. In summary, none of the existing approaches come up with a comprehensive AR/VR BLS training environment that fulfills all stated requirements. Furthermore, there is no existing work that provides a detailed comparative usability analysis to identify benefits and limitations of AR-/ and VR-based solutions for BLS training in terms of efficiency, effectiveness, and user satisfaction compared with traditional methods (R6 which relates to RQ2). 

\section{Solution Overview}
\label{section:solOverview}

\begin{figure}[h]
    \centering
    \includegraphics[width=.8\linewidth]{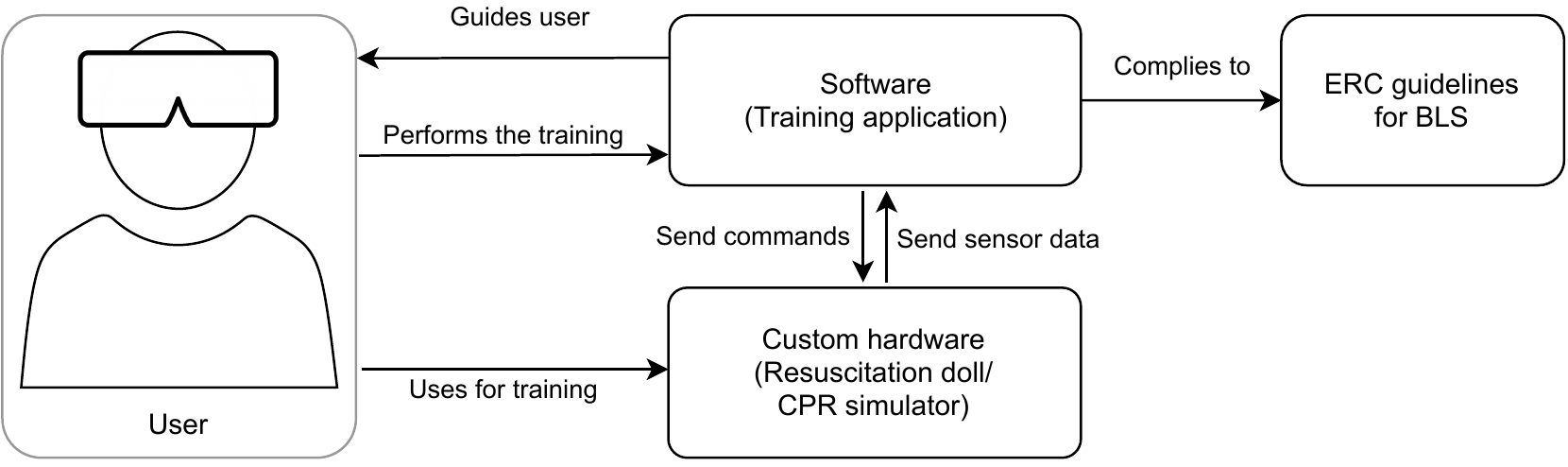}
    \caption{High-level Solution Overview}
    \label{fig:solOverview:HighLevel}
\end{figure}

To address the first research question (RQ1) motivated in Section 1, we have designed a novel AR- and VR-based BLS training environment which tackles the described requirements R1-R5. A high-level architectural overview of our training environment is shown in Figure \ref{fig:solOverview:HighLevel}.

In the following, each component of the architectural overview will be described in more detail. In Section \ref{section:solOverview:BLSSequence}, we first present the ERC guideline compliant process, which builds up the basis for creating our BLS training environment. With that in mind, we present in Section \ref{section:solOverview:Software} the concept for the software part of the training environment, followed by Section \ref{section:solOverview:Hardware}, where we present the concept of the custom hardware for realizing an interactive and haptic training environment.

\subsection{BLS Sequence}\label{section:solOverview:BLSSequence}
\begin{figure}
    \centering
    \includegraphics[width=\linewidth]{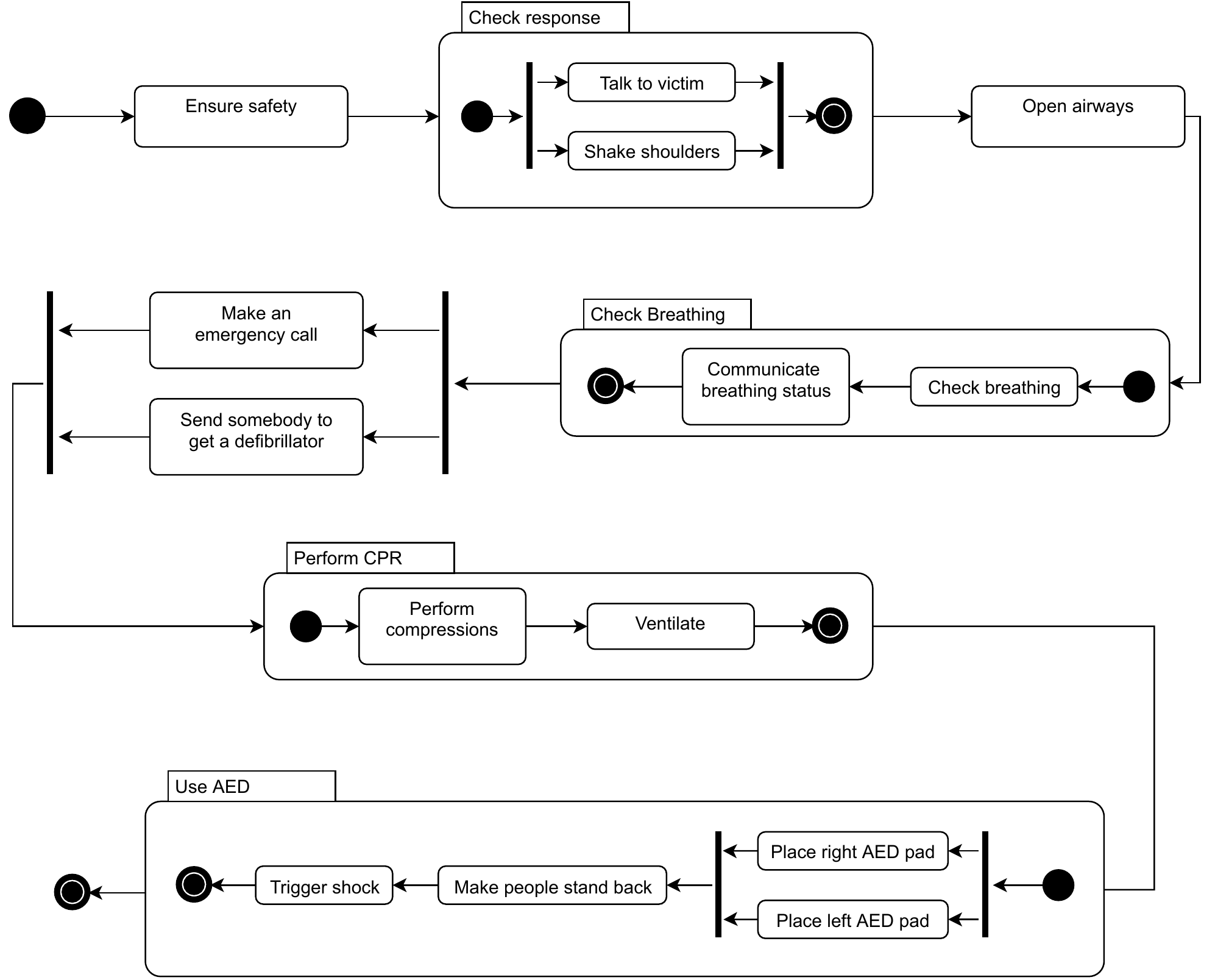}
    \caption{BLS Process}
    \label{fig:eval:blsProcess}
\end{figure}

As a basis for our training, we use the ERC guidelines for BLS based on \cite{olasveengen2021european} (R1). With them, we derived the process shown as a UML activity diagram in Figure \ref{fig:eval:blsProcess}.

First, the user has to ensure that the scene is safe and nobody is in danger. After that, it needs to be checked whether the victim is responding. The user should talk to the victim and shake their shoulders to see if there is a reaction. If the victim is not reacting, the airways need to be opened, so that the breathing can be checked afterward. If there are other helpers present, the breathing status should be communicated. After that, the user should make an emergency call (or ask somebody else to do so) and send somebody to get a defibrillator. In a real accident, the victim should never be left alone. Thus, if no helpers are nearby an AED is not used. In the next step, cardiopulmonary resuscitation (CPR) has to be performed. That contains two parts, compressions and ventilations. For the compressions, the rate (compressions per minute) and the depth are evaluated. After 30 compressions, the victim needs to be ventilated. If a defibrillator (AED) is available it should be used immediately. First, the AED pads need to be placed. Then, the AED analyses the heart rhythm. During that time, nobody should touch the victim. So the user has to communicate to other helpers, that they should stand back. After analyzing, the AED will display whether a shock is necessary. If that is the case, the user needs to trigger the shock. In our training process, that is the last step to execute, in real-life, the CPR should be continued until the ambulance arrives.

\subsection{Software}\label{section:solOverview:Software}
As in AR applications the users still see their environment, we decided to create a scenario close to real resuscitation training. The user is in a safe space (e.g. a training center or an office) and stands in front of a resuscitation manikin simulating the victim. We then use augmented reality to explain what to do and give feedback on the executed steps. Everything shown to the user should be correctly placed around the manikin. So we use the manikin as an anchor to sync the position of the virtual objects with the real world.

For VR applications we are not bound to the place the user is in. We can create whatever environment we want. To keep it simple, we decided to use a scenario where the victim is in an encapsulated garden. Here, we do not need a detailed resuscitation manikin as for AR, but we can use a simplified version. Despite the environment and the manikin, the application itself should be similar for AR and VR. Thus, in the following, we present the concept for the applications independent from the used technology.

\begin{figure}
    \centering
    \includegraphics[width=\linewidth]{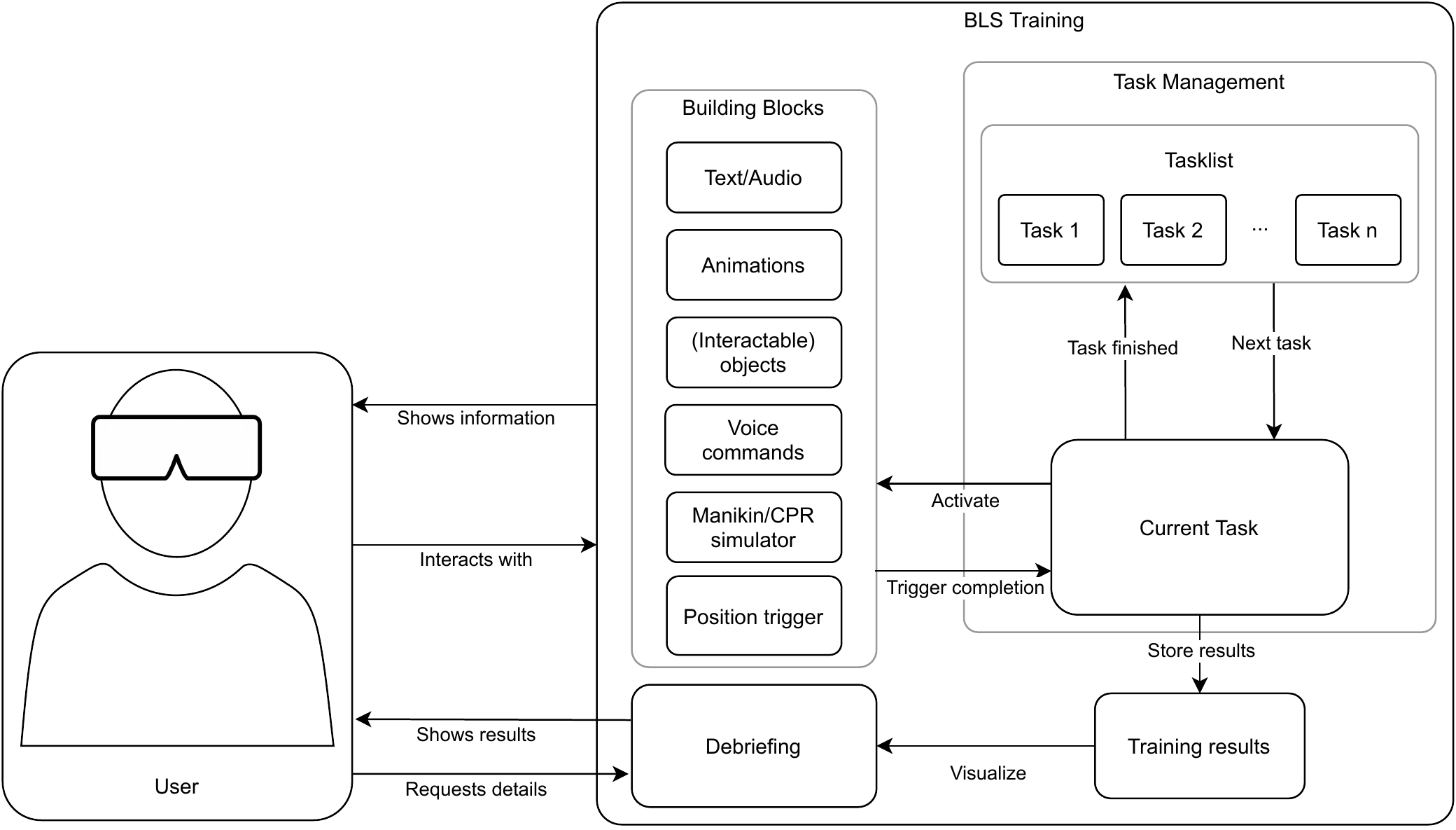}
    \caption{Architectural Overview}
    \label{fig:solOverview:ApplicationOverview}
\end{figure}

Figure \ref{fig:solOverview:ApplicationOverview} shows the overview of the BLS Training architecture. 
\paragraph{Task Management}
To allow the application to keep track of which steps the user executed, and how well it was done, we developed a \textit{Task Management}. That consists of a \textit{Task List} which we derive from the process described in the previous section. Every step in the process corresponds to a \textit{Task} the user needs to execute. The training starts with the first task and guides the user sequentially through the process. After one task was finished, the next task is selected. With that, we can record how well a task was executed, and automatically assess the user's performance (R2).

\paragraph{Building Blocks}
As the task management only realizes the basic logic of the process, we created several \textit{Building Blocks} which can be activated by a task and realize the interaction of the user with the application.

\begin{itemize}
    \setlength\itemsep{.5em}
    \item[] \textit{Text/Audio.} We needed a possibility to display static information to the user. That can be done in multiple ways. For conventional training, there exist texts or images to describe a certain step, or a trainer explains and demonstrates it. For our applications, we combined all those possibilities to retrieve the highest possible learning effect. We created \textit{Texts} describing the current step, how to execute it, and why it needs to be executed. These texts have been derived from the ERC guidelines for BLS. As it can be tiring for a user to only read text, and to simulate a trainer explaining the content, we provide \textit{Audio} files of the text being read. In the case users missed something or have problems understanding the language, we additionally show the text itself. In addition to the texts, we show corresponding images depicting the task. 
    
    \item[] \textit{Animations.} Until now, we adapted techniques used in real life so we can use them in our application. But as we are in a digital environment we can control, we also created \textit{Animations} showing how the described steps are executed. With that, a user can walk around and watch the animation from the spot they like to fully understand how to execute a step properly. 
    
    \item[] \textit{(Interactable) Objects}. In real accidents, multiple helpers are likely to be present at the accident scene. Thus it is also important, that users learn how to communicate with them and where tasks can be delegated to someone else. We provide an avatar representing other helpers. As the users cannot directly interact with it, we call it a (static) \textit{Object}. Additionally, we needed some \textit{Interactable Objects} the user can directly interact with. For our BLS training, we created \textit{broken glasses, a mobile phone, and a defibrillator with its electrode pads}. 
    
    \item[] \textit{Voice Commands.} To evaluate if a user correctly communicated with bystanding helpers, we integrated the possibility to recognize \textit{Voice Commands} which corresponds to the information and tasks given to them. 
    
    \item[] \textit{Manikin/CPR simulator.} Chest compressions are a central aspect of the BLS sequence. To increase immersion and realism, we created a haptic \textit{Manikin/CPR Simulator}. In Section \ref{section:solOverview:Hardware}, we explain the simulator's hardware concept in more detail. 
    While performing compressions, it is substantial that the rate and depth are correct. So we display them both, so the user can easily see how they are doing. 
    
    \item[] \textit{Position Trigger}. When interacting with the manikin, the user's head or hands need to be in a specific position. For example, when ventilating the victim, the user's head is right above the victim's head. To retrieve that information, we created \textit{Position Triggers} which raise an event after the correct position is reached.
    
\end{itemize}
To let the users know, if they are doing well, we give them real-time feedback (R3) while they execute the training. Whenever a task was executed correctly, the user gets notified by a sound cue. In addition, during CPR, we also display the current compression depth and rate.

With that, we described our concept for the \textit{learning} part of the application where we offer explanations, descriptions, and guidance through the training. 

In addition to that, we developed a \textit{training} mode (R4). In the \textit{training} mode, explanations and guiding elements are omitted so the users have to do everything on their own. 

\paragraph{Debriefing}
When a training session has finished, the system presents an overview of the training results. Results show the list of tasks in combination with their completion result, i.e. whether the task has been completed correctly or not. Further, the system provides hints for improvement where appropriate, such as when a task has been completed successfully, or a time constraint associated with the corresponding task has been exceeded. In addition to the real-time feedback during the execution, results for CPR (average compression rate and depth and whether the chest was always fully released) are also presented during the debriefing. The application can also give specific improvement suggestions, for instance when the compression rate could be optimized in one or the other direction.
To not overwhelm the user, we created two levels of detail. First, an overview of all the tasks, and second, more detailed insight into a single task.

\subsection{Hardware}\label{section:solOverview:Hardware}
As working on/with the victim is an essential part of BLS, we need some kind of hardware to train that. That is especially important for performing chest compressions. Without real-life hardware, the users would perform compressions in the air or on the ground without realistic resistance. In addition to that, we could not precisely assess the rate and depth of the compressions. For conventional training, there are resuscitation manikins available to buy. Standard manikins which allow chest compressions and ventilation are mostly used in conventional training. But, as we want to evaluate the rate and depth of the compressions, we need sensors to be integrated into the manikin. 

\textit{Manikins with sensors} are also commercially available, such as the Laerdal Resuci Anne\footnote{\url{https://laerdal.com/us/products/simulation-training/emergency-care-trauma/resusci-anne-simulator/}}. They are mostly used to train medical professionals and thus consist of many complex sensors. 
As these commercially available professional manikins are usually closely tied to their own companion tools and apps, accessing their sensor data is not easily possible. Thus, we have explored an alternative solution where we created our own CPR manikin, which is equipped with only those sensors we need for our application (R5).

\begin{figure}
    \centering
    \includegraphics[width=\linewidth]{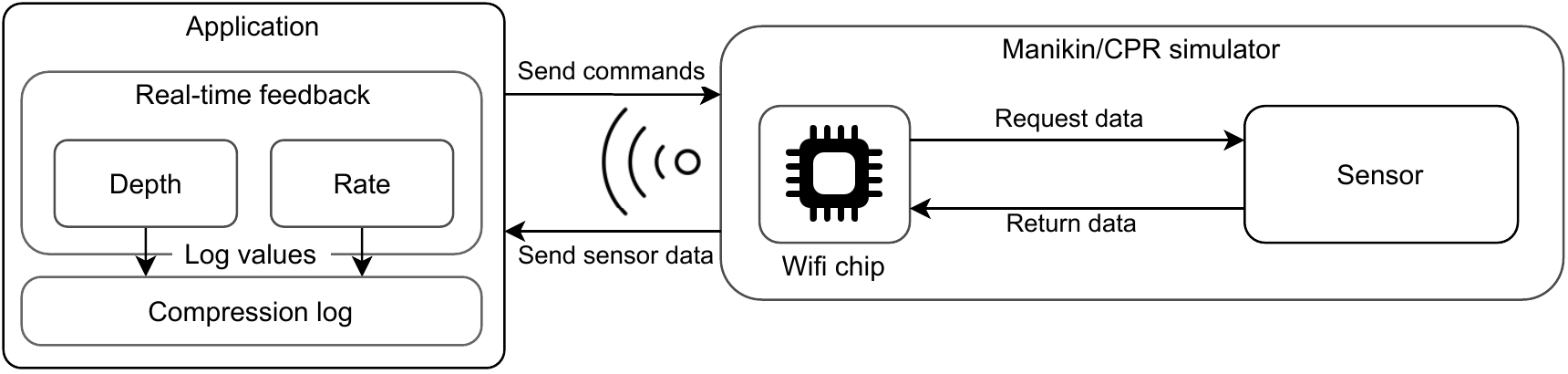}
    \caption{Sensor Communication}
    \label{fig:solOverview:Hardware:SensorCommunication}
\end{figure}

Figure \ref{fig:solOverview:Hardware:SensorCommunication} shows an overview of the connection between the CPR simulator and the application. The application can connect to a wifi chip and send commands e.g. to start/stop sending data from a sensor. The chip then reads the sensor's value and sends it back to the application. Here, the user should be provided with real-time feedback on the compression rate and depth. Additionally, for later analysis, the received values are logged. 

\paragraph{AR}\label{section:solOverview:Hardware:AR}
Since in AR applications the users' real environment is perceivable, we utilize a real manikin for the training. As a starting point, we employed a plastic manikin and inserted springs into the chest to allow compressions. We installed sensors in both the mannequin's chest and head. These are used for measuring compression depth during CPR and for recognizing a head tilt during airway clearance.

\paragraph{VR}\label{section:solOverview:Hardware:VR}
For VR, the compression simulation can be simpler. As the users do not see the real world, we do not need a correct representation of a resuscitation manikin. In addition, every user interaction happens in the virtual environment, so we do not need a realistic manikin. Instead, for the VR hardware, we only need a device the user can push and which can measure the compressions similar to the manikin for the AR application, omitting all components used to increase the immersion. 

\section{Implementation}
\label{section:impl}
This section describes the implementation of our AR- and VR-based BLS training environment. First, in Section \ref{section:impl:Hardware}, we describe how we built the hardware components with which the user interacts, followed by Section \ref{section:impl:Software}, where we present the software implementation of the two applications. In Section \ref{section:impl:Software:Walkthrough}, we present a walkthrough of both applications and explain which of the developed building blocks were used to create the training applications. Finally, in Section \ref{section:impl:Debriefing} we present how we realized the debriefing mode for our training.

\subsection{Hardware}\label{section:impl:Hardware}

\subsubsection{AR}\label{section:impl:Hardware:AR}
\begin{figure}
    \begin{center}  
    \subfloat[Resuscitation Manikin Schematic\label{fig:implementation:Hardware:DollSchematic}]{
    \includegraphics[width=.45\linewidth]{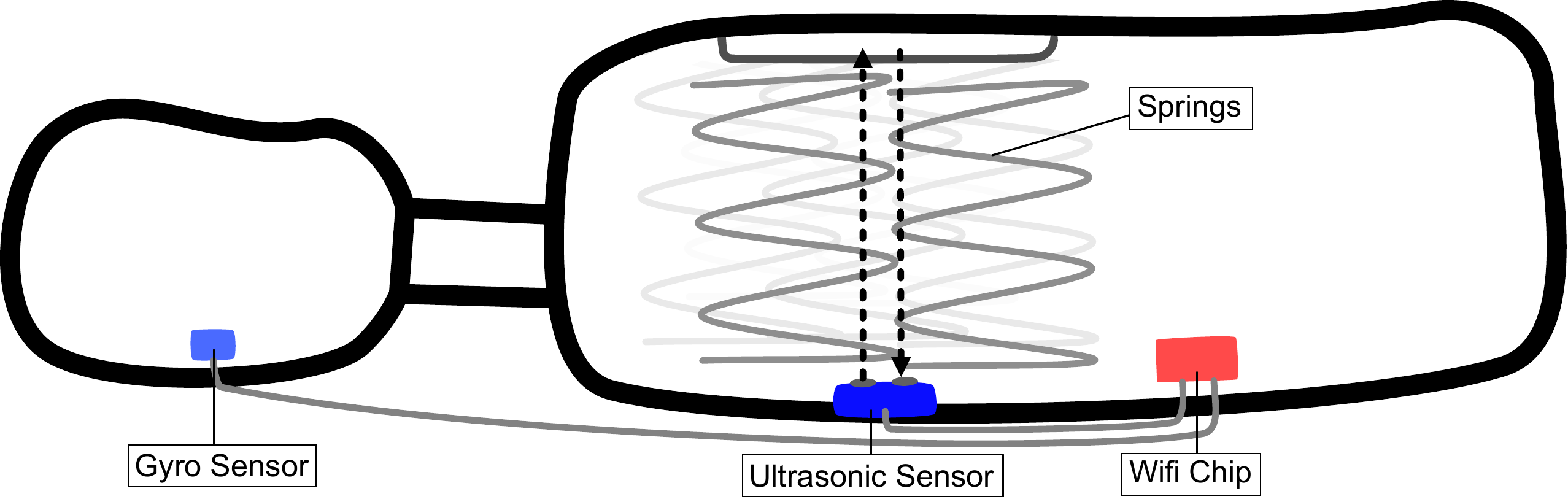}
    }\hspace*{3pt}
    \subfloat[Top View\label{fig:implementation:Hardware:Mesut1}]{
    \includegraphics[width=.45\linewidth]{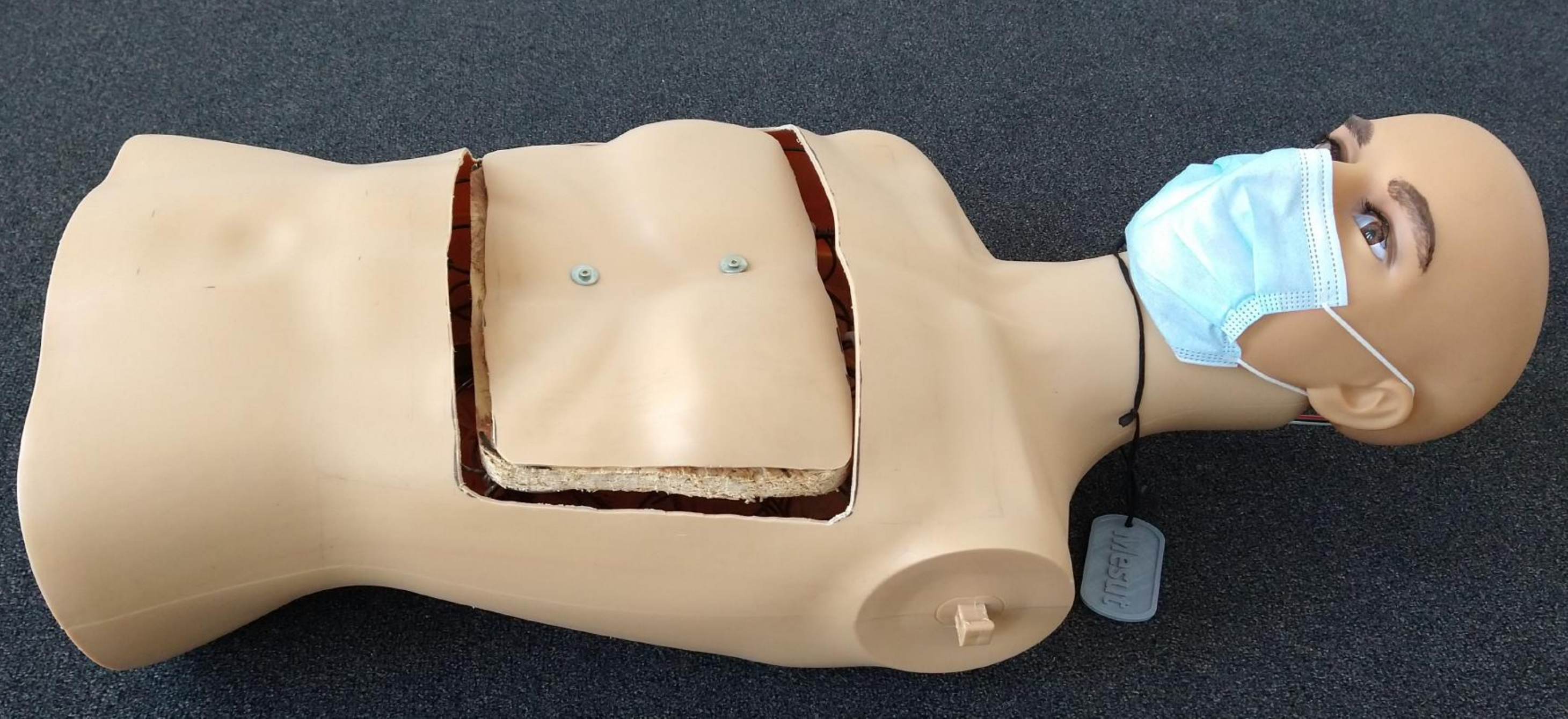}
    }
    \end{center}
    
    \begin{center}
    \subfloat[Bottom View\label{fig:implementation:Hardware:Mesut2}]{
    \includegraphics[width=.45\linewidth]{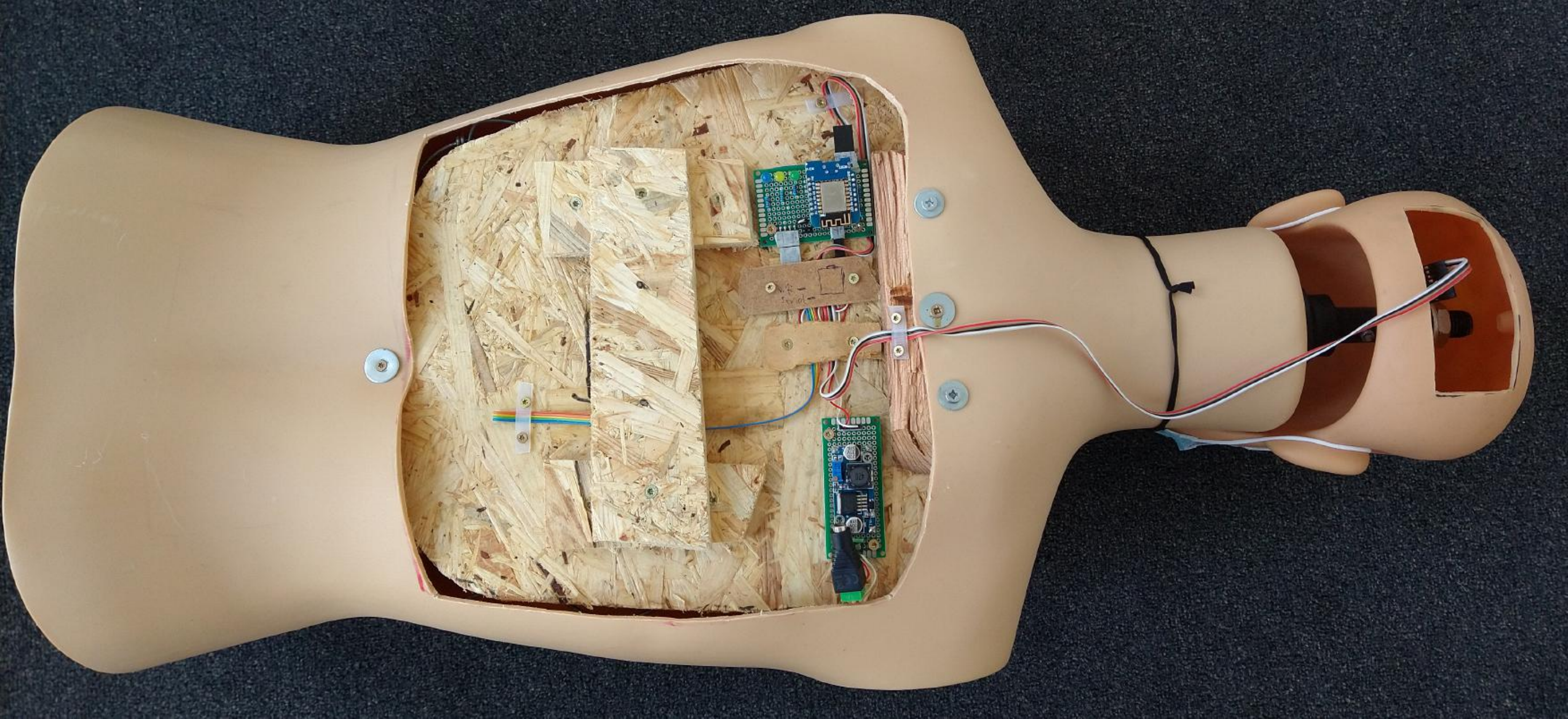}
    }\hspace*{3pt}
    \subfloat[Manikin as Seen by User\label{fig:implementation:Hardware:Mesut3}]{
    \includegraphics[width=.45\linewidth]{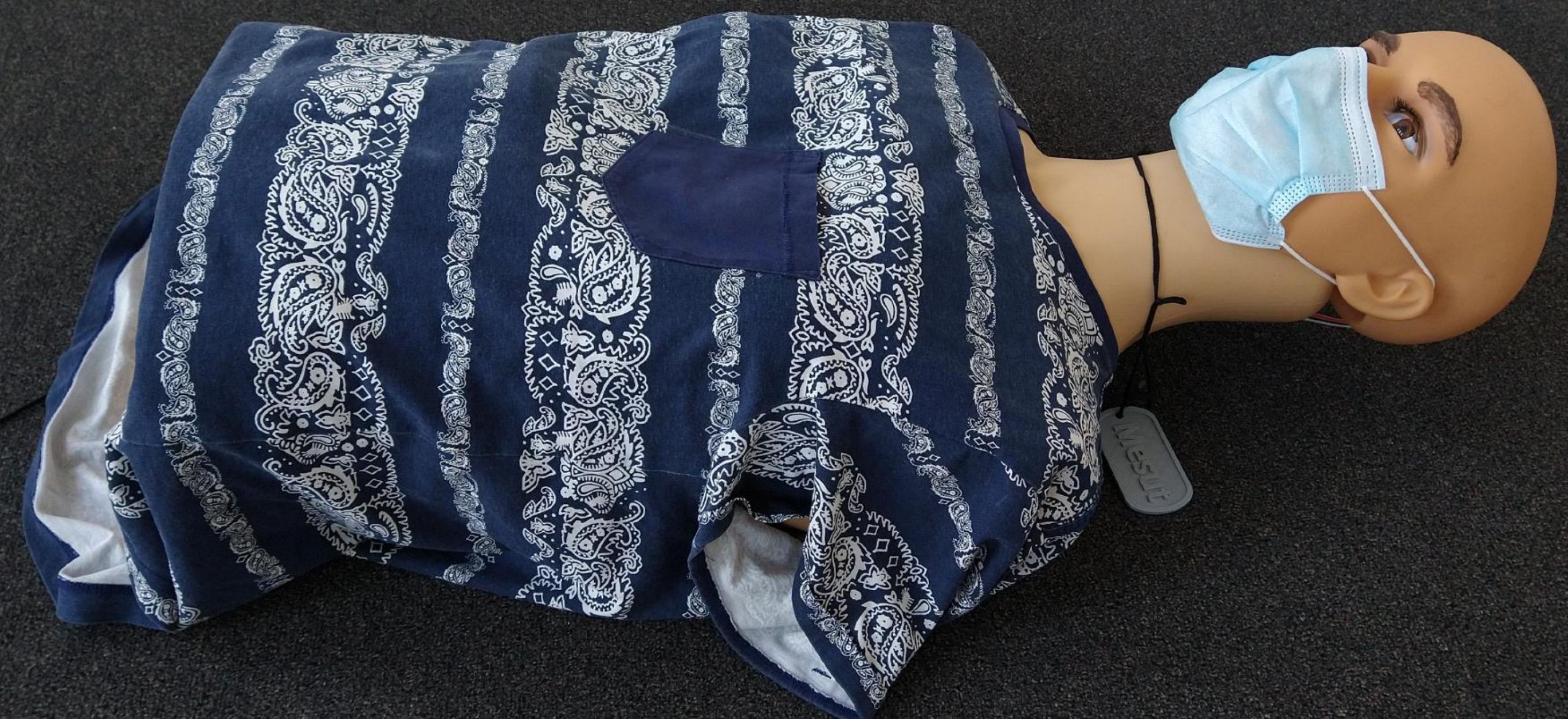}
    }
    \end{center}
    \caption{Custom Resuscitation Manikin}
    \label{fig:implementation:Software:AR:Objects}
\end{figure}

The resuscitation manikin is a customized display dummy. Figure \ref{fig:implementation:Hardware:DollSchematic} shows the main components, Figures \ref{fig:implementation:Hardware:Mesut1}-\ref{fig:implementation:Hardware:Mesut3} show the final result of our custom manikin.

First, we cut out a part of the chest and filled the space with springs, so that the chest can be pushed down by the user. To measure the depth, we use an ultrasonic sensor which measures the distance between the ground and the chest top. To retrieve the tilt angle of the head (a user needs to overextend the neck to open the victim's airways), we attached a gyro sensor to it. Both sensors are connected to an \texttt{ESP-8266} wifi chip. The chip hosts a simple server, to which the application can connect. It can then send simple commands to control the chip. For example, the application can request data for both sensors or tell the chip to stop sending data. When data is requested, the chip checks the corresponding sensor and sends its value together with a timestamp to the application. 

\subsubsection{VR}\label{section:impl:Hardware:VR}
\begin{figure}
    \begin{center}  
    \subfloat[Top View\label{fig:implementation:Hardware:VR:SimTop}]{
    \includegraphics[width=.3\linewidth]{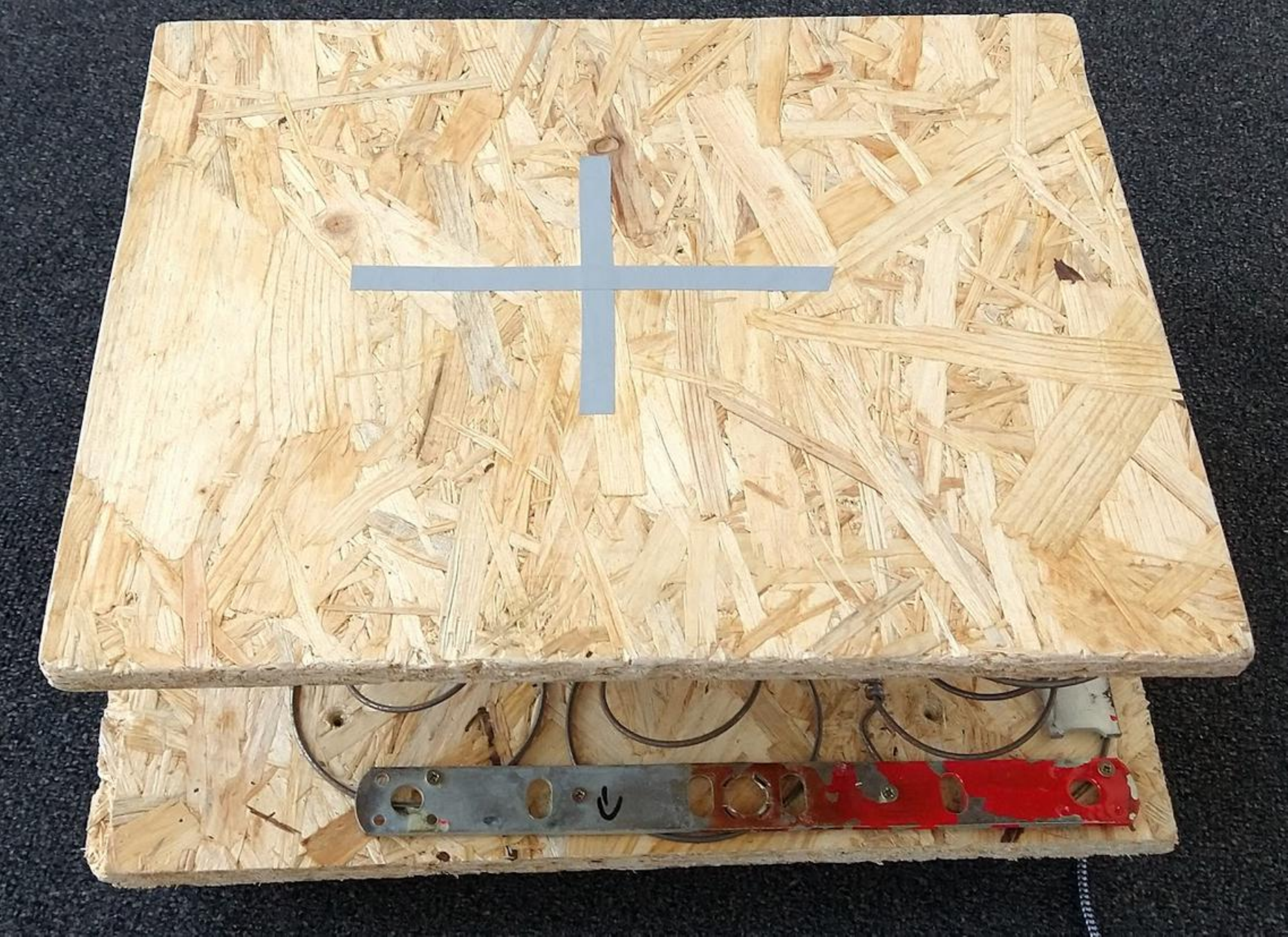}
    }\hspace*{3pt}
    \subfloat[Bottom View\label{fig:implementation:Hardware:VR:SimBottom}]{
    \includegraphics[width=.3\linewidth]{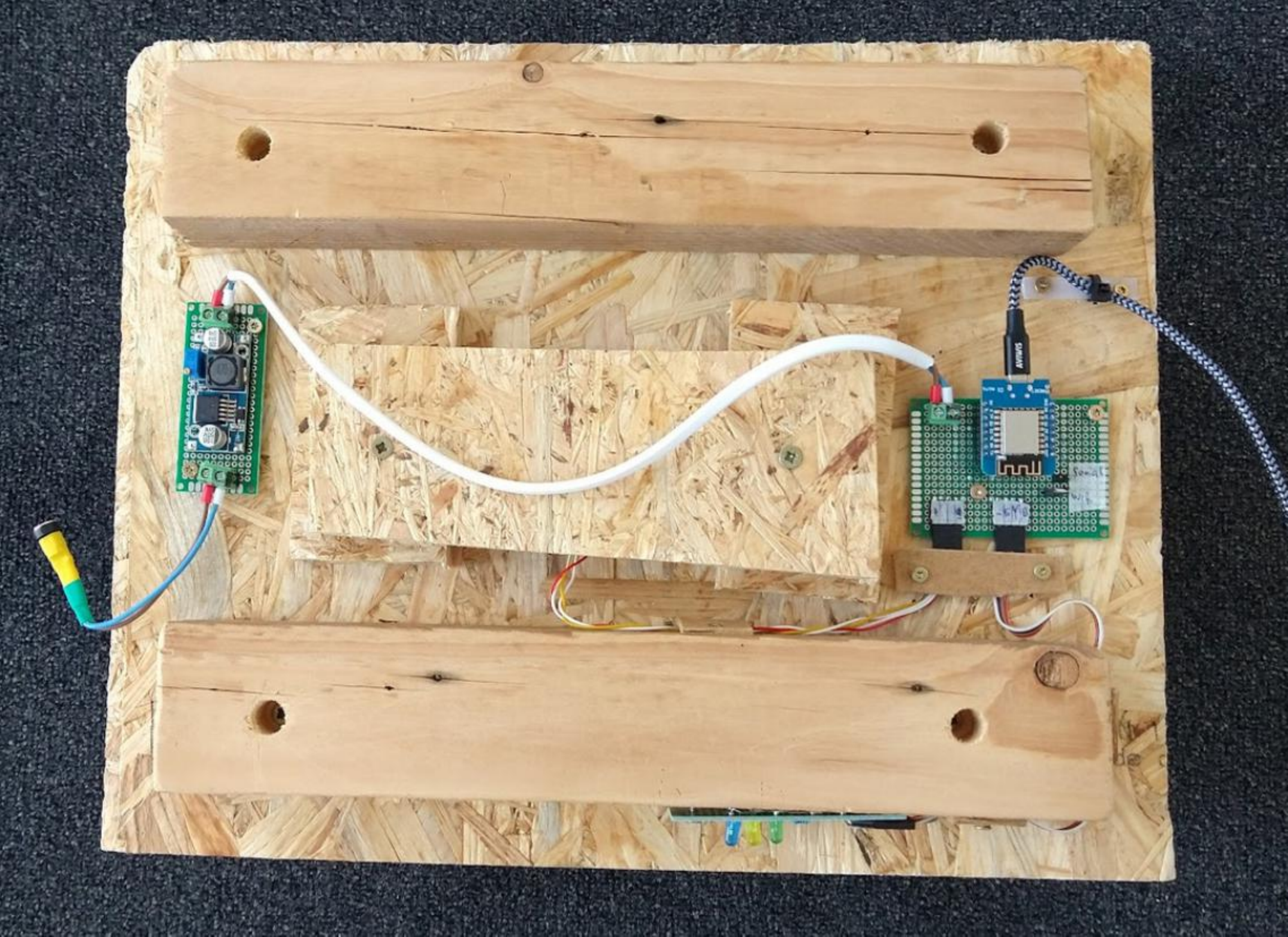}
    }\hspace*{3pt}
    \subfloat[Side View\label{fig:implementation:Hardware:VR:SimSide}]{
    \includegraphics[width=.3\linewidth]{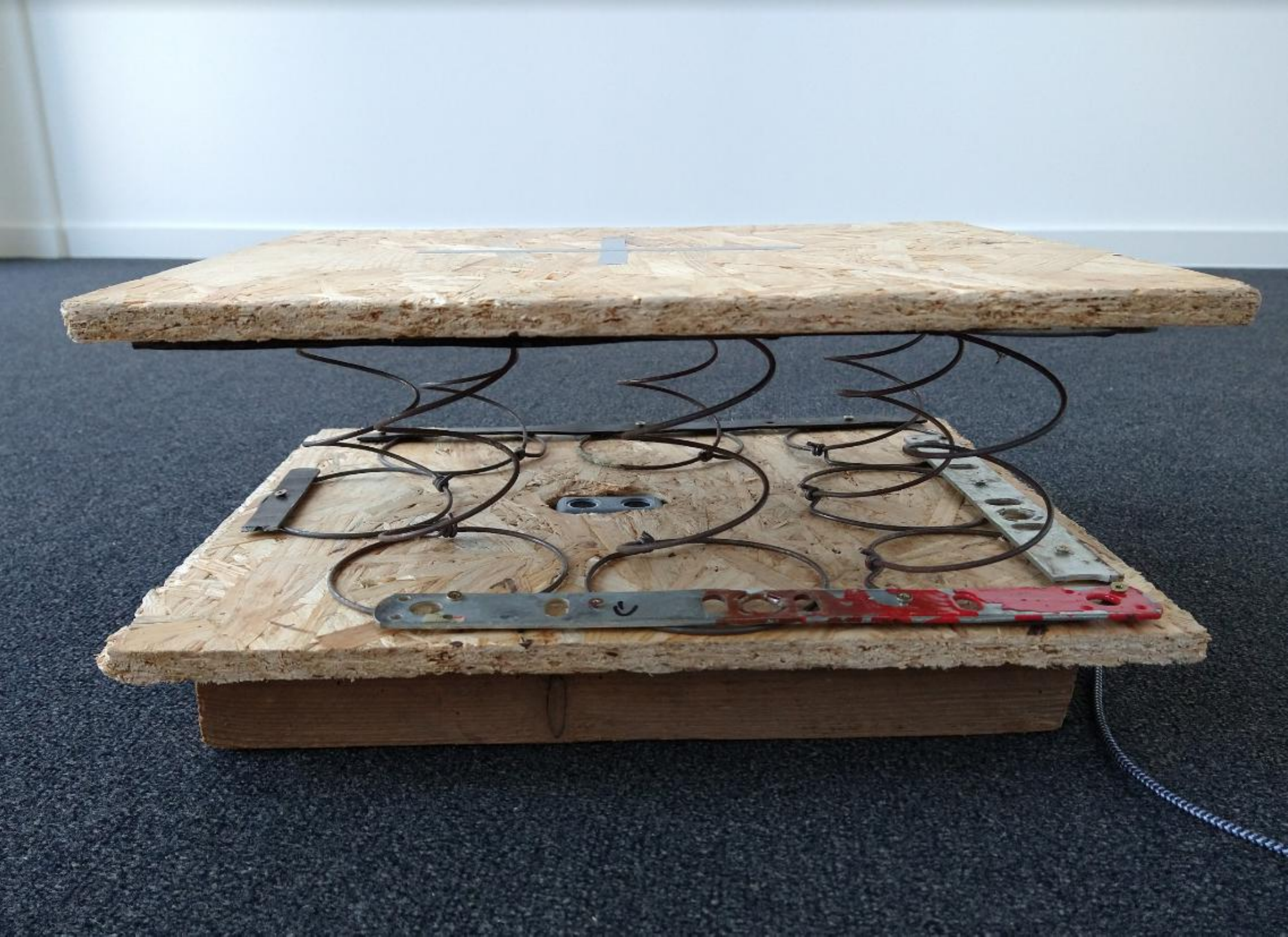}
    }
    \end{center}
    \caption{Custom Resuscitation Manikin}
    \label{fig:implementation:Hardware:VR:Sim}
\end{figure}

For the VR application, we do not need a realistic resuscitation manikin, as the user only sees the virtual world. Taking this into account and additionally to provide an independent solution for the VR side, we created a simpler CPR simulator dedicated to the VR-based BLS training environment. Figure \ref{fig:implementation:Hardware:VR:Sim} shows the simulator used for the VR application. Also, the sensor for the head's angle is not needed anymore as that is also tracked within the application. So we only need an ultrasonic sensor for the compressions and the \texttt{ESP-8266} wifi chip to communicate with the application. As the applications can control when and which sensor data is sent, we could use the same code as for the AR application. The VR application will not request data from the gyro sensor, so it does not matter if there is no sensor connected.

\subsection{Software}\label{section:impl:Software}
To develop our applications, we used the Unity\footnote{\url{https://unity.com/}} game engine. We developed two independent applications, one in AR and one in VR, which have basically the same functions but use different technologies to create them.

\subsubsection{AR}\label{section:impl:Software:AR}
The AR application was developed for the \textit{Microsoft Hololens 2}\footnote{\url{https://www.microsoft.com/en-us/hololens}}. 
We were using Unity version 2019.4.3f1 together with the Mixed Reality Toolkit (MRTK)\footnote{\url{https://github.com/Microsoft/MixedRealityToolkit-Unity}} version 2.4.0.
The MRTK offers a set of components and features that facilitate the development of mixed reality applications. For instance, via the \textit{InputSystem}\footnote{\url{https://docs.microsoft.com/en-us/windows/mixed-reality/mrtk-unity/features/input/overview?view=mrtkunity-2021-05}} provided by the MRTK, head, gaze, and gesture inputs that a user executes on HoloLens can be easily retrieved. This is necessary to facilitate the recognition of user actions which is required for the automated assessment capabilities of the training system. To achieve advanced tracking techniques beyond the scope of features offered by the MRTK and the HoloLens as such, Vuforia Engine\footnote{\url{https://developer.vuforia.com/}} version 9.4.6 has been utilized. Vuforia has been chosen as a technology because it offers great compatibility with Unity and provides the right amount of functionality even in its free version. Vuforia provides a set of advanced recognition features that allow recognition and tracking of real-world objects. For the training system developed
here, the most important feature offered by Vuforia is the recognition of image targets. Image targets are based on flat images, for instance, a simple QR code, that Vuforia can detect in the real environment. Image targets are created via a web interface in the Vuforia Engine Developer Portal and are stored locally within a device target database that can be imported into a Unity project. The image targets can then be added to a scene within Unity. We used image tracking to synchronize the virtual with the real world. To achieve this, an image marker is placed on the belly of the manikin. When the application starts, that image needs to be scanned by looking at it from a short distance while wearing the HoloLens. After the position is set, we can remove the image marker from the manikin so it does not distract the users. 

We developed the application based on the concept described in Section \ref{section:solOverview:Software} and by realizing the architecture depicted in Figure \ref{fig:solOverview:ApplicationOverview}.

\paragraph{User}
The User wears the Hololens 2 with which they interact with the training application. The application itself consists of several components. First, there are the \textit{Building Blocks} that provide the functionality to interact with the user.

\begin{figure}
    \centering
    \includegraphics[width=.9\linewidth]{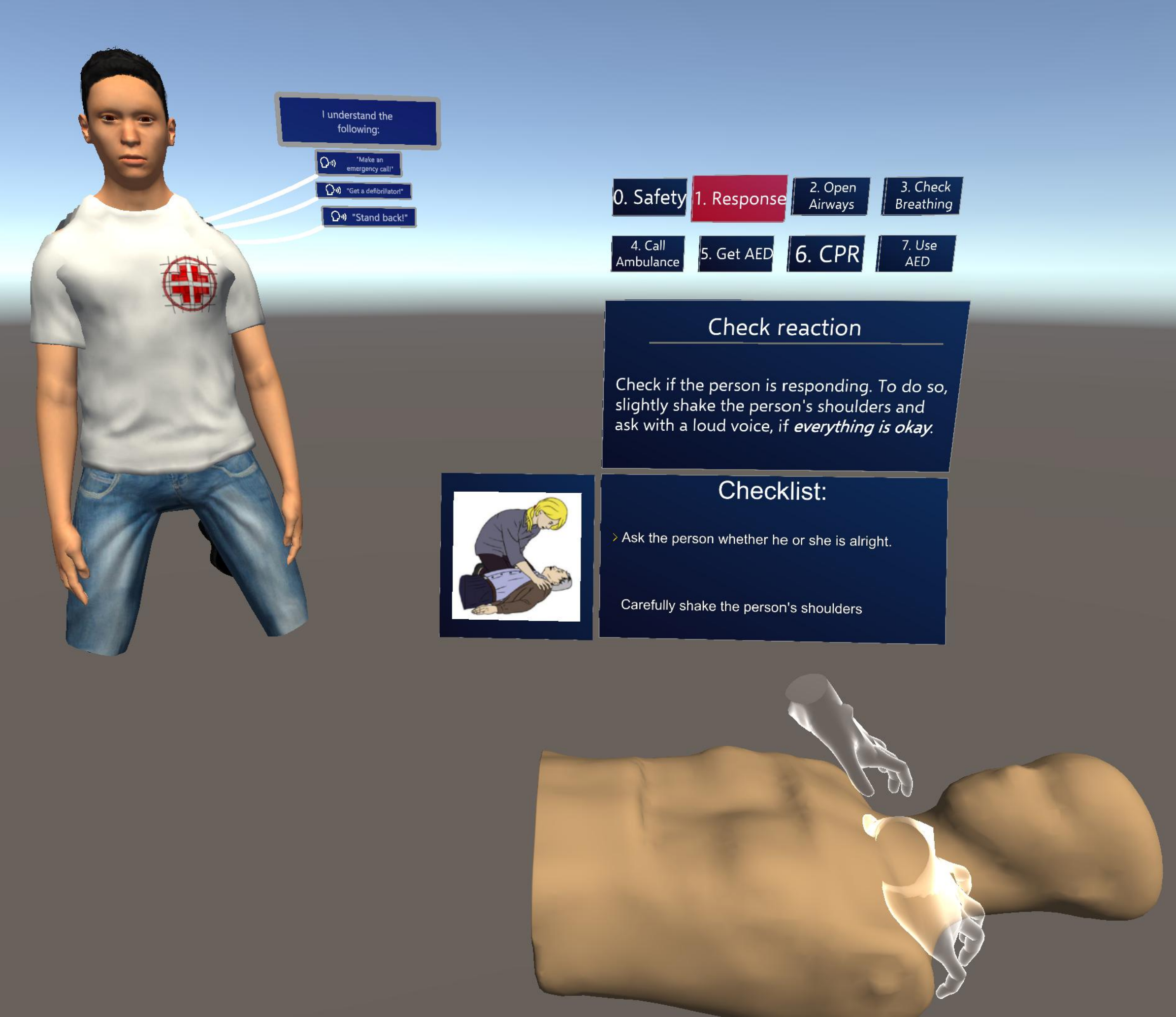}
    \caption{AR Scene as Displayed to the User}
    \label{fig:implementation:Software:AR:AROverview}
\end{figure}

\paragraph{Text/Audio}
To convey content to the user, we implemented different elements.
To provide a \textit{Text} to the user, we created simple description panels. In this dashboard, we can display arbitrary text using the \textit{TextMeshPro}\footnote{\url{https://docs.unity3d.com/Manual/com.unity.textmeshpro.html}} package. Next to the text, there is a place to additionally integrate images as 2D \textit{sprites}\footnote{\url{https://docs.unity3d.com/Manual/Sprites.html}}. The text content and the image can be changed by a script. Additionally, we provide a checklist that was created based on the tasks we defined for the user. It shows which tasks to do and which tasks are already done. 
To provide the user a better overview of the current, previous, and next steps in the BLS sequence, we show a breadcrumb-like overview of the tasks during the whole procedure.
The dashboards and breadcrumbs can be seen in the top right corner of the scene displayed in Figure \ref{fig:implementation:Software:AR:AROverview}.

As a next step, we developed a system to read the displayed text as \textit{audio} to the user. We decided to record the texts in high quality rather than using a text-to-speech system to increase the immersion and better simulate real-life training. Inside Unity, audio files are used as \textit{AudioClips}\footnote{\url{https://docs.unity3d.com/ScriptReference/AudioClip.html}}. The audio tracks are played when the corresponding task starts and the displayed text is updated so that it fits with the recording.

\paragraph{Animations}
To show the user \textit{Animations} about what to do, we first needed a possibility to create such animations. As it is very time-consuming to create all animations by hand, we exploited the HoloLens' head, hand, and finger tracking capabilities in order to record the movement of the head, the hands, and the fingers by saving their position and rotation in fixed time intervals. We then converted these recordings into Unity \textit{animations}\footnote{\url{https://docs.unity3d.com/Manual/AnimationSection.html}} where we used the recorded keyframes to create smooth animations. The animations are then mapped on head and hand meshes. In the bottom right corner of Figure \ref{fig:implementation:Software:AR:AROverview}, a screenshot of an animation for shaking the victim's shoulders is shown.

\paragraph{(Interactable) Objects}

\begin{figure}
    \centering
    \includegraphics[width=.9\linewidth]{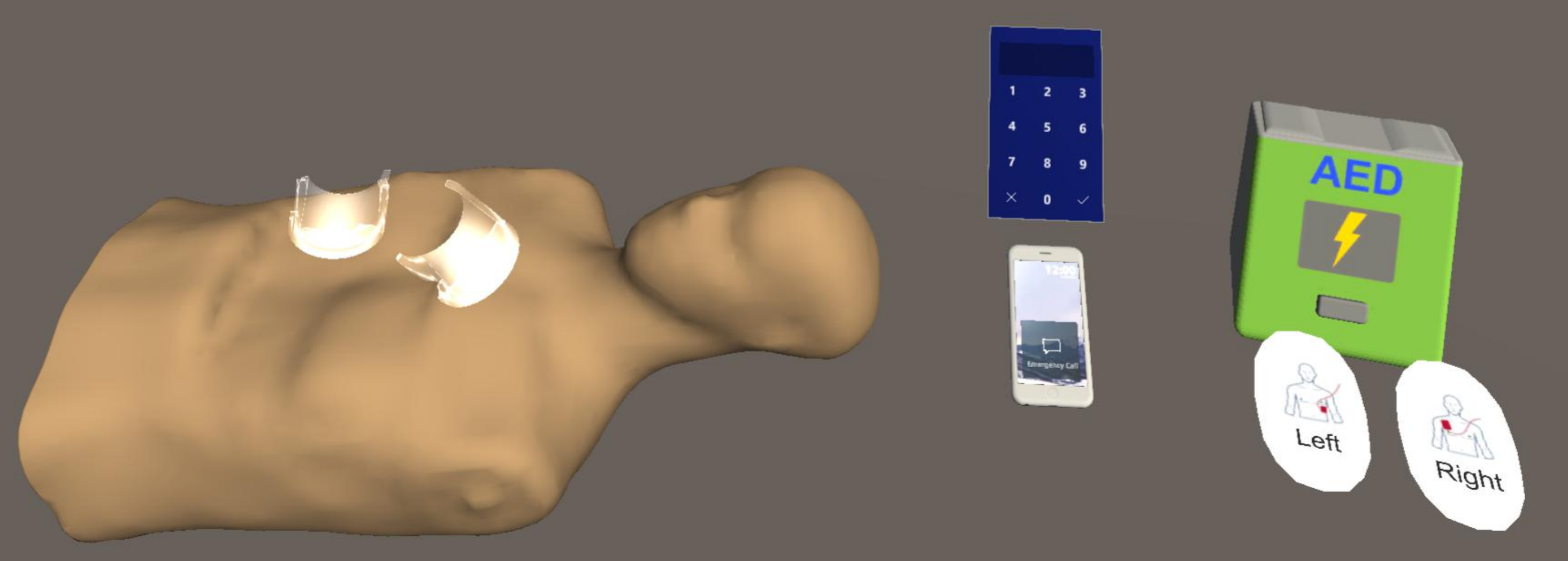}
    \caption{Interactable Objects}
    \label{fig:Implementation:Software:AR:InteractableObjectsOverview}
\end{figure}

To simulate by-standing helpers, we used a simple human model, as can be seen on the left side of Figure \ref{fig:implementation:Software:AR:AROverview}. The bystander is used to encourage the use of voice commands.
Additionally, there are multiple virtual objects, with which a user can interact directly. They are shown in Figure \ref{fig:Implementation:Software:AR:InteractableObjectsOverview}.
First, we added a broken glass that needs to be removed at the start of the training. The glass needs to be placed in a nearby dustbin and disappears when being dropped there. To allow the user to grab a 3D mesh, we use MRTK's \textit{NearInteractionGrabbable}\footnote{\url{https://docs.microsoft.com/en-us/dotnet/api/microsoft.mixedreality.toolkit.input.nearinteractiongrabbable?view=mixed-reality-toolkit-unity-2020-dotnet-2.7.0}} component by adding the corresponding script to the mesh. 
Next, we needed a phone to call the emergency service. The phone is grabbable and interactive. When pressing the message button, a number pad shows up, where the user needs to enter the emergency number. For that, we used and modified the \textit{pressable buttons}\footnote{\url{https://docs.microsoft.com/en-us/windows/mixed-reality/mrtk-unity/features/ux-building-blocks/button?view=mrtkunity-2021-05}} provided by the MRTK.
Finally, we created a defibrillator (AED) with its electrode pads. The pads can be grabbed and moved. When the pads are moved closer to their correct location on the manikin's chest, they snap to the right position. The button on the AED can be pressed to confirm delivering the shock.

\paragraph{Voice Commands}
As we want to train the communication with real bystanders, we needed to implement voice commands. For that, we used MRTK's \textit{SpeechInputHandler}\footnote{\url{https://docs.microsoft.com/en-us/dotnet/api/microsoft.mixedreality.toolkit.input.speechinputhandler?view=mixed-reality-toolkit-unity-2020-dotnet-2.7.0}} component, where we can define which keyphrases can be recognized, and what happens as a response when they are recognized. By utilizing the \textit{SpeechInputHandler} component, voice commands are implemented as static keyphrases, which turned out to be the most reliable solution for voice recognition in the current version of the HoloLens. To guide the user to which keyphrases are available, we added hints to the AR environment using MRTK's \textit{tooltip}\footnote{\url{https://docs.microsoft.com/en-us/windows/mixed-reality/mrtk-unity/features/ux-building-blocks/tooltip?view=mrtkunity-2021-05}} component (see Fig \ref{fig:implementation:Software:AR:AROverview}).

\paragraph{Manikin}
In Section \ref{section:impl:Hardware:AR}, we already described how we created the resuscitation manikin. To give real-time feedback for the chest compressions in the application, there are two elements to consider. First, we have the compression rate, and second the compression depth. The manikin measures how far the chest is pushed multiple times every second and sends those values with a timestamp to the application.

\begin{figure}
    \begin{center}  
    \subfloat[Sensor Data of User Push\label{fig:implementation:Software:AR:UserPushData}]{
    \includegraphics[width=.45\linewidth]{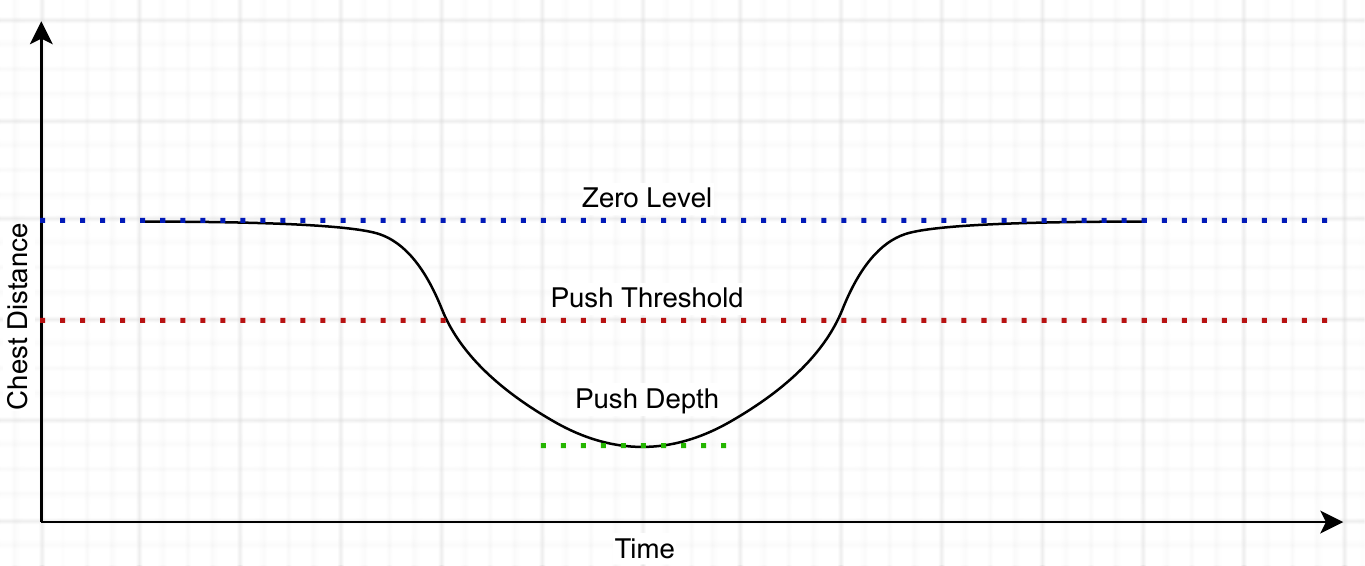}
    }\hspace*{3pt}
    \subfloat[Compression Display\label{fig:implementation:Software:AR:CompressionDisplay}]{
    \includegraphics[width=.45\linewidth]{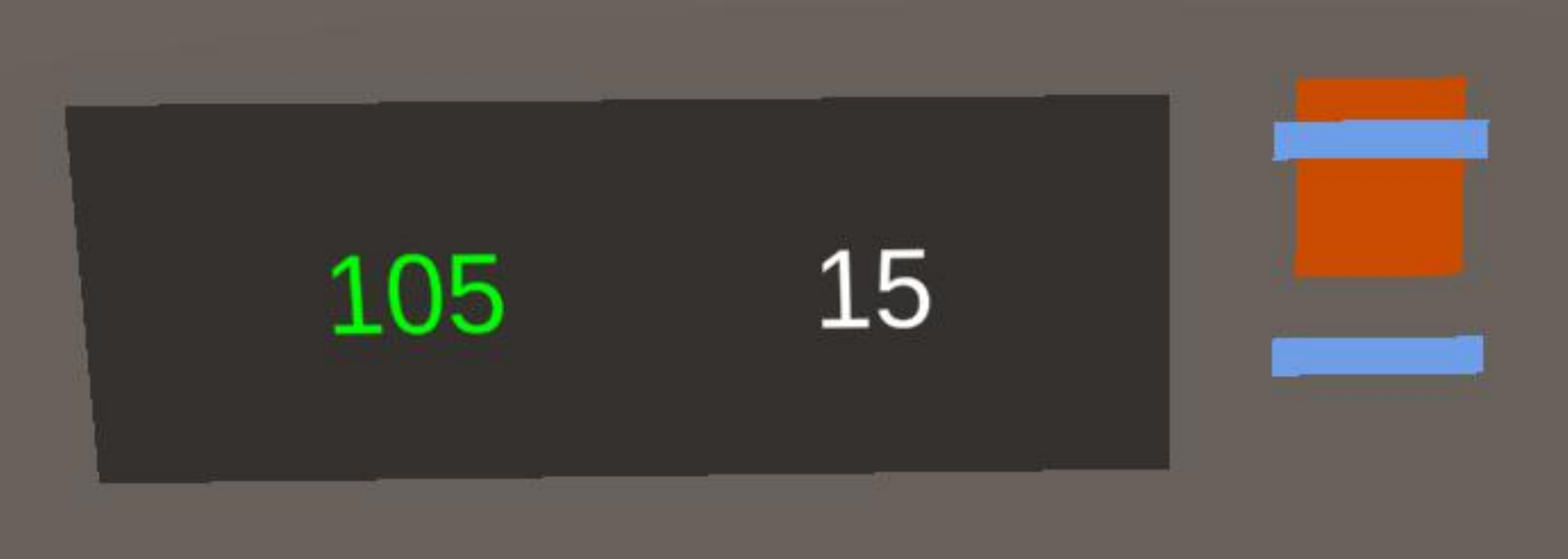}
    }
    \end{center}
    \caption{Using Sensor Data to Evaluate Compressions}
    \label{fig:implementation:Software:AR:Compressions}
\end{figure}

Figure \ref{fig:implementation:Software:AR:UserPushData} shows how we use the incoming data. The black line is the measured distance from the top of the chest to the ground. When the application starts, we measure the default distance when the chest is not pushed. That value is set as the zero level. To find out whether the chest was pushed, we check if the distance is lower than a fixed threshold (in our case 3cm). To calculate the compression rate, we consider the moment when the threshold is reached. A push is finished, when the depth gets again higher than the threshold. The depth for a push is calculated from the lowest distance measured during a push. Therefore, we subtract the lowest distance from the zero level to get how deep the user pushed the chest. As a measure for the compression rate, we calculate the compressions per minute. Therefore, we measure the time between two pushes and calculate how many pushes per minute that would be.
To give feedback to the user, we provide three types of information. Figure \ref{fig:implementation:Software:AR:CompressionDisplay} shows the display we provide to the user. On the left side, we show the current compression rate (averaged over four pushes so the number changes smoothly). In the middle, we show the number of compressions already performed. On the right, we display the current compression depth by using Unity's \textit{slider}\footnote{\url{https://docs.unity3d.com/2018.3/Documentation/ScriptReference/UI.Slider.html}}. The red bar indicates how deep the chest is pushed. The lower blue bar indicates the desired depth and the top blue bar indicates the zero level. With that, we can provide real-time feedback for the compression. We also log the compression rate and depth values to analyze the user's performance.

\begin{figure}
    \centering
    \includegraphics[width=.5\linewidth]{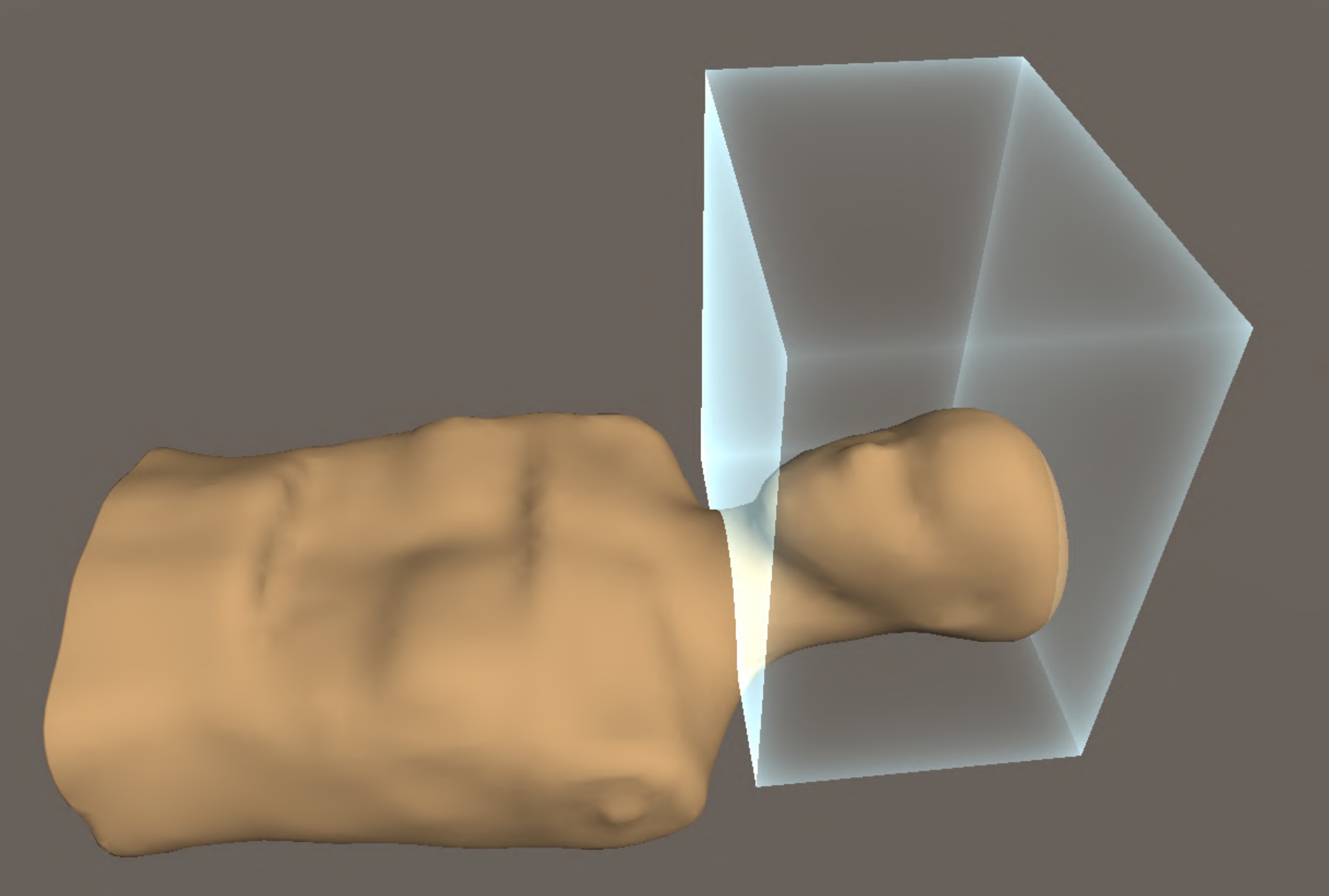}
    \caption{3D Volume for Position Tracking}
    \label{fig:implementation:Software:AR:3DVolume}
\end{figure}

\paragraph{Position Trigger}
We needed to track the position of the user's head and hands. For that, we utilize the GazeProvider component and the hand joint utilities provided by the MRTK's Input System\footnote{\url{https://docs.microsoft.com/en-us/windows/mixed-reality/mrtk-unity/features/input/gaze?view=mrtkunity-2021-05}}. We create a 3D volume like the one shown in Figure \ref{fig:implementation:Software:AR:3DVolume}, allowing us to respond to events such as when users place their hands on the manikin's head during airway clearance.

\paragraph{Task Management}

\begin{figure}
    \centering
    \includegraphics[width=.8\linewidth]{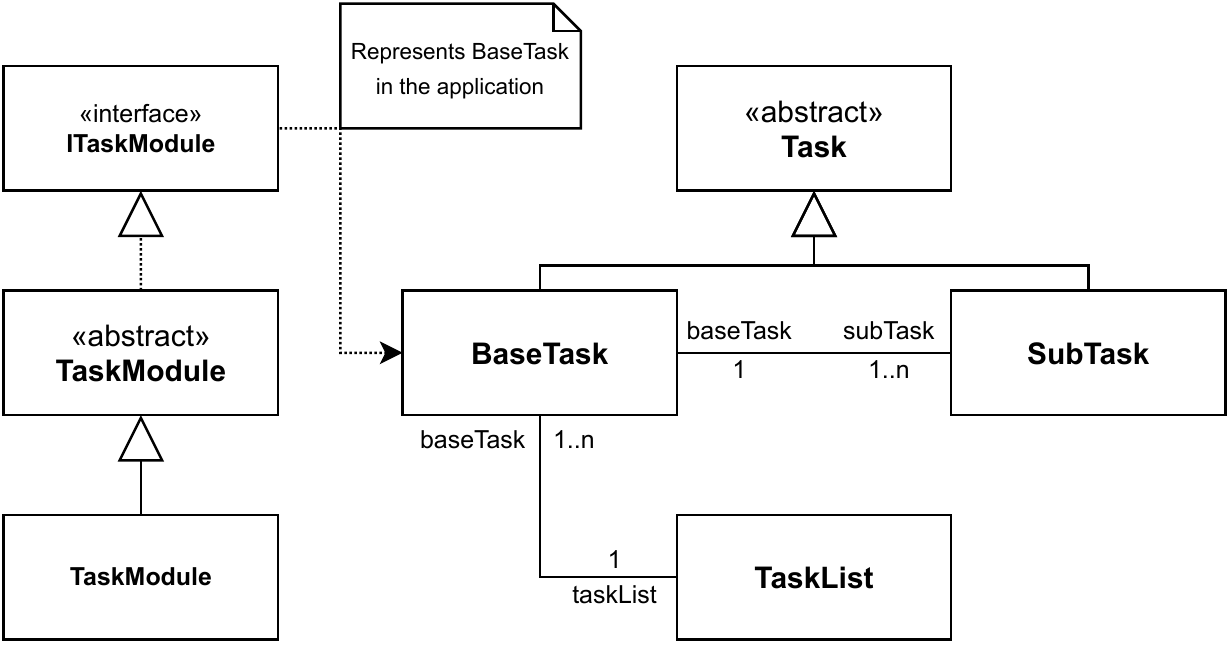}
    \caption{Task Management Classes}
    \label{fig:implementation:Software:AR:TaskSystem}
\end{figure}

The core workflow of a training session is defined by the task management. Every step of the training procedure is defined as a \textit{Task}. \textit{BaseTasks} are complex tasks, i.e. they are composited of one or more \textit{SubTasks} (Figure \ref{fig:implementation:Software:AR:TaskSystem}). A subtask, in contrast, is an atomic action that the user of the system performs, i.e. it is not further composited of more refined subtasks. As an example, utilization of the AED is one step of the BLS/AED sequence, thus it is defined as a base task within the system. This step consists of several subtasks, such as placing the electrode pads of the AED in the correct locations on the victim’s body, ensuring that nobody touches the victim, and delivering a shock.

Each base task holds a list with its associated subtasks. The base tasks, in turn, are arranged within a \textit{Tasklist}. Hence, the tasklist describes the sequence of steps of the final training procedure.
To connect the tasks to elements of the application, each base task is assigned to an individual \textit{TaskModule}. A task module is an encapsulated entity responsible for the assessment of exactly one base task. It gets informed when the user did everything to complete the (sub) task and triggers the loading of the next task.
As the requirements for each task module can vary, we created a generic task module implementation such that any concrete implementations can inherit and, if necessary, override its basic functionality (Figure \ref{fig:implementation:Software:AR:TaskSystem}).

\subsubsection{VR}\label{section:impl:Software:VR}
\begin{figure}
    \centering
    \includegraphics[width=.8\linewidth]{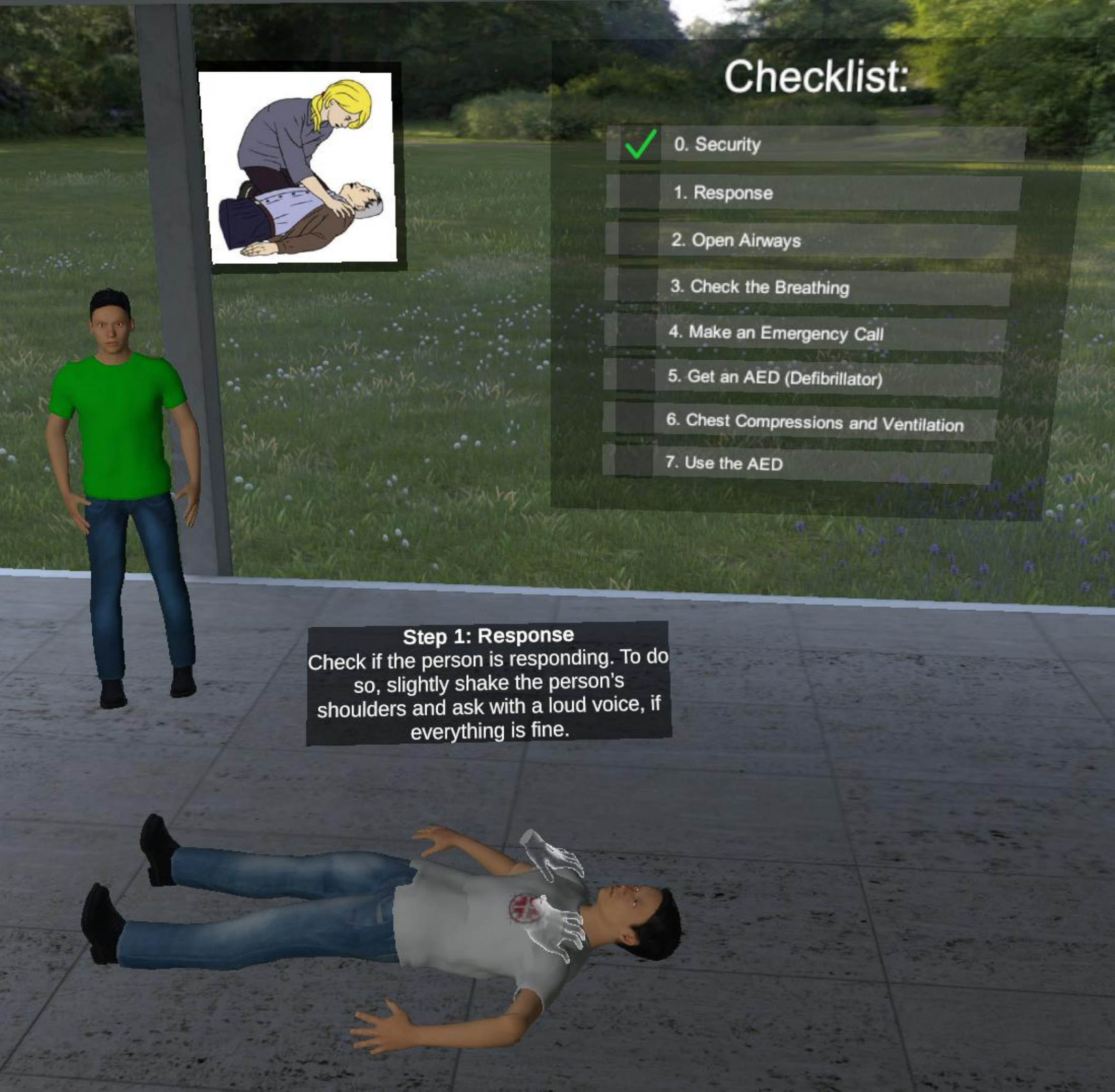}
    \caption{VR Scene as Displayed to the User}
    \label{fig:implementation:Software:VR:VRSceneOverview}
\end{figure}

The VR application was built with Unity 2019.1.7f1 in combination with SteamVR\footnote{\url{https://valvesoftware.github.io/steamvr_unity_plugin/}} version 2.3.2. SteamVR offers a variety of functionality to support VR development. It, for example, provides some basic components like a preconfigured VR camera or objects representing the controller movement. Additionally, applications build with SteamVR work on different VR Headsets. For our development, we used the Valve Index\footnote{\url{https://www.valvesoftware.com/en/index}} VR headset. Whereas we use a real-life manikin as the victim in the AR application, we need a virtual victim for the VR variant. For that, we use a 3D model of a human lying on the ground which can be seen at the bottom in Figure \ref{fig:implementation:Software:VR:VRSceneOverview}. To increase realism, the model uses physics and can be grabbed and moved by the user. As, during CPR, the user has to push the victim's chest, the model needs to be aligned with the real world so when the user pushes on the simulator, it looks for them as they push the chest. So we decided to attach the model to specific points on the ground so that the user can interact with the victim, but it stays at roughly the same position. 

\paragraph{User}
To interact with the application, the user wears the Valve Index VR headset and its corresponding controllers.
Further, we also created all needed components as described in Section \ref{section:solOverview:Software} and depicted in Figure \ref{fig:solOverview:ApplicationOverview}.

\paragraph{Text/Audio}

To display \textit{Text} in the VR application, a description panel is placed near the ground behind the victim (Figure \ref{fig:implementation:Software:VR:VRSceneOverview}), in the middle. We use the \textit{TextMeshPro}\footnote{\url{https://docs.unity3d.com/Manual/com.unity.textmeshpro.html}} package, where arbitrary text can be inserted via script. The checklist in the background shows which big tasks are already completed. Next to the checklist, images as 2D \textit{sprites}\footnote{\url{https://docs.unity3d.com/Manual/Sprites.html}} can be displayed.
The \textit{Audio} descriptions are also integrated into the VR application. For that, we could use the same audio files we recorded for the AR version.

\paragraph{Animations}
Similarly, for the \textit{Animations} within the VR application, we reused the Unity animations we recorded for the AR variant. At the bottom of Figure \ref{fig:implementation:Software:VR:VRSceneOverview}, there is a screenshot of the shoulder shaking animation.

\begin{figure}
    \centering
    \includegraphics[width=.8\linewidth]{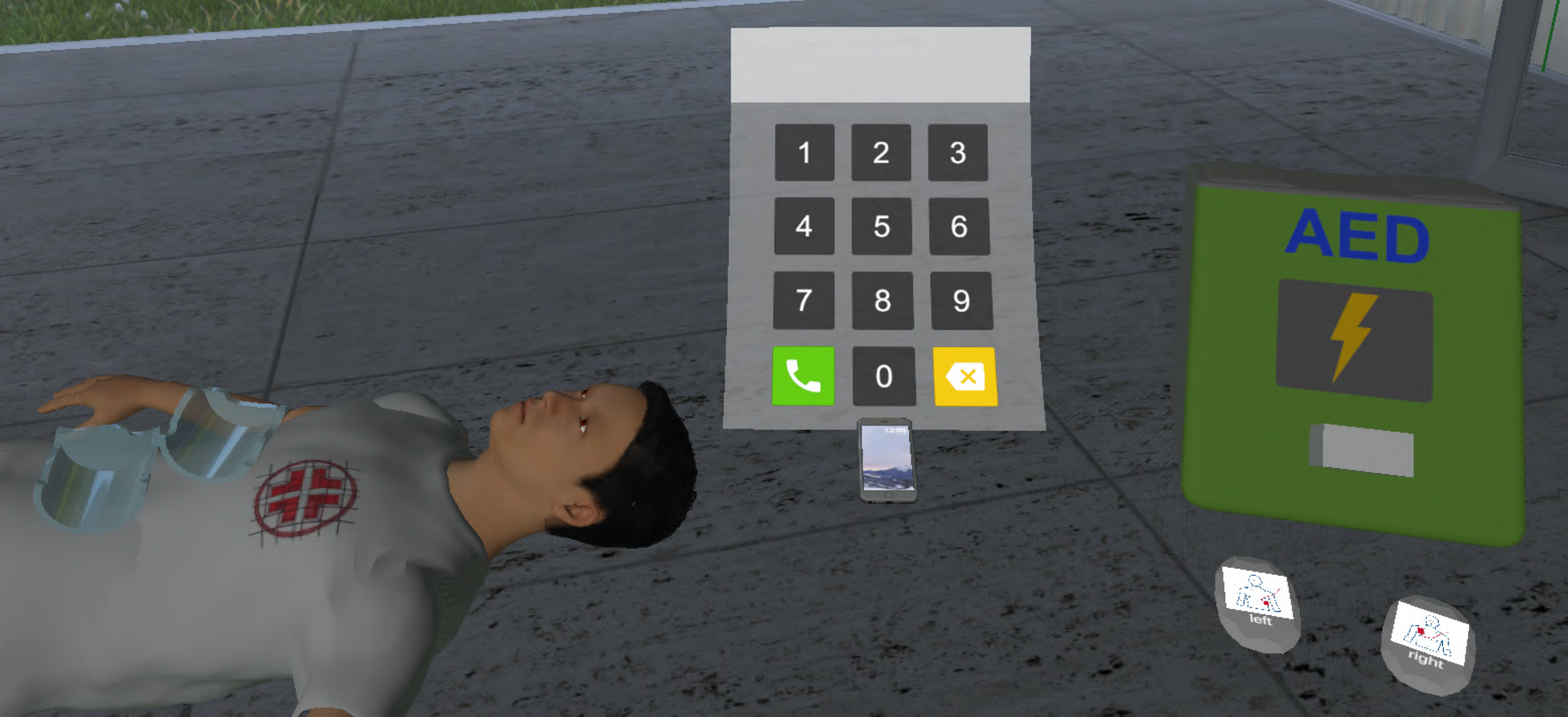}
    \caption{Interactable Objects}
    \label{fig:Software:VR:InteractableObejctsOverview}
\end{figure}

\paragraph{(Interactable) Objects}
The bystander in the VR application can be seen on the left side in Figure \ref{fig:implementation:Software:VR:VRSceneOverview}. It is a simple human model that can be interacted with via voice commands.
The objects with which the user can directly interact are shown in Figure \ref{fig:Software:VR:InteractableObejctsOverview}.
We created broken glasses the users need to put away at the beginning. They are shown on the left. We used the same base model as for the AR application but needed to make small adjustments to the model and create a new material to make it look good in the VR application. To allow the user to grab and move the glass, we use SteamVR's \textit{Throwable}\footnote{\url{https://valvesoftware.github.io/steamvr_unity_plugin/api/Valve.VR.InteractionSystem.Throwable.html}} script and added physics to the models. 
The phone's model is the same as in the AR application, but we had to recreate the number pad. The phone itself is also a \textit{Throwable}, but when the user grabs it, we display a number pad to dial the number. Using the \textit{HoverButton}\footnote{\url{https://valvesoftware.github.io/steamvr_unity_plugin/api/Valve.VR.InteractionSystem.HoverButton.html}} from SteamVR's input system for each number, the user can input the number by pushing them. The phone is shown in the middle of Figure \ref{fig:Software:VR:InteractableObejctsOverview}.
Finally, we needed the AED. Here, we could use the same models as we did for the AR application. We only needed to change the interaction. The pads became \textit{Throwables} with a physics simulation and the button became a \textit{hover button}.

\begin{figure}
\begin{minipage}{.4\linewidth}
    \centering
    \includegraphics[width=.9\linewidth]{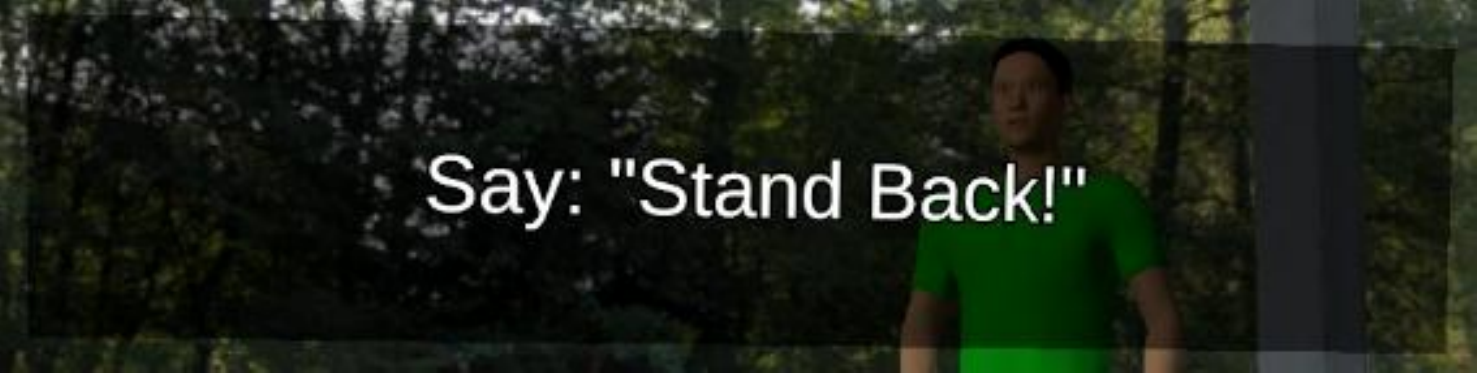}
    \caption{Voice Command Hint}
    \label{fig:implementation:Software:VR:VoiceCommandHint}
\end{minipage}
\begin{minipage}{.4\linewidth}
    \centering
    \includegraphics[width=.9\linewidth]{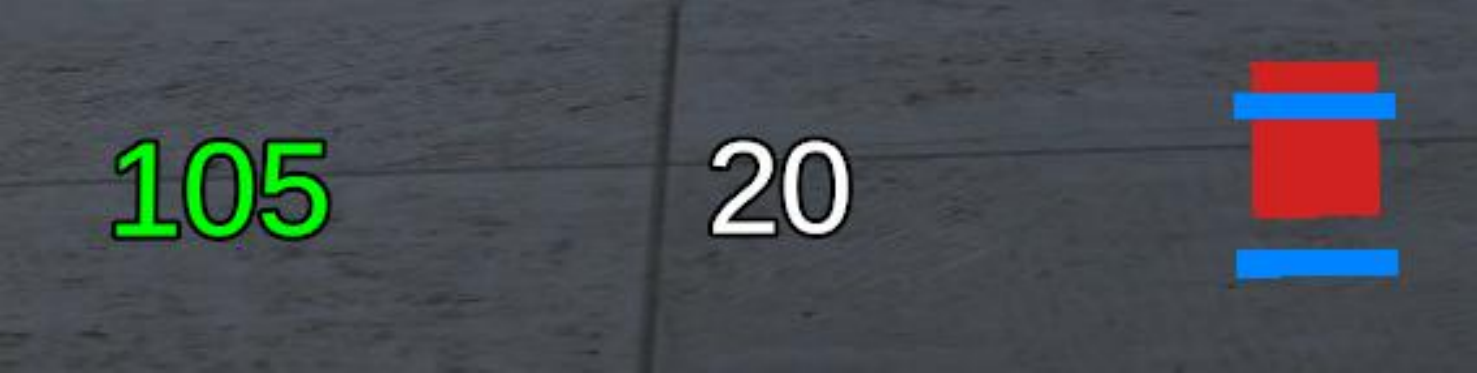}
    \caption{Sensor Data of User Push}
    \label{fig:implementation:Software:VR:UserPushData}
\end{minipage}
\end{figure}

\paragraph{Voice Commands}
To enable voice commands, we used Unity's \textit{PhraseRecognizer}\footnote{\url{https://docs.unity3d.com/ScriptReference/Windows.Speech.PhraseRecognizer.html}} class. With that, we can define which key phrases should be recognized and get notified when they were said. To show the user the keyphrase to say, we add a hint like the one shown in Figure \ref{fig:implementation:Software:VR:VoiceCommandHint} when needed.

\paragraph{Manikin/CPR Simulator}
We already presented the CPR simulator we used in the VR application in Section \ref{section:impl:Hardware:VR}.
To give real-time feedback to the user while performing compressions, we could reuse the components of the AR application (Figure \ref{fig:implementation:Software:VR:UserPushData}). Two text fields show the rate and how many compressions were done, and the indicator on the right shows the depth.

\begin{figure}
\begin{center}
    \subfloat[3D Volume for Head Position Tracking\label{fig:implementation:Software:VR:3DVolume}]{\includegraphics[height=120pt]{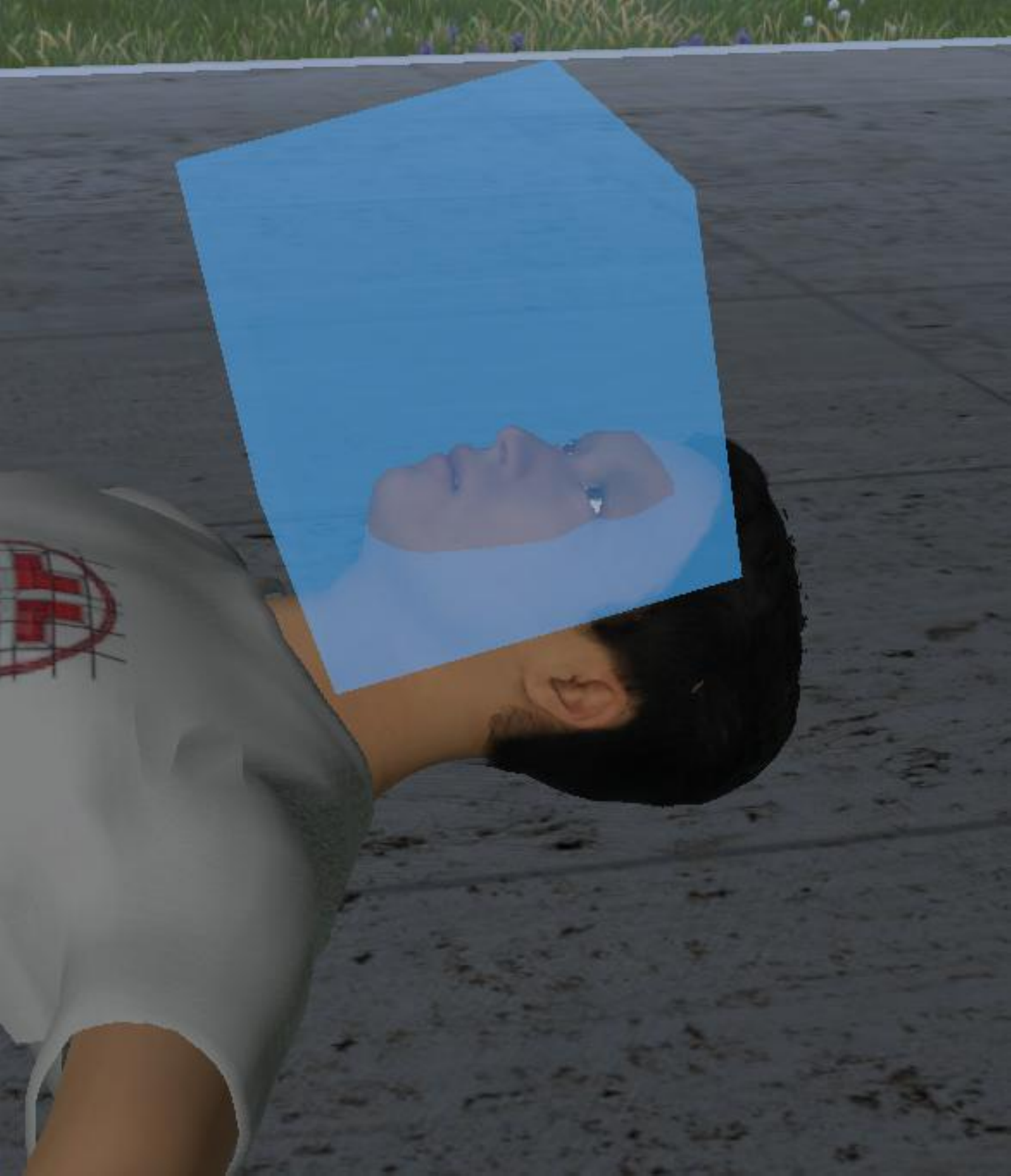}}
    \hspace*{30pt}
    \subfloat[3D Volume for Hand Position Tracking\label{fig:implementation:Software:VR:HandTracking}]{\includegraphics[height=120pt]{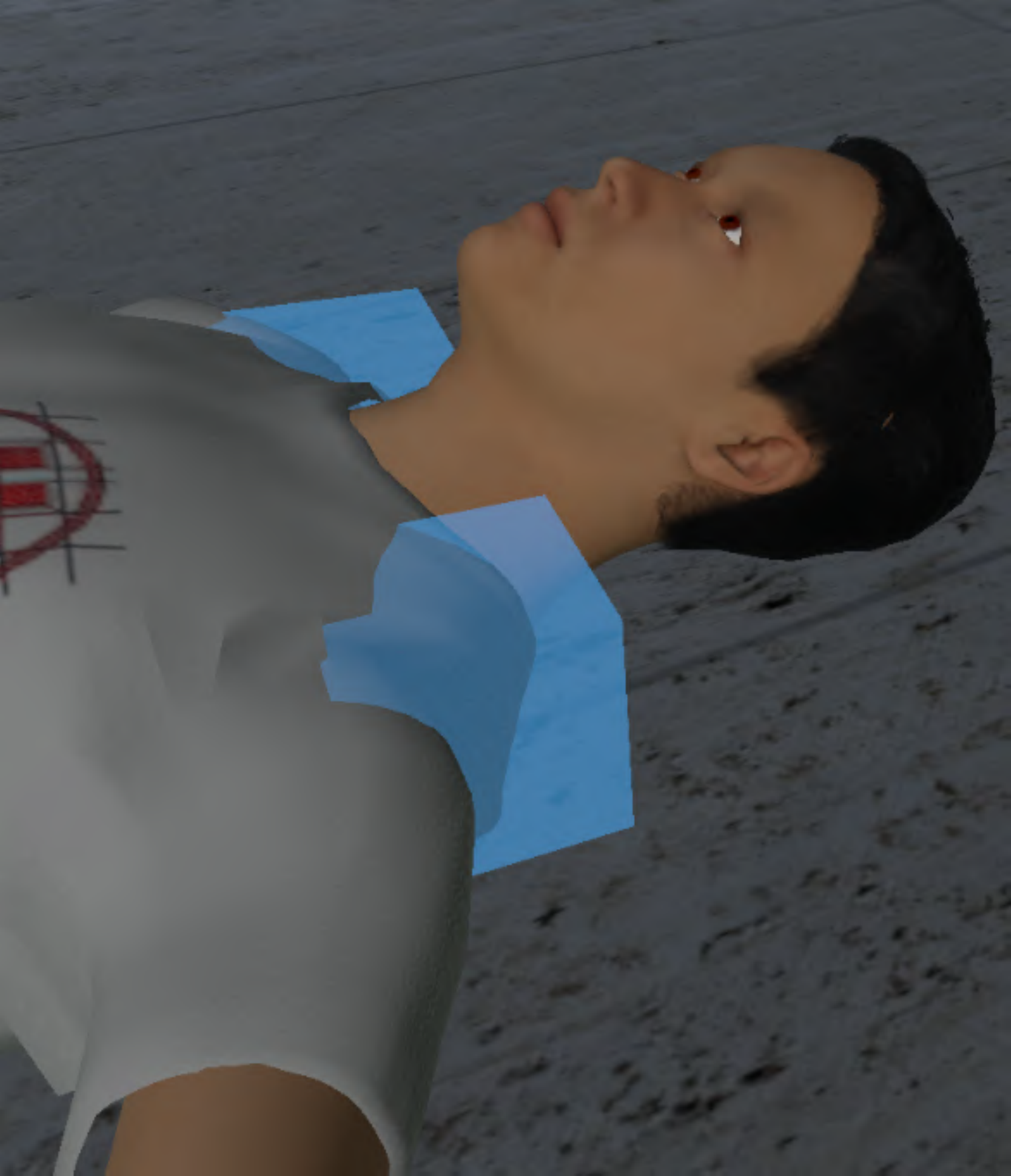}}
    \end{center}
    \caption{Position Tracking}
    \label{fig:implementation:Software:VR:PositionTracking}
\end{figure}

\paragraph{Position Trigger}
To track the user's head position, we created simple 3D volumes (Figure \ref{fig:implementation:Software:VR:3DVolume}) which can detect whenever an object enters it. Then, we can check whether the object was the user's head and trigger the next steps. That is similar to how the position trigger worked in AR. But for the hands, we used a different concept. In AR, we could only track a hand's position, but we could not track whether the user grabs e.g. the victim's shoulder. However, in VR, based on the Valve Index hand controllers we can track all interactions being made with the environment. So when the user needs to interact with the victim (e.g. shake shoulder, see Figure \ref{fig:implementation:Software:VR:HandTracking}) we can place small 3D volumes there and add SteamVR's \textit{Throwable} script which allows us to get notified when the user grabs/releases it. 

\paragraph{Task Management} 
As the AR and VR applications only differ in the way they interact with the user while the underlying application stays the same, we used the same task system for the AR as well as the VR application.

\subsection{Training Walkthrough}\label{section:impl:Software:Walkthrough}

\begin{figure}
    \begin{center}  
    \subfloat[Scene Safety\label{fig:implementation:Software:ARWalktrough:0}]{
    \includegraphics[width=.3\linewidth]{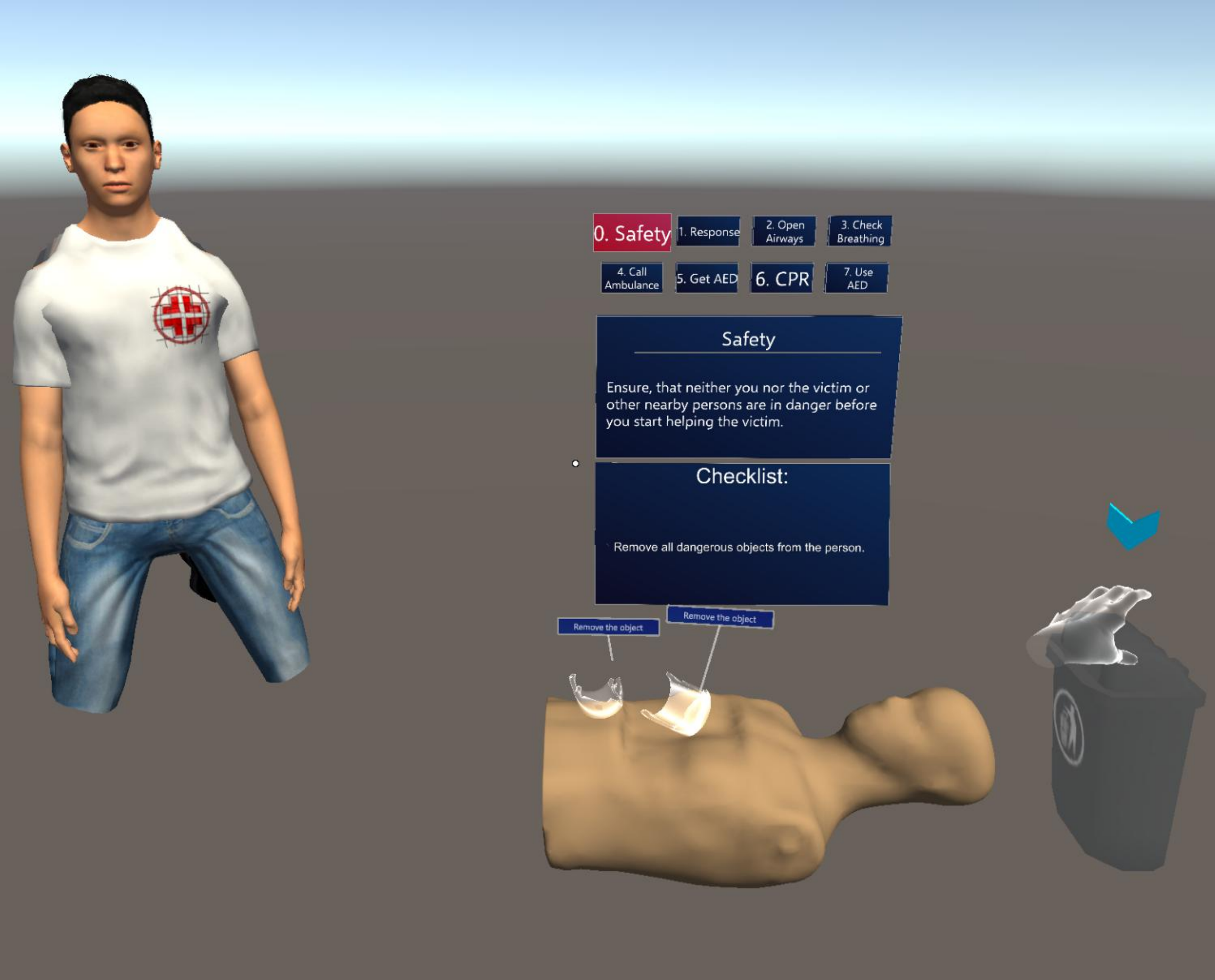}
    }\hspace*{3pt}
    \subfloat[Check Response\label{fig:implementation:Software:ARWalktrough:1}]{
    \includegraphics[width=.3\linewidth]{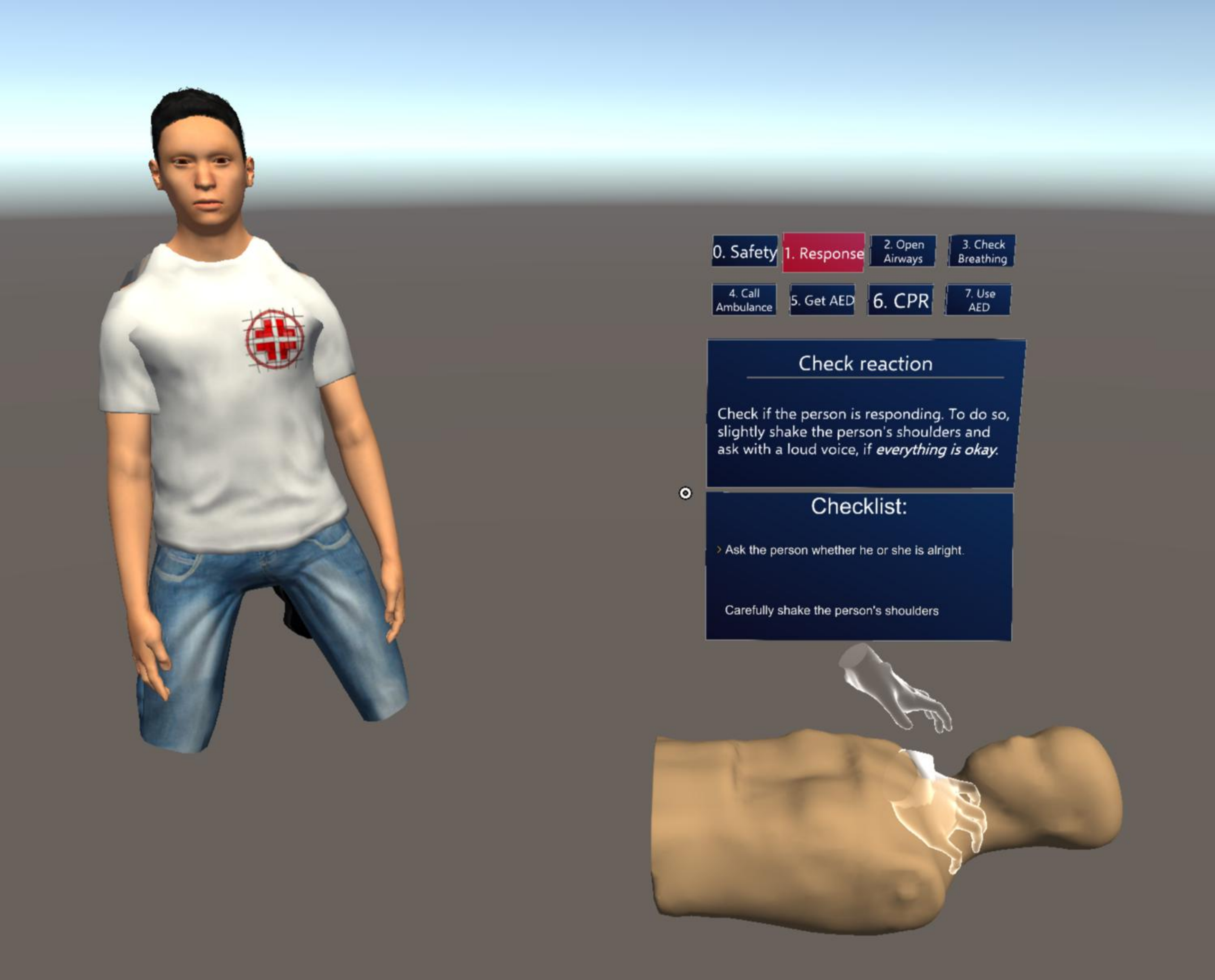}
    }\hspace*{3pt}
    \subfloat[Open Airways\label{fig:implementation:Software:ARWalktrough:2}]{
    \includegraphics[width=.3\linewidth]{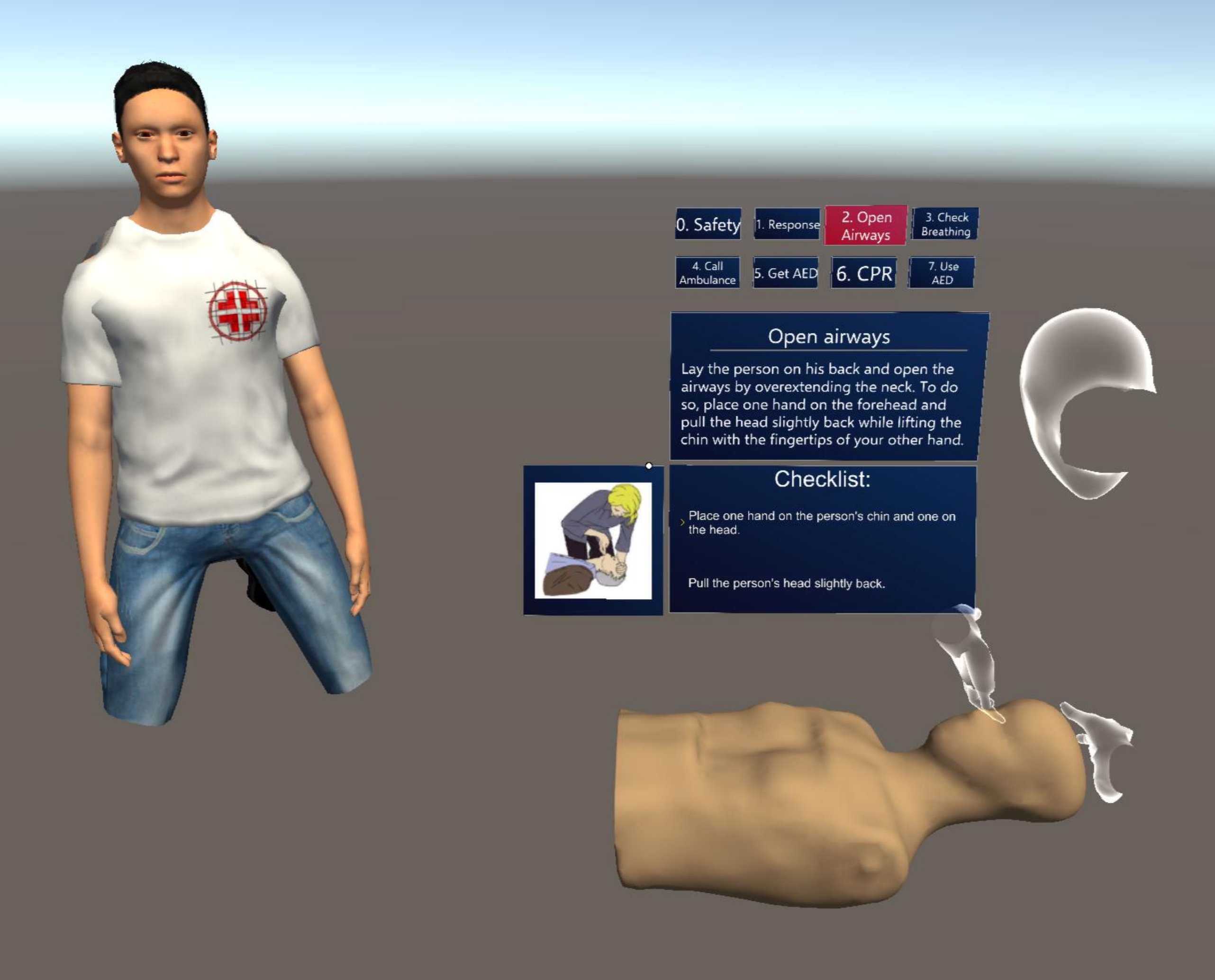}
    }
    \end{center}
    
    \begin{center}  
    \subfloat[Check Breathing\label{fig:implementation:Software:ARWalktrough:3}]{
    \includegraphics[width=.3\linewidth]{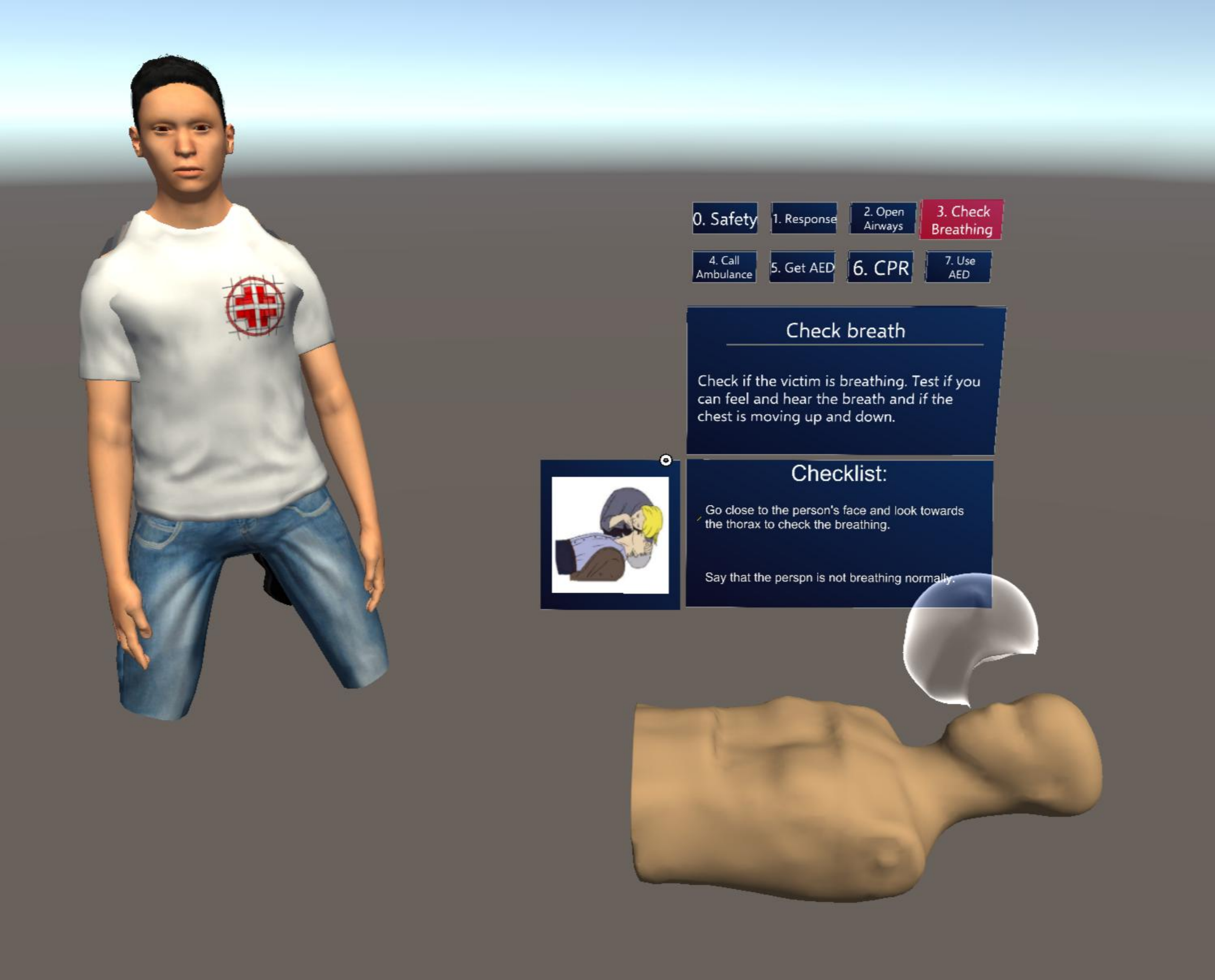}
    }\hspace*{3pt}
    \subfloat[Call Ambulance\label{fig:implementation:Software:ARWalktrough:4}]{
    \includegraphics[width=.3\linewidth]{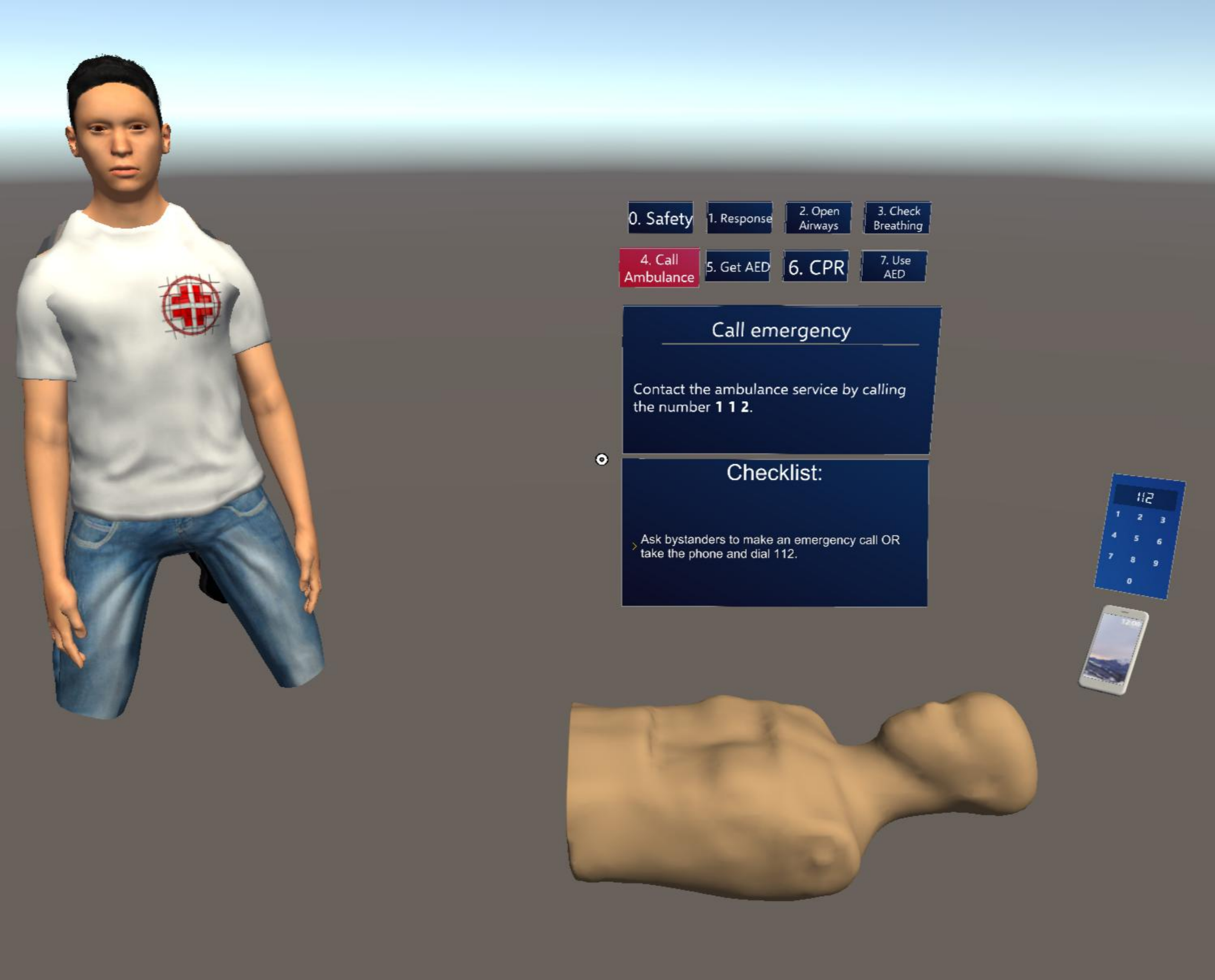}
    }\hspace*{3pt}
    \subfloat[Get AED\label{fig:implementation:Software:VRWalktrough:5}]{
    \includegraphics[width=.3\linewidth]{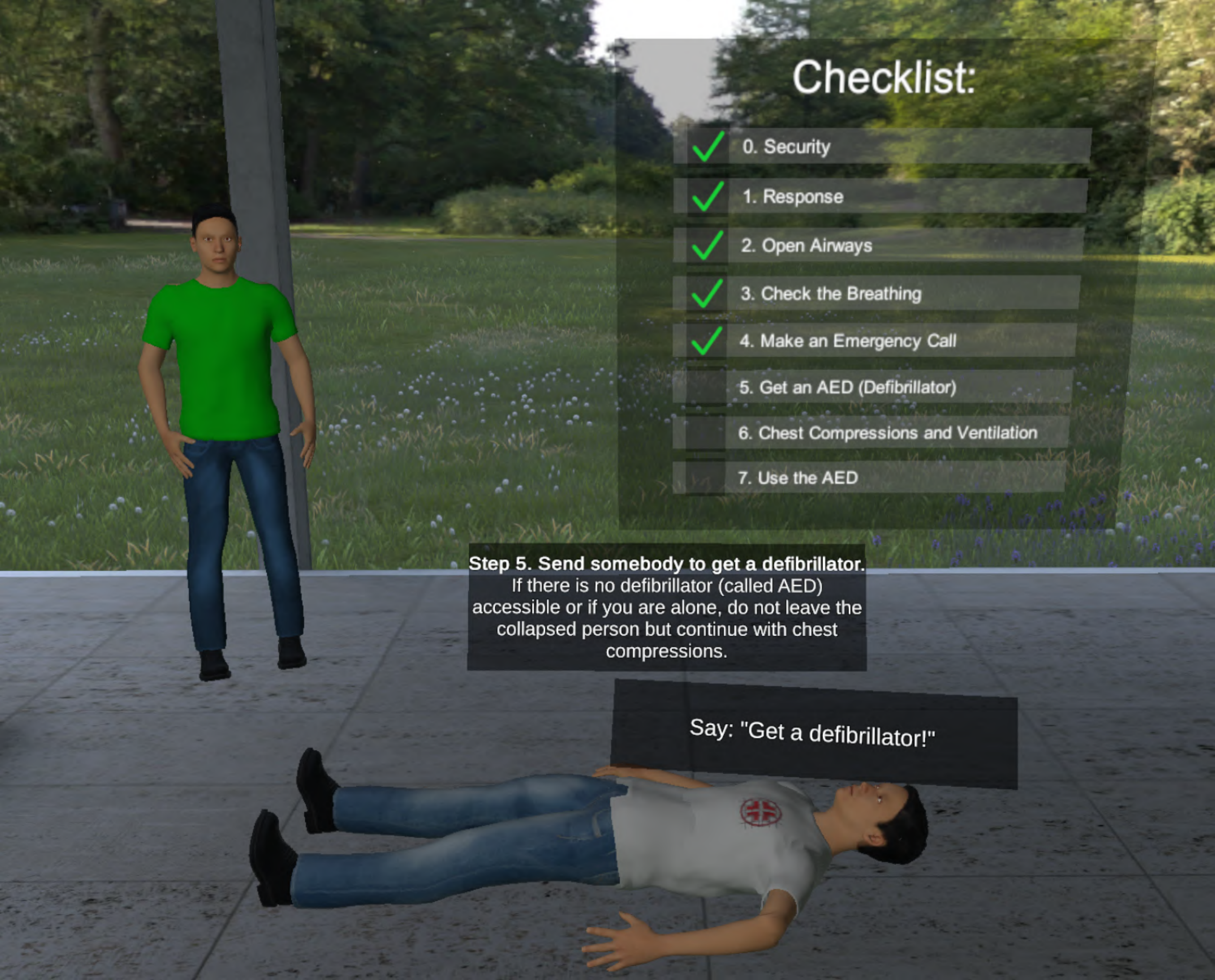}
    }
    \end{center}
    
    \begin{center}  
    \subfloat[Chest Compressions\label{fig:implementation:Software:VRWalktrough:6.1}]{
    \includegraphics[width=.3\linewidth]{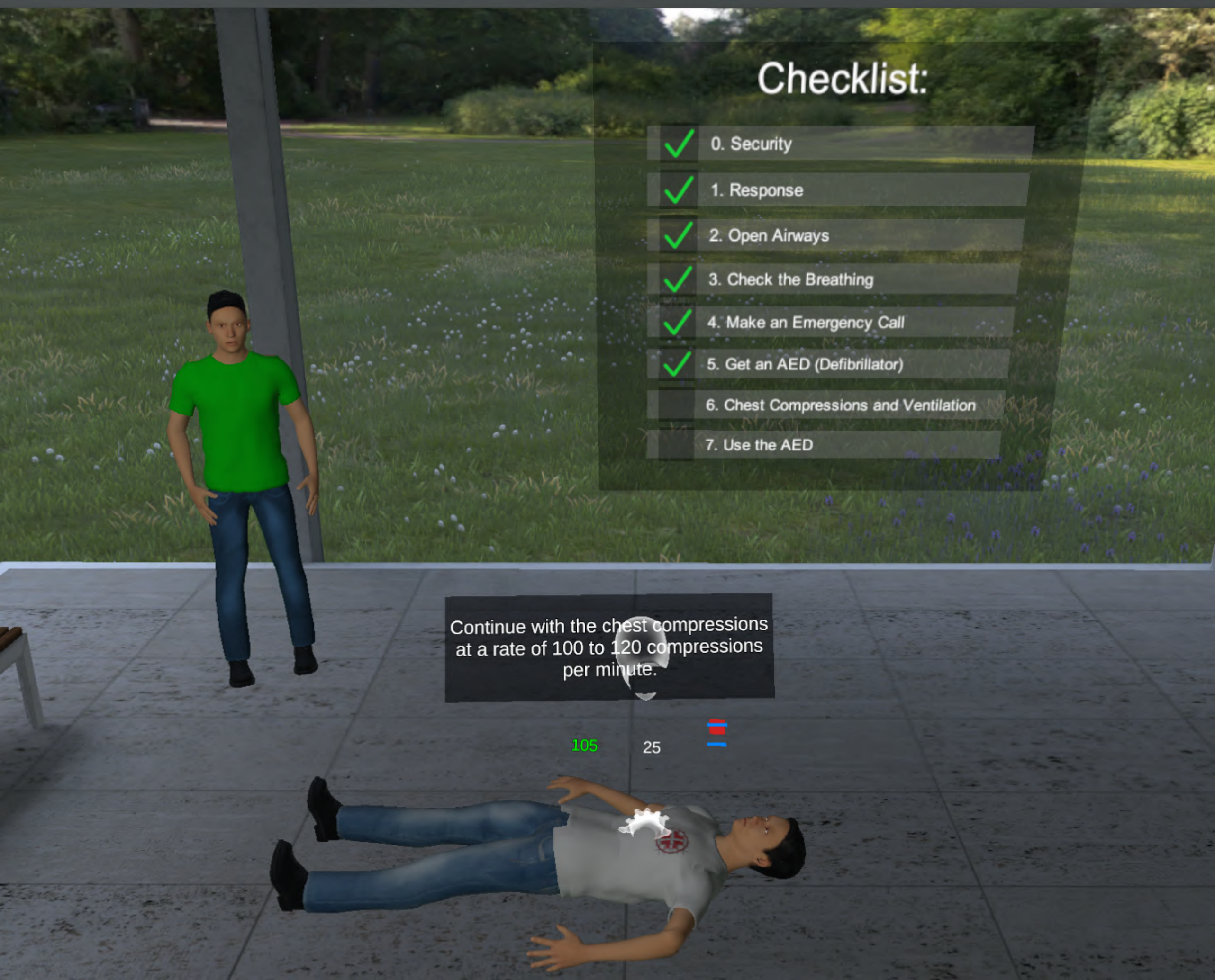}
    }\hspace*{3pt}
    \subfloat[Ventilation\label{fig:implementation:Software:VRWalktrough:6.2}]{
    \includegraphics[width=.3\linewidth]{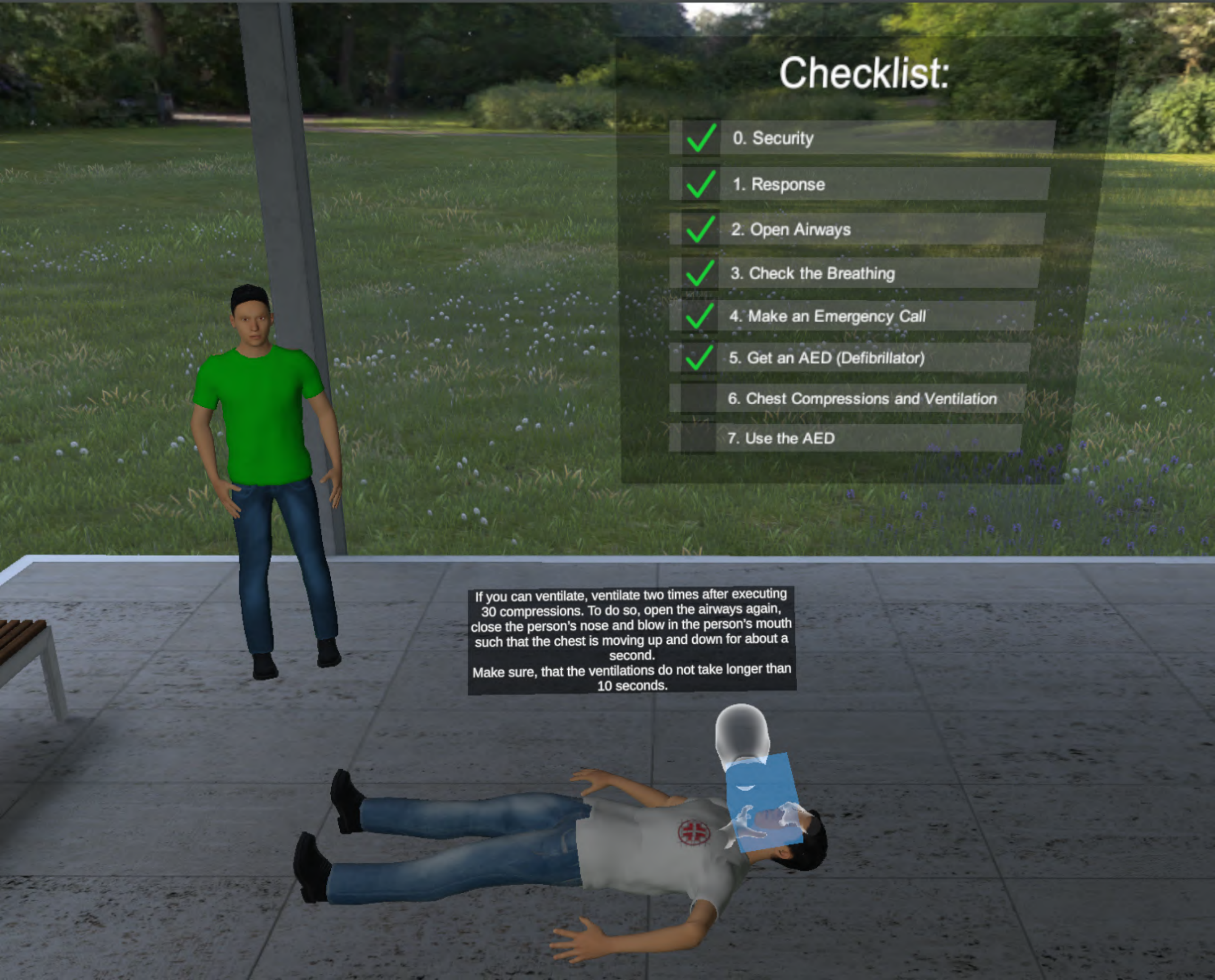}
    }\hspace*{3pt}
    \subfloat[Use AED\label{fig:implementation:Software:VRWalktrough:7}]{
    \includegraphics[width=.3\linewidth]{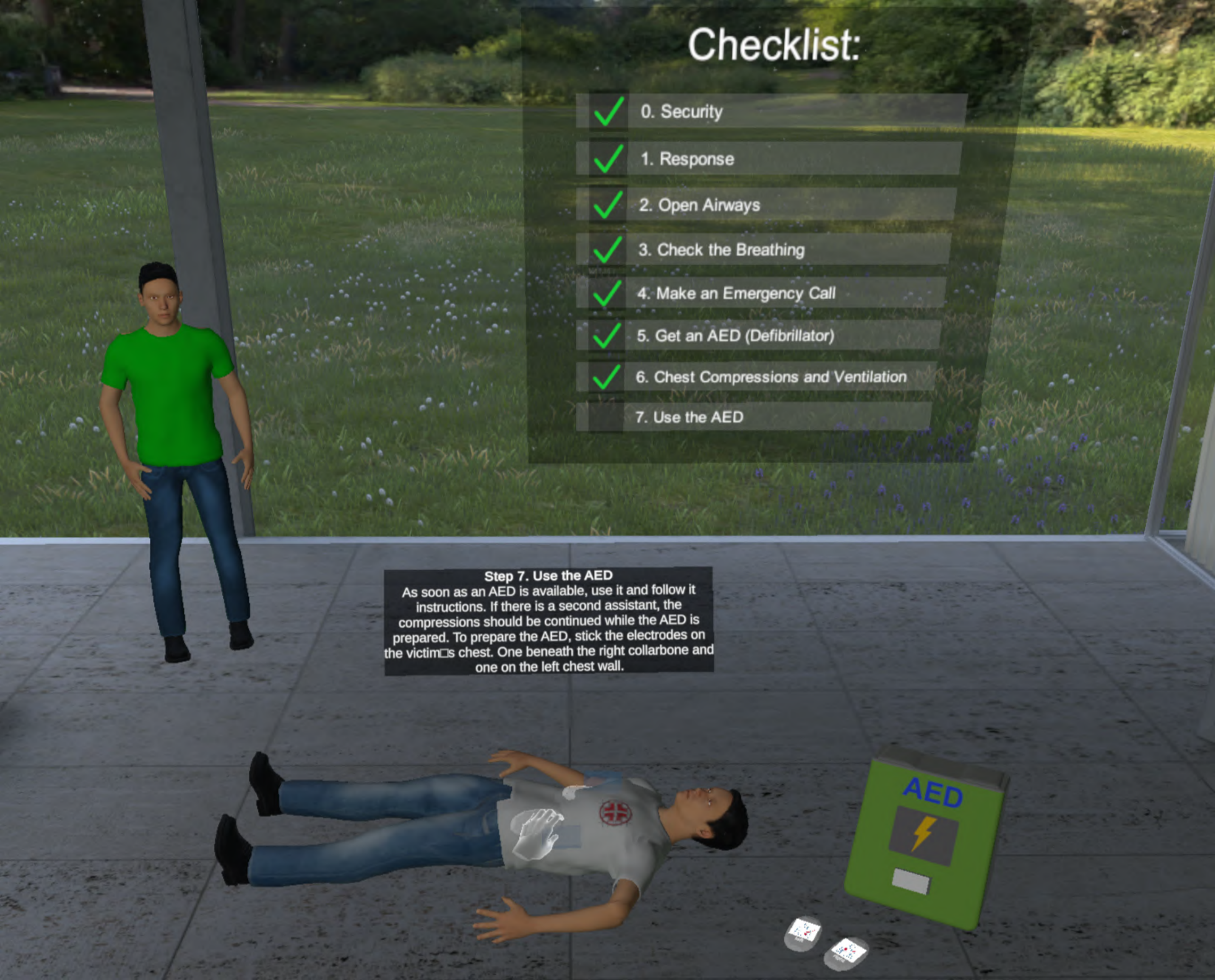}
    }
    \end{center}
    
    \caption{AR/VR Walkthrough}
    \label{fig:implementation:Software:ARWalktrough}
\end{figure}

Using those building blocks, we created the whole training application which guides the users through the BLS sequence. Figure \ref{fig:implementation:Software:ARWalktrough} shows a full walkthrough of all tasks in the AR as well as in the VR training. Note that the 3D model of the manikin in the AR training is not shown to the user. During the training, all virtual objects are placed on and around the real manikin. The model is just shown here for illustration purposes.
Figures \ref{fig:implementation:Software:ARWalktrough:0} - \ref{fig:implementation:Software:ARWalktrough:4} are taken from the AR application, whereas Figures \ref{fig:implementation:Software:VRWalktrough:5}-\ref{fig:implementation:Software:VRWalktrough:7} are taken from the VR application.
Changing the applications is only done here to illustrate the walkthrough of both applications. The users did not change the training in the middle, they stayed in their application until the end.
In all steps, description panels, images, and audio files are involved as we use them to explain the basics of every step. 

In the beginning, the scene safety needs to be ensured (Figure \ref{fig:implementation:Software:ARWalktrough:0}). Here we placed the broken glasses on the victim's body. They are interactable objects and need to be dragged into the dustbin on the right side. Animations of the hands give an additional hint of what the user needs to do. 
To check the victim's response (Figure \ref{fig:implementation:Software:ARWalktrough:1}), we check if the user's hands are at the victim's shoulders with our position trigger. Animated hands show the movements to be done. In addition, the user needs to ask the victim if everything is fine. So we also use voice commands in this step. 
When the airways need to be opened (Figure \ref{fig:implementation:Software:ARWalktrough:2}) we use a position trigger to recognize if the hands are placed on the victim's head. For the head tilt, we use the measurements of the manikin's gyro sensor to calculate the head's angle. We also show animations for the placement as well as for the tilting. In the VR application, the user is moving the head of the virtual manikin, so we can measure the tilt here.
While checking the breathing (Figure \ref{fig:implementation:Software:ARWalktrough:3}) we use a position trigger to check if the user's head is above the victim's mouth and the user is looking towards the victim's chest. Here, an animated head also shows the movement. Next, the breathing status needs to be communicated. So we use voice commands to recognize if the user correctly announces the status. 
When the ambulance needs to be called (Figure \ref{fig:implementation:Software:ARWalktrough:4}), the user can grab the phone (interactable object) to dial the number. Alternatively, the bystander can be asked to call the ambulance. In this case, we also use voice commands.
In the next step, sending someone to get an AED (Figure \ref{fig:implementation:Software:VRWalktrough:5}), we only use voice commands.
When performing chest compressions (Figure \ref{fig:implementation:Software:VRWalktrough:6.1}), we use the manikin's ultrasonic sensor to measure the compression depth and to display the rate and depth to the user. Animations show how the compressions need to be performed. 
After that, the victim needs to be ventilated (Figure \ref{fig:implementation:Software:VRWalktrough:6.2}). Here we check if the user's head is correctly placed over the victim's head as is shown by an animation. 
For the final step, using the AED (Figure \ref{fig:implementation:Software:VRWalktrough:7}), we have the AED pads as interactable objects which the user needs to place and the AED's button which needs to be activated at the end. The pad placement is also shown by an animation. Before triggering the shock, it needs to be ensured that nobody touches the victim. As the user needs to communicate that, we use voice commands.

\subsection{Debriefing}\label{section:impl:Debriefing}
\begin{figure}
    \begin{center}  
    \subfloat[Debriefing Overview\label{fig:implementation:Debriefing:Overview}]{
    \includegraphics[width=.7\linewidth]{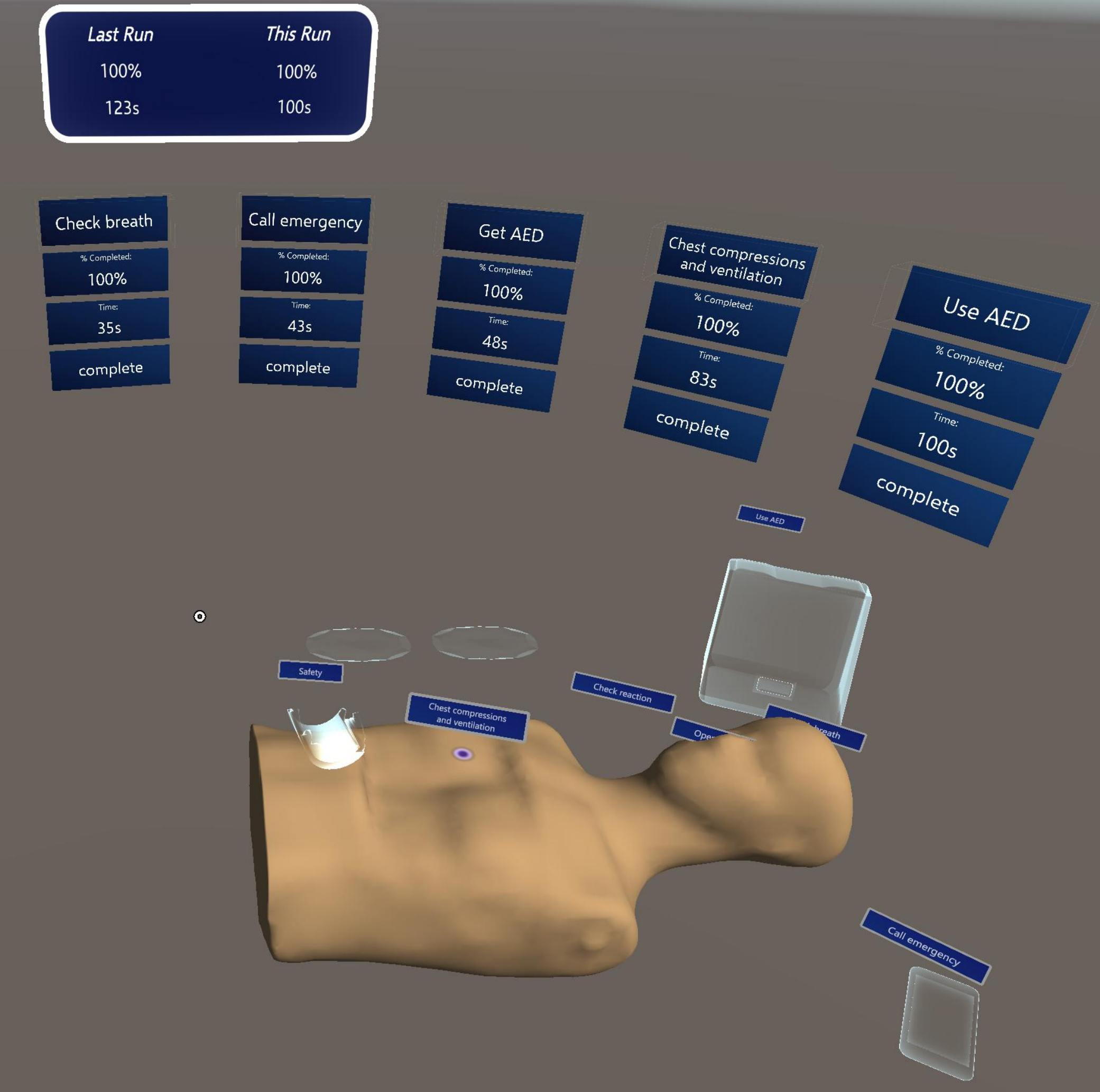}
    }
    \end{center}
    
    \begin{center}
    \subfloat[Debriefing CPR\label{fig:implementation:Debriefing:CPR}]{
    \includegraphics[width=.7\linewidth]{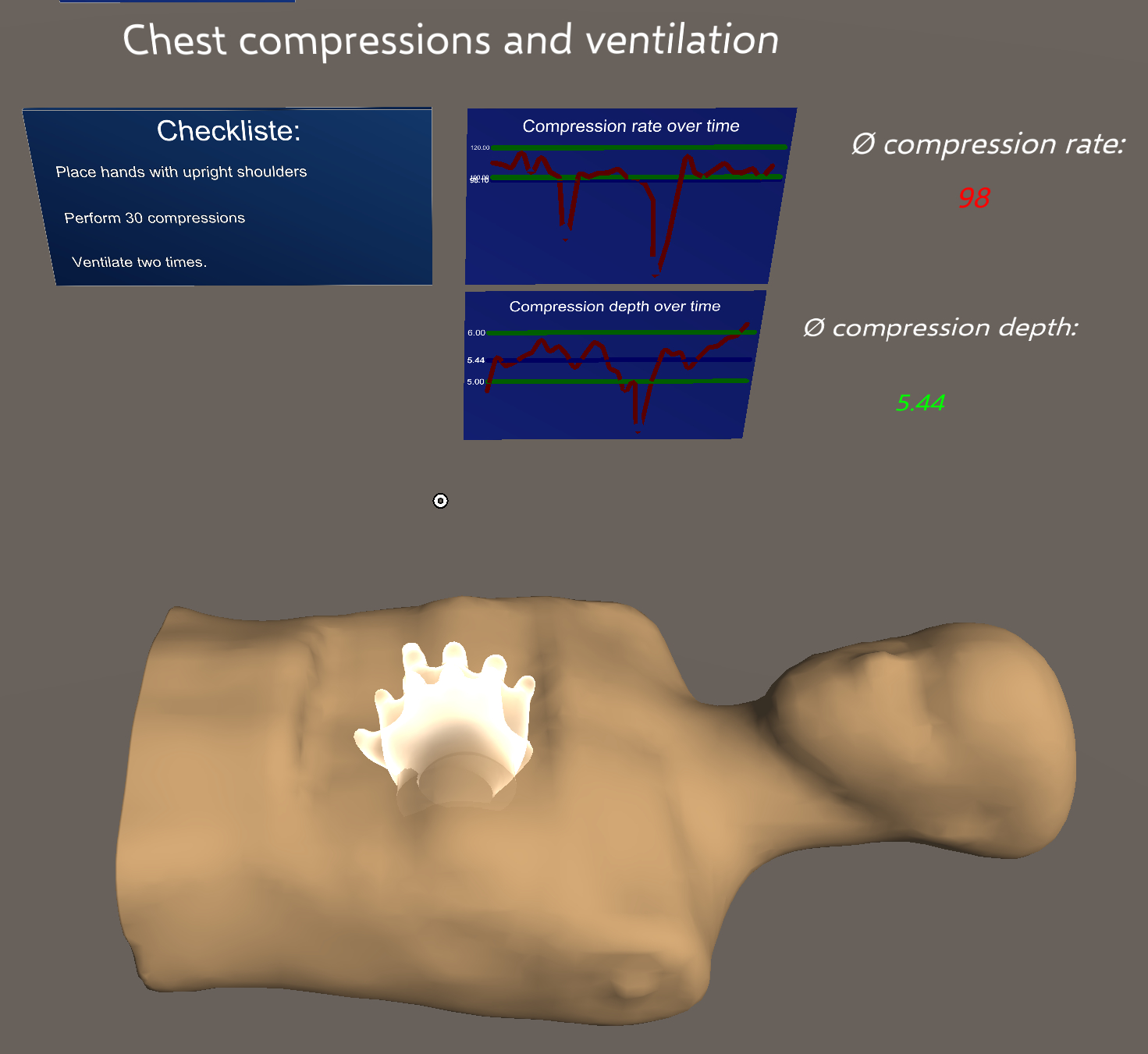}
    }
    \end{center}
    \caption{Debriefing}
    \label{fig:implementation:Debriefing}
\end{figure}

When the users perform the training, we log their performance. With that, we get the duration of how long it took to complete a task, and how (well) the task was completed. During the debriefing, the user sees the scene depicted in Figure \ref{fig:implementation:Debriefing:Overview}.

Here, we present all the tasks with some basic information about the user's performance. 
First, we give an overview of the overall performance compared to the last training. We show how many percent of the tasks were executed correctly and how long the user needed to complete the training. Beneath that, we show the same information but on task-level. At the top, we show how many percent of the task was executed, i.e. how many subtasks were executed correctly. That is also summarized at the bottom. In between them, we show the duration the user needed to complete the task. In addition to that, the user can zoom into the results of a single task, either by gazing at the panel of the corresponding task in the overview or by gazing at the location where the task has been executed on the manikin. To enable that, we use MRTK's \textit{Interactbable}\footnote{\url{https://docs.microsoft.com/en-us/windows/mixed-reality/mrtk-unity/features/ux-building-blocks/interactable?view=mrtkunity-2021-05}} script, which triggers loading the details using the \textit{InteractableOnFocusReceiver}\footnote{\url{https://hololenscndev.github.io/MRTKDoc/api/Microsoft.MixedReality.Toolkit.UI.InteractableOnFocusReceiver.html}}. In the VR version, users aim at the spots using a laser pointer. When the user activates the controller's trigger, we cast a ray\footnote{\url{https://docs.unity3d.com/ScriptReference/Physics.Raycast.html}} along the laser pointer and load the details for the task where the ray hits. All other components are the same for the AR and VR versions. The zones mentioned above can be seen at the bottom of the screenshot. To facilitate re-experiencing of the performed tasks, we decided to also include the animations used in the learning mode into the debriefing. However, if desired, the animations can be hidden. For the CPR, we have got additional values for the compressions (see Figure \ref{fig:implementation:Debriefing}). We show them in two different ways. On the right side, we show an average of the compression depth and rate whereas, in the middle, we display graphs showing how the values changed over time. Both graphs can be enlarged for better readability by focusing them or selecting them with the laser pointer. The graphs are calculated dynamically. We display the user's performance (red line) together with the upper and lower threshold (green lines). The graphs themselves are generated using Unity's \textit{LineRenderer}\footnote{\url{https://docs.unity3d.com/Manual/class-LineRenderer.html}}.

\section{Evaluation}
\label{section:Evalution}

To answer the second research question (RQ2) motivated in the introduction, we have conducted a usability evaluation to analyze the efficiency, effectiveness, and user satisfaction of our AR- and VR-based BLS training environment compared to traditional training methods. In Section \ref{section:Evalution:setupProcedure}, we describe the used setup and procedure. Following that, we present our participants and their background in Section \ref{section:Evaluation:Participants}. We display the results we gathered during our usability experiment in Section \ref{section:Evalution:results}. Those results are discussed in Section \ref{section:Evalution:discussion}. Finally, in Section \ref{section:Evalution:threatsToValidity} we describe the threats to the validity of our study.

\subsection{Setup and Procedure}\label{section:Evalution:setupProcedure}

\begin{figure}
    \centering
    \includegraphics[width=\linewidth]{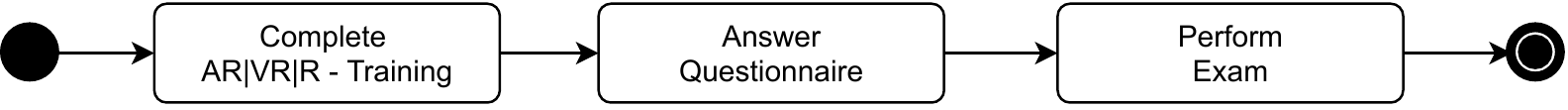}
    \caption{User Study Procedure}
    \label{fig:eval:userStudyProcess}
\end{figure}

\begin{figure}
    \begin{center}  
    \subfloat[AR\label{fig:experiment:participantsOverview:AR}]{
    \includegraphics[width=.23\linewidth]{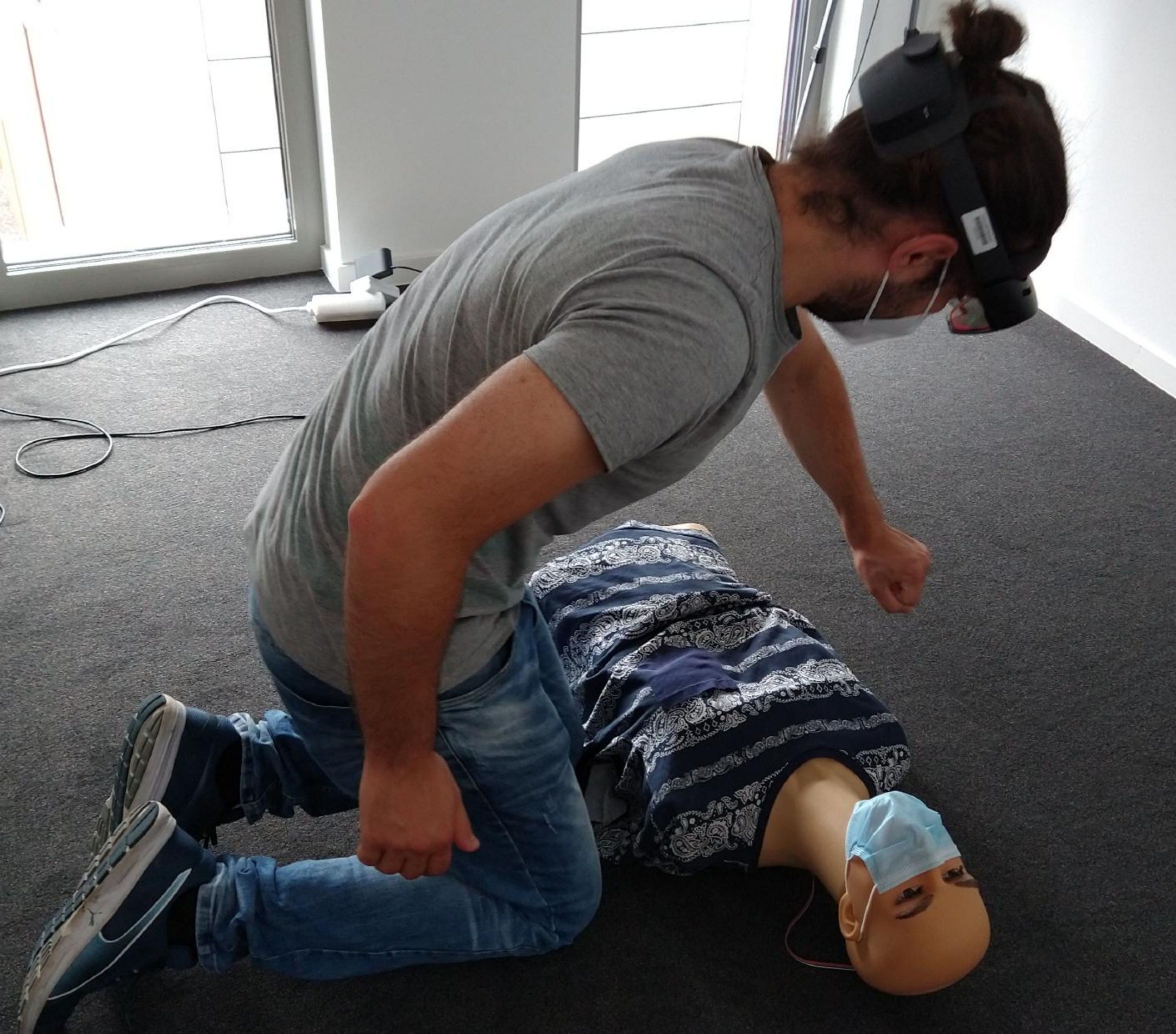}
    }\hspace*{1pt}
    \subfloat[VR\label{fig:experiment:participantsOverview:VR}]{
    \includegraphics[width=.23\linewidth]{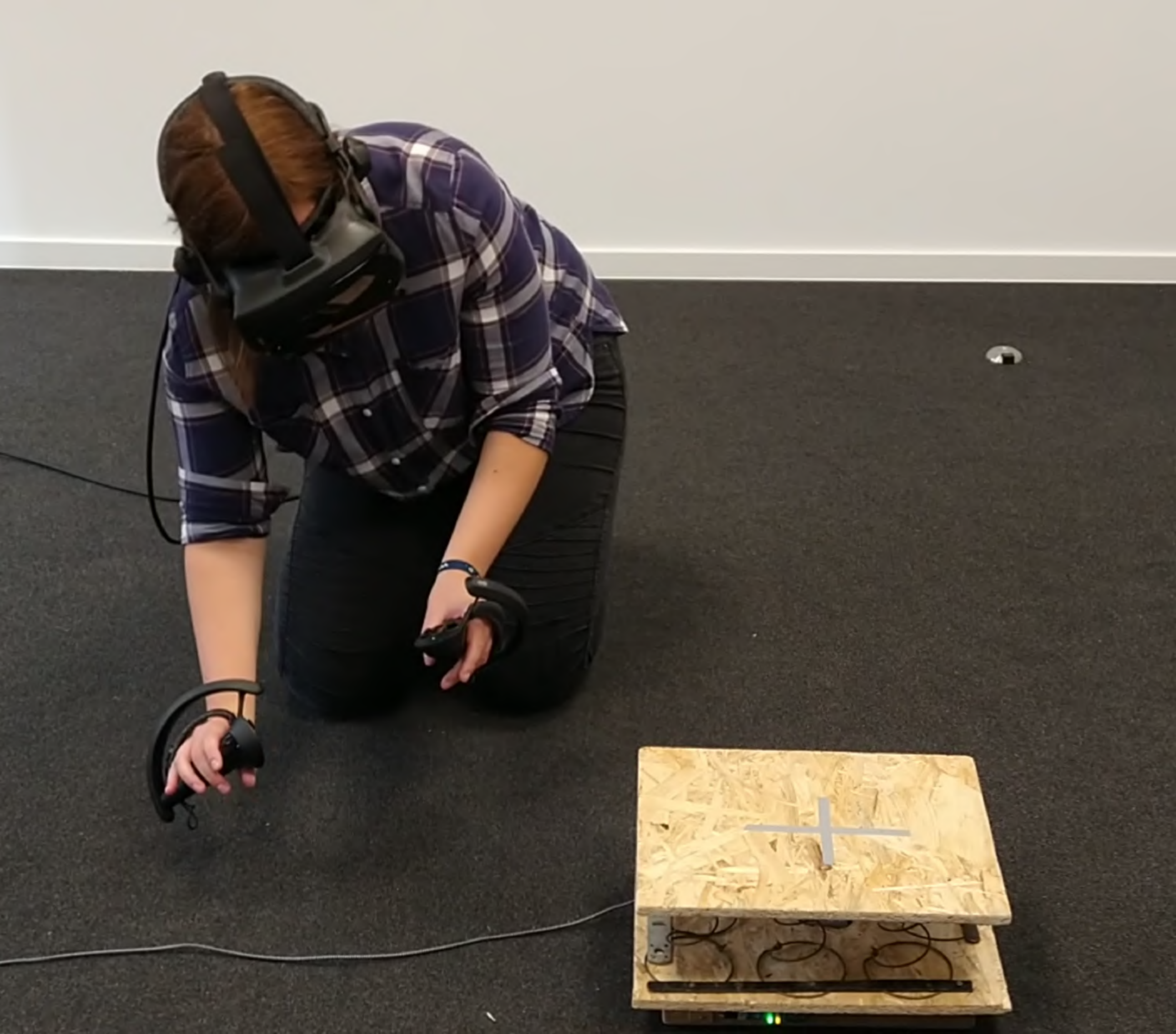}
    }\hspace*{1pt}
    \subfloat[R\label{fig:experiment:participantsOverview:R}]{
    \includegraphics[width=.23\linewidth]{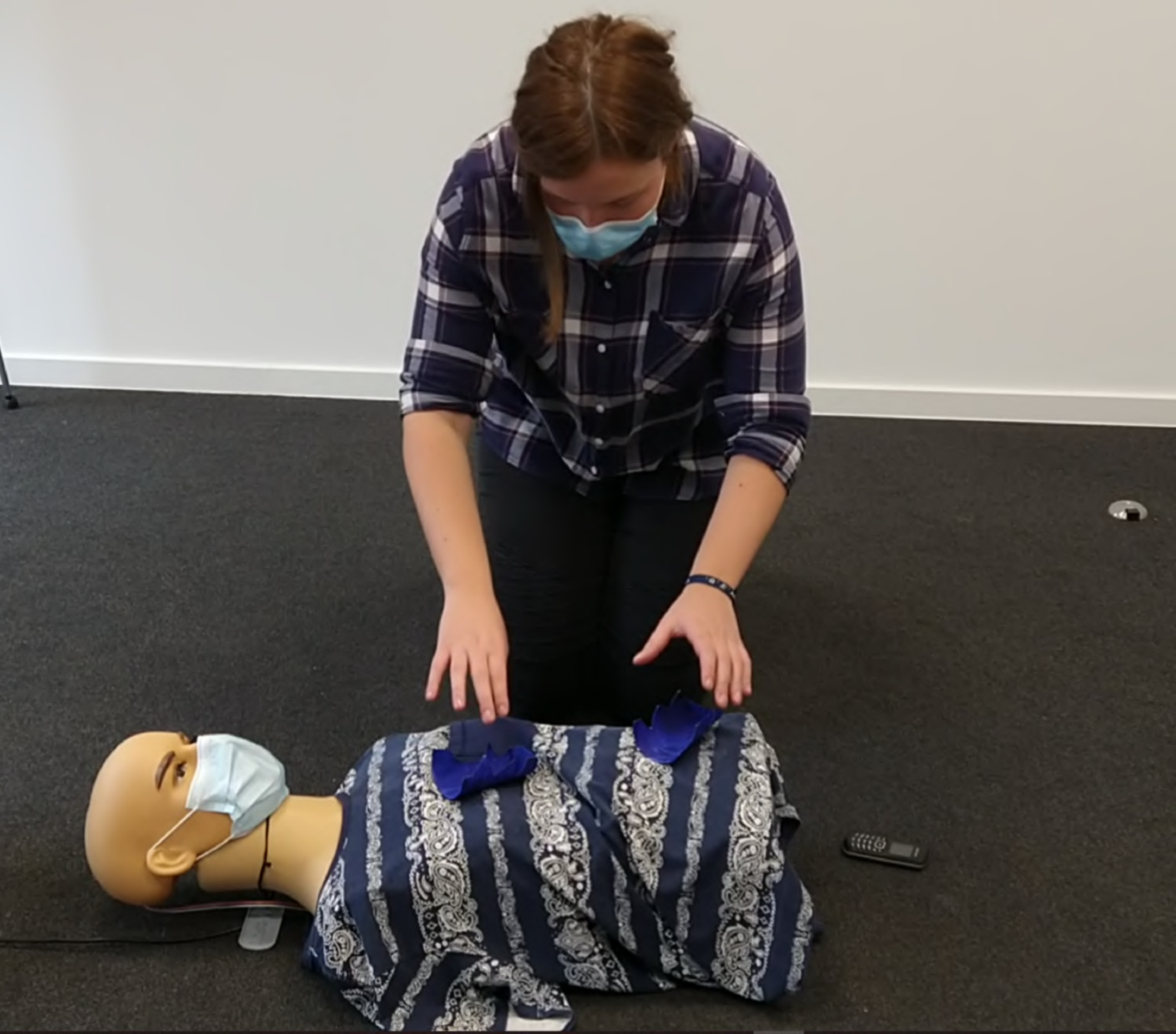}
    }\hspace*{1pt}
    \subfloat[Exam setup\label{fig:experiment:participantsOverview:exam}]{
    \includegraphics[width=.23\linewidth]{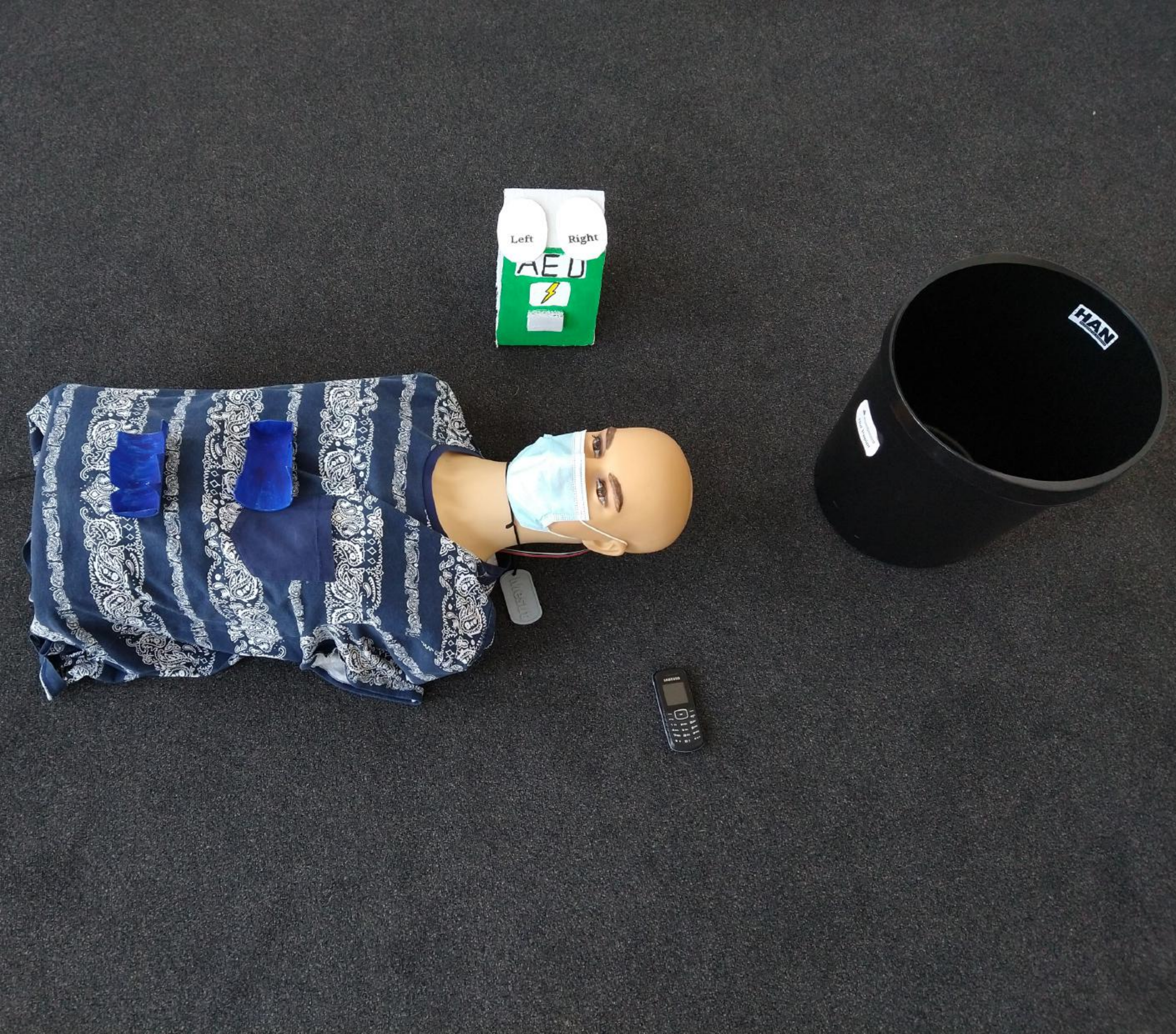}
    } 
    \end{center}
    \caption{Usability Experiment}
    \label{fig:experiment:participantsOverview}
\end{figure}

To answer RQ2, we conducted a user study that targets comparing our AR and VR training with real-life (R) training. For the R training, the trainer explained the tasks on the manikin and the trainee performed them afterward. In this context, please note that we had no certified trainer, but rather an expert familiar with the BLS training process. The content was the same as in the digital trainings, i.e. the trainer described the same steps and used similar descriptions as the AR and the VR application. 
For the usability experiment, we followed the process depicted in Figure \ref{fig:eval:userStudyProcess}. 
First, the users completed their assigned training (Figure \ref{fig:experiment:participantsOverview}). Next, they were asked to answer a questionnaire. We used the SUS questionnaire \cite{brooke1996sus} to get insight into the systems' overall usability, the NASA TLX score \cite{hart1988development} to evaluate the workload, and the presence questionnaire by Usoh et al. \cite{DBLP:journals/presence/UsohCAS00}. Here, we adjusted the questions so that they fit our scenario. 

Finally, the users performed an exam. That means they have to perform the whole BLS sequence without any help. We measured the time and noted whenever a user completed a step, so we know in which order the tasks were executed and how long it took. Additionally, we recorded the sensor values of the manikin to analyze the compression rate and depth. We then calculate a score that indicates how well the sequence was executed. Note that we do not consider the duration here.

To compare the execution of the BLS sequence, we introduced a simple scoring system. Every time the user executes an activity correctly, they get the corresponding score (Figure \ref{table:evaluation:scores}). The score distribution was implemented in a similar way to Strada et al. \cite{strada2019holo}.

\begin{figure}
\centering
  \begin{tabular}{ | l | r | }
  \hline
    \textbf{Task name}          &\textbf{Score}\\\hline
    Ensure Safety               &       2           \\ \hline
    Check Response              &       1           \\ \hline
    Open Airways                &       1           \\ \hline
    Make an emergency call      &       2           \\ \hline
    Send somebody to get an AED &       2           \\ \hline
    Perform compressions        &     0-4           \\ \hline
    Ventilate                   &       2           \\ \hline
    Place AED pads              &  1 for each pad   \\ \hline
    Make people stand back      &       1           \\ \hline
    Trigger shock               &       1           \\
    \hline
  \end{tabular}
  \caption{Scores}\label{table:evaluation:scores}
\end{figure}
For the compressions, the average rate (compressions per minute) and the average depth are evaluated. The average rate points have the following intervals. 95$<$avg. rate$<$125: 2 points, 80$<$avg. rate$<$95: 1 point, 125$<$avg. rate$<$140: 1 point, all other values: 0 points. For the average depth, 5$<$avg. depth$<$6 results in 2 points, all other values result in 0 points.

By adding the points for executing the tasks, we get an intermediate score. As the correct execution order is important for most tasks, we weigh the score. For that, we count the number of tasks that were executed at the correct time (i.e. the task the user executed before the current task is also its predecessor in the activity diagram). If the order is 100\% correct, the user gets the full score, if the order is 0\% correct, the user gets 50\% of the intermediate score. All values in between are mapped accordingly.

\subsection{Participants}\label{section:Evaluation:Participants}
For our user study, we had 21 participants, with ages ranging from 22 to 54. We had 5 female and 16 male participants. The participants were distributed randomly on the training types so that 7 participants performed the AR, VR, and R training. 19 participants had done first aid training before, but for most of them, the last training was 5 or more years ago. Only 2 persons regularly participate in first aid training every two years. Most AR participants had no to little experience with AR, whereas most VR participants had at least a little experience with VR applications.

\subsection{Results}\label{section:Evalution:results}

\begin{figure}
    \begin{center}  
    \subfloat[SUS Score\label{fig:eval:susScoreAll}]{\includegraphics[height=150pt]{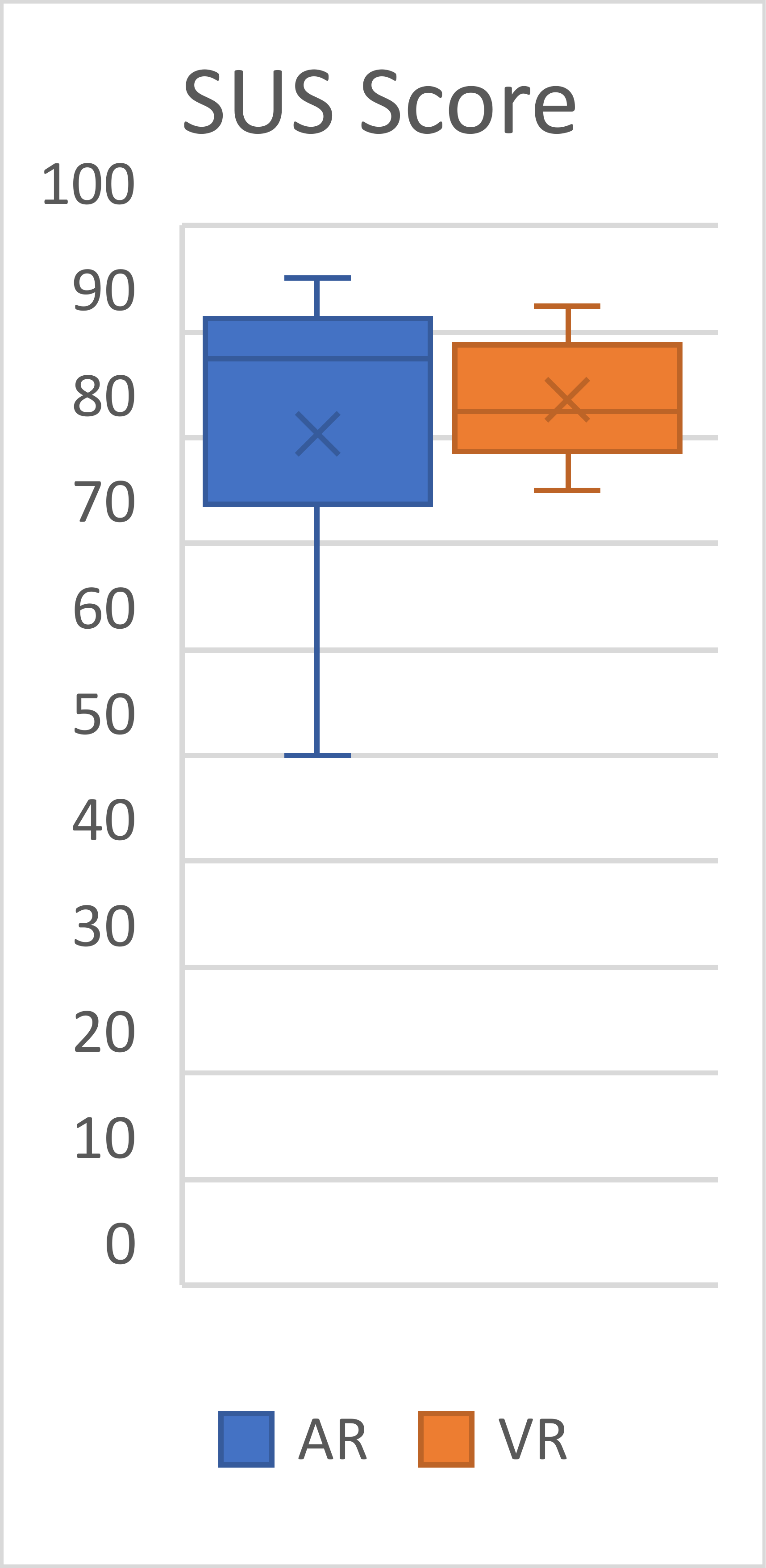}}\hspace*{5pt}
    \subfloat[TLX Score\label{fig:eval:TLXScore}]{\includegraphics[height=150pt]{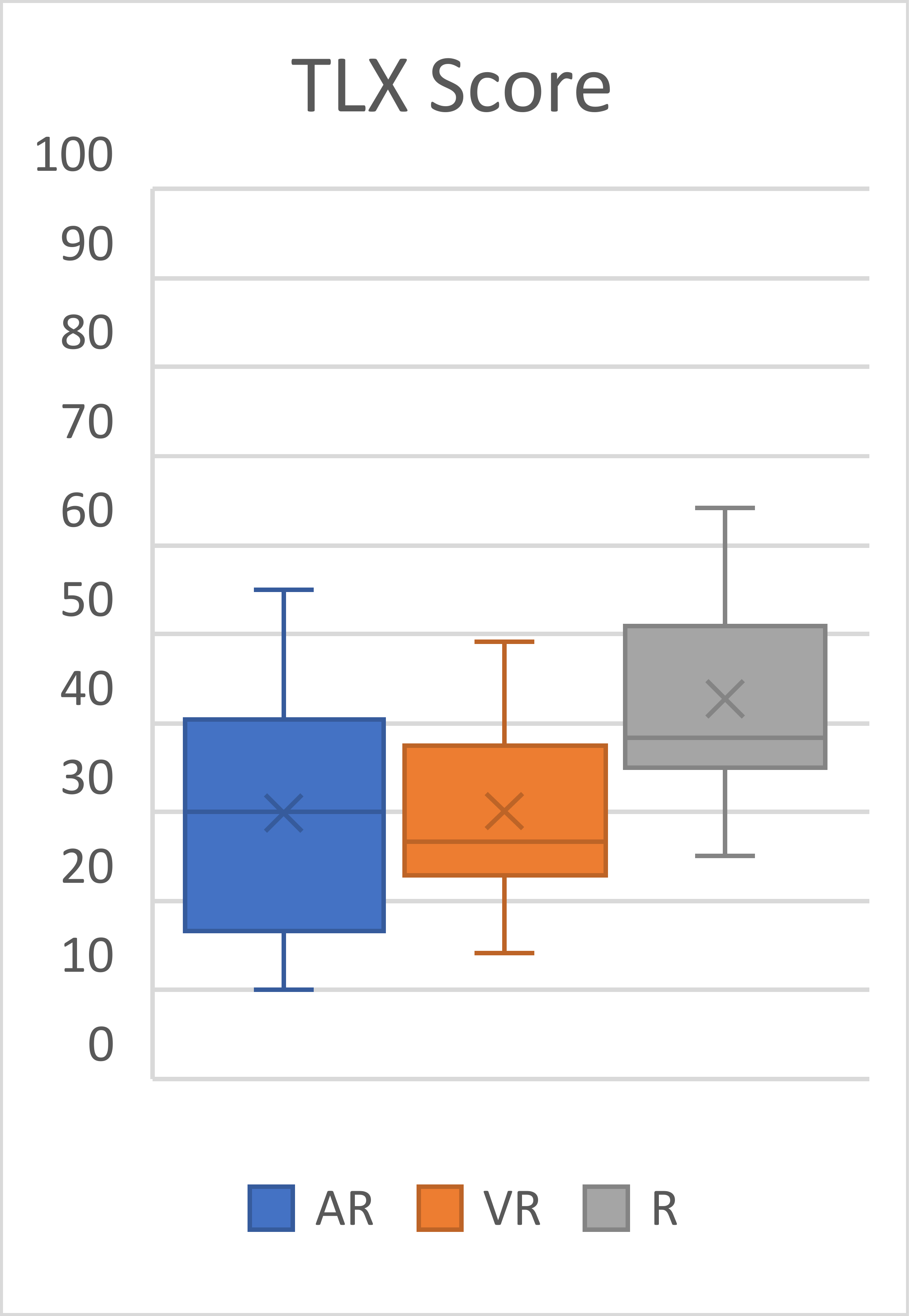}}\hspace*{5pt}
    \subfloat[Presence Score\label{fig:eval:PresenceScore}]{\includegraphics[height=150pt]{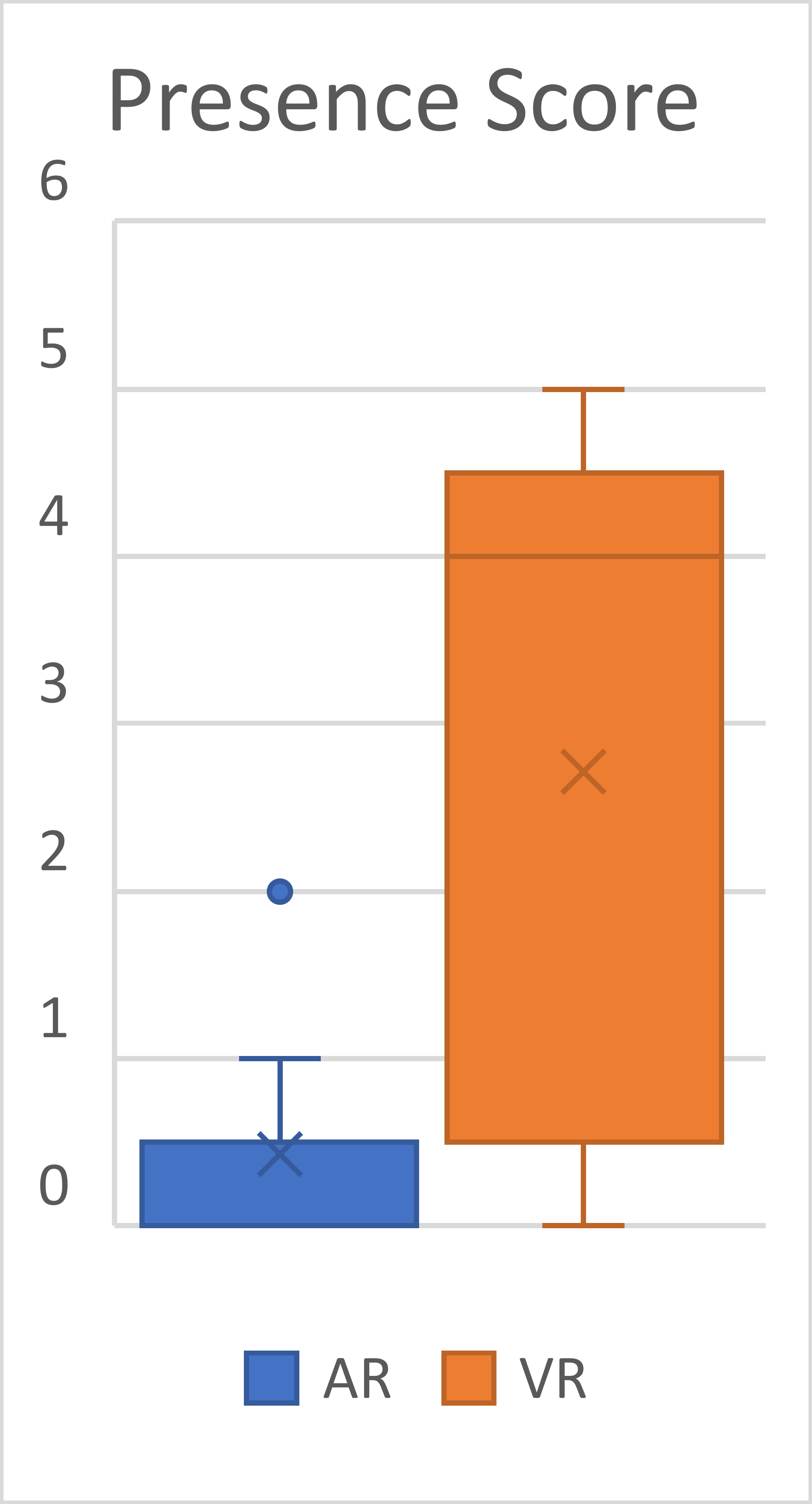}}  
    \end{center}
    \caption{Questionnaire Scores}
    \label{fig:eval:questionnaireScores}
\end{figure}
Figure \ref{fig:eval:susScoreAll} shows the SUS scores for the AR and VR training applications. The mean scores are $80.36$ for the AR and with $83.57$ slightly higher for the VR training. But considering the medians, the AR application is rated higher ($87.5$) than the VR application ($82.5$). It is also worth noting that the scores for the AR training are broader distributed on the scale, whereas the scores for the VR training are close together.

Figure \ref{fig:eval:TLXScore} shows the results of the NASA-TLX \cite{hart1988development} questionnaire. Here, we can see that, on average, the workloads of the AR ($29.88$) and the VR training ($30.12$) are almost the same. A small difference can be seen at the median which is $30$ for the AR and $26.67$ for the VR application. Again, the values for the AR application are more distributed than the VR ones. A bigger difference can be seen when compared to the R training, where the mean ($42.74$), as well as the median ($38.33$), are higher. 

Finally, we consider the presence score proposed in \cite{DBLP:journals/presence/UsohCAS00}.  Here, when using the VR application users had a stronger feeling to be present at an accident scene than the users of the AR application.


\begin{figure}
    \begin{center}  
    \subfloat[Training Score\label{fig:eval:trainingScore}]{\includegraphics[height=150pt]{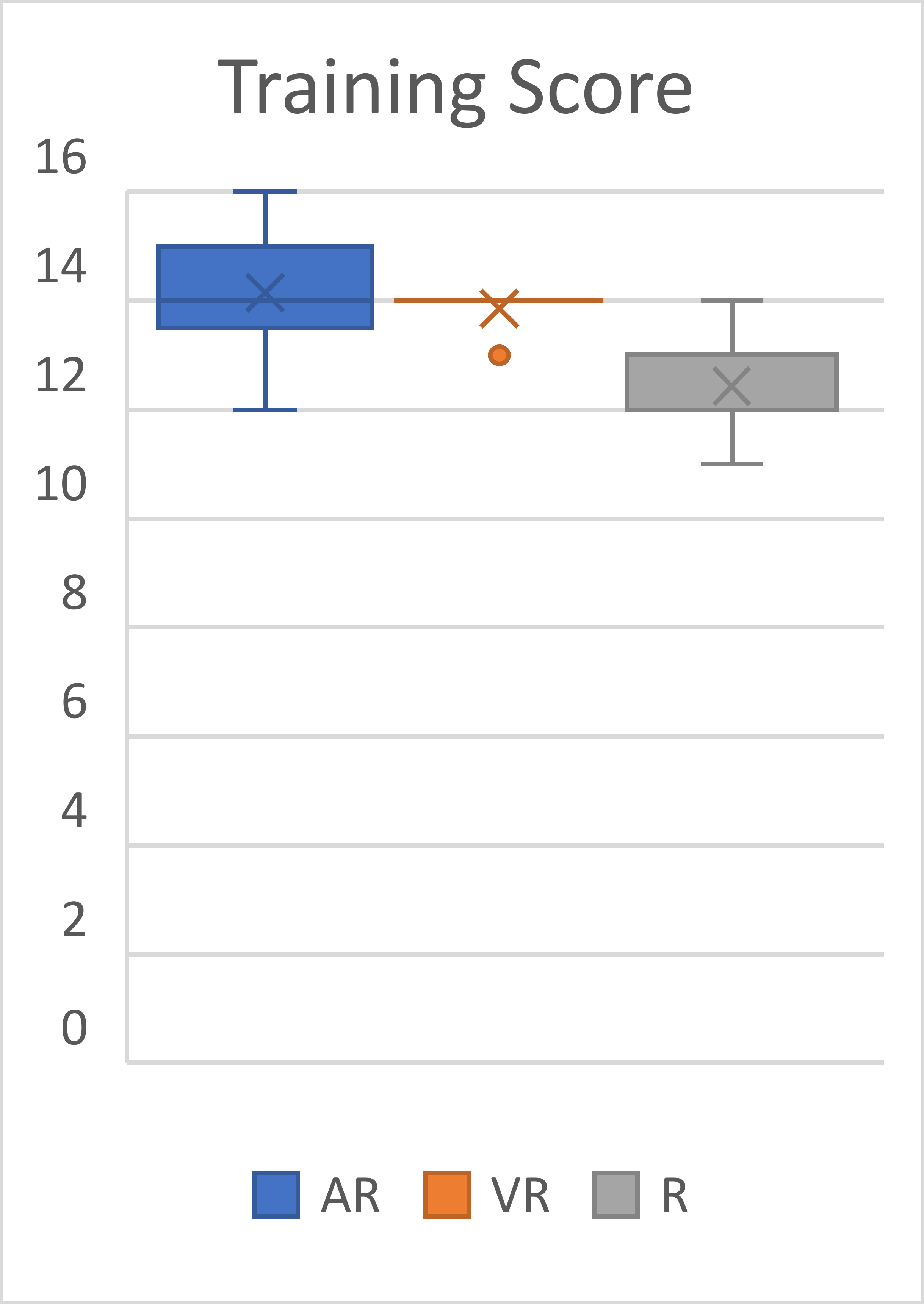}}\hspace*{5pt}
    \subfloat[Exam Score\label{fig:eval:examScore}]{\includegraphics[height=150pt]{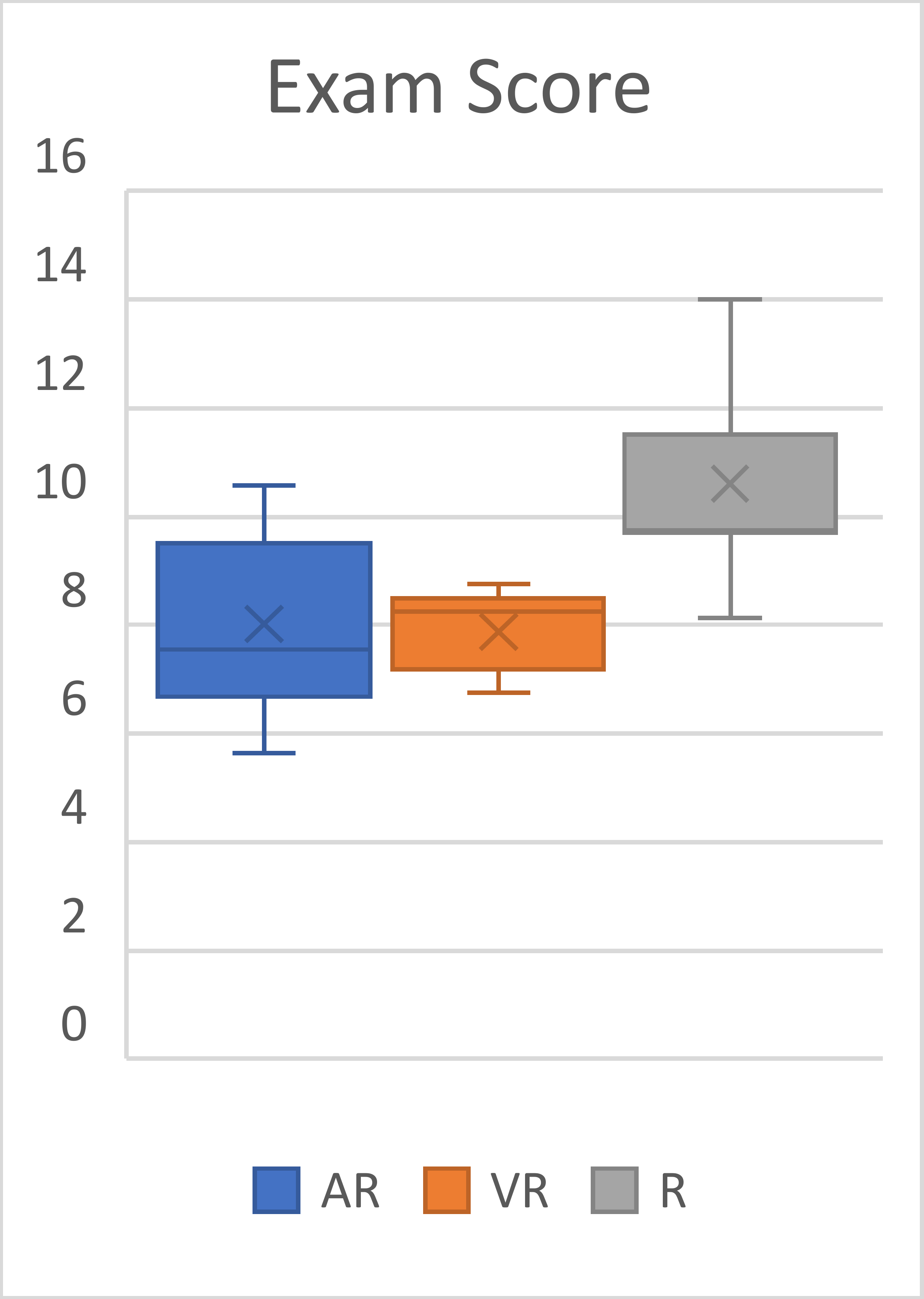}}
    \end{center}
    \caption{Scores}
    \label{fig:eval:scores}
\end{figure}
Figure \ref{fig:eval:scores} shows the scores the users got when executing the training and the exam. Figure \ref{fig:eval:trainingScore} shows the scores in the training. Here, the average score differs slightly for the AR/VR trainings (VR: $14.14$, AR: $13.86$) but they have the same median ($14$). For the real-life training, the mean ($12.43$) and the median ($12$) are lower. The scores for the VR training are less distributed than the scores for the other trainings.

Considering the exam scores of Figure \ref{fig:eval:examScore}, we see that the score of participants of the AR  (mean: $8.01$, median $7.56$) and VR training (mean: $7.88$, median: $8.25$) are close to each other, while the participants of R trainings have higher scores (mean: $10.6$, median: $9.75$). Again, the values for the VR training participants are less distributed than the others.

\begin{figure}
    \begin{center}  
    \subfloat[Training Duration\label{fig:eval:trainingDuration}]{\includegraphics[height=150pt]{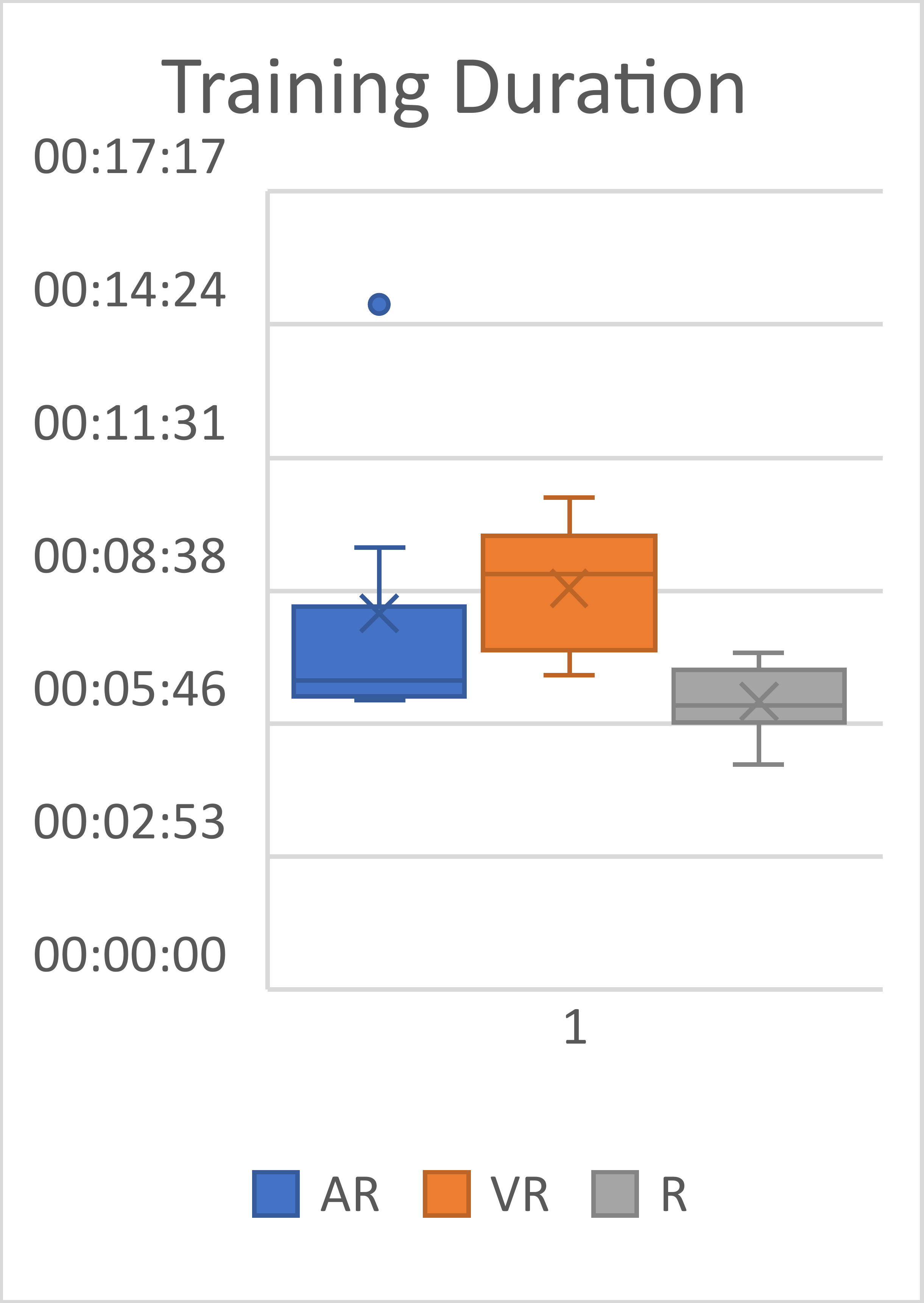}}\hspace*{5pt}
    \subfloat[Exam Duration\label{fig:eval:examDuration}]{\includegraphics[height=150pt]{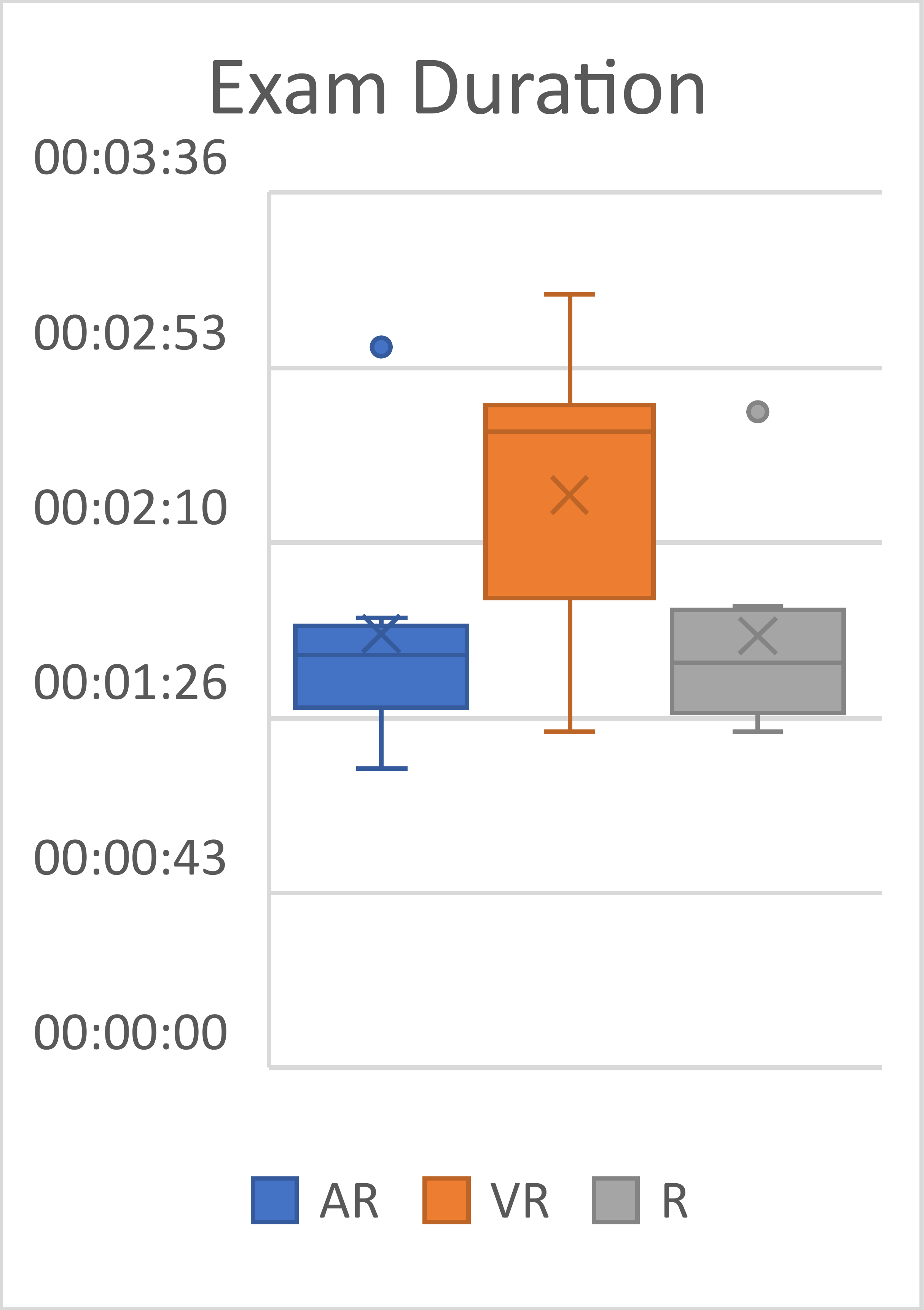}}
    \end{center}
    \caption{BLS Durations}
    \label{fig:eval:blsDurations}
\end{figure}
Figure \ref{fig:eval:blsDurations} shows the durations for the trainings and exams. Considering the trainings (Figure \ref{fig:eval:trainingDuration}) it can be seen that the R training was the fastest (mean: 6:15 mins, median: 6:09 mins). With on outlier (14:51 mins), the AR training's duration was in the middle (mean: 8:09, median: 6:42) and the VR training took slightly longer (mean: 8:41, median: 9:01).

For the exam, the participants of the AR and R training were almost equally fast. AR participants needed 1:47 mins on average, the median is 1:42. Participants of the R training needed 1:47 mins on average and had a median of 1:40 mins. Both had one outlier with 2:58 mins (AR) and 2:42 mins (R). Participants of the VR training needed more time, 2:21 mins on average with a median of 2:37 mins.


\begin{figure}
    \begin{center}  
    \subfloat[Training CPR Rate Average\label{fig:eval:trainingRate}]{\includegraphics[height=150pt]{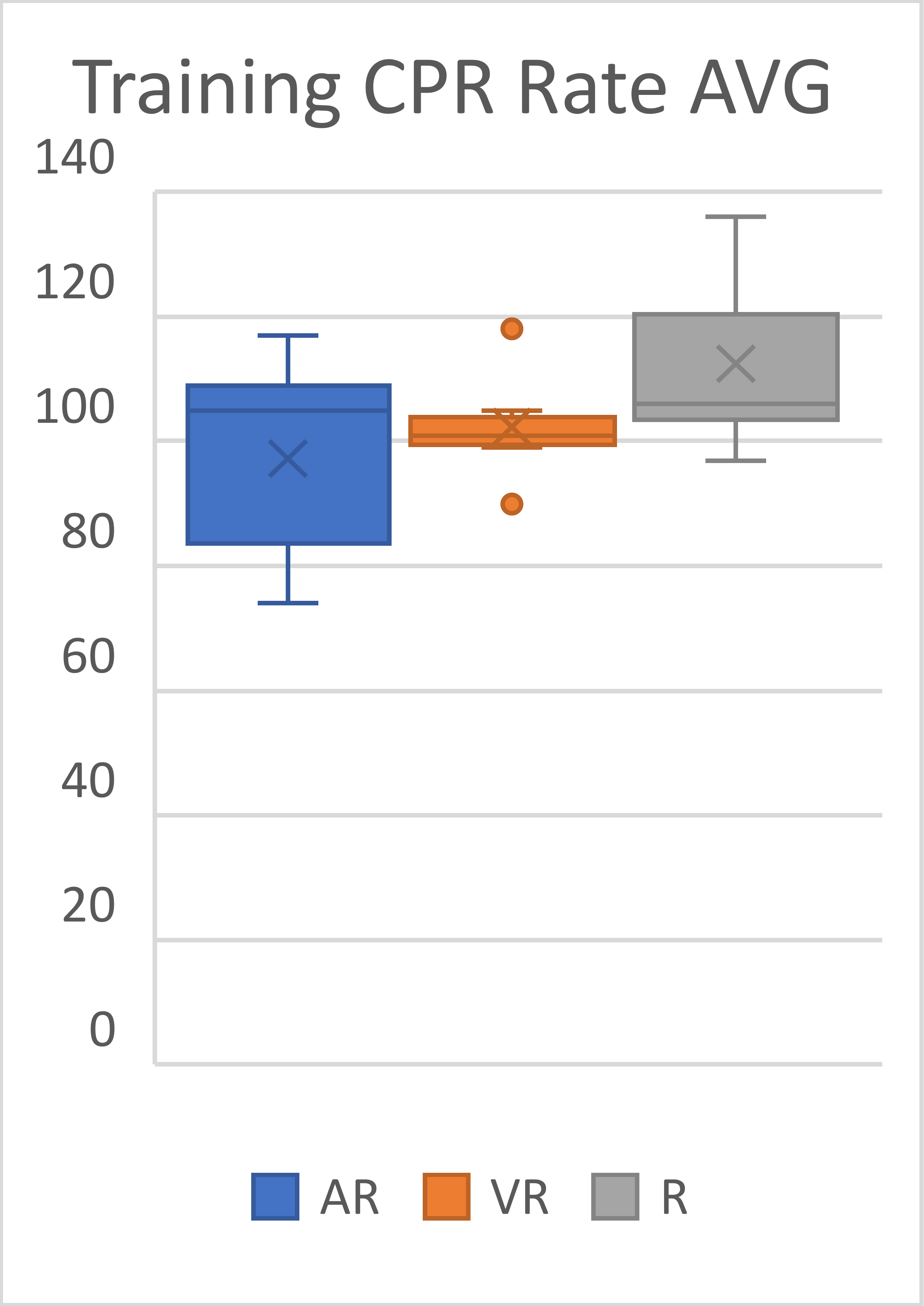}}\hspace*{5pt}
    \subfloat[Exam CPR Rate Average\label{fig:eval:examRate}]{\includegraphics[height=150pt]{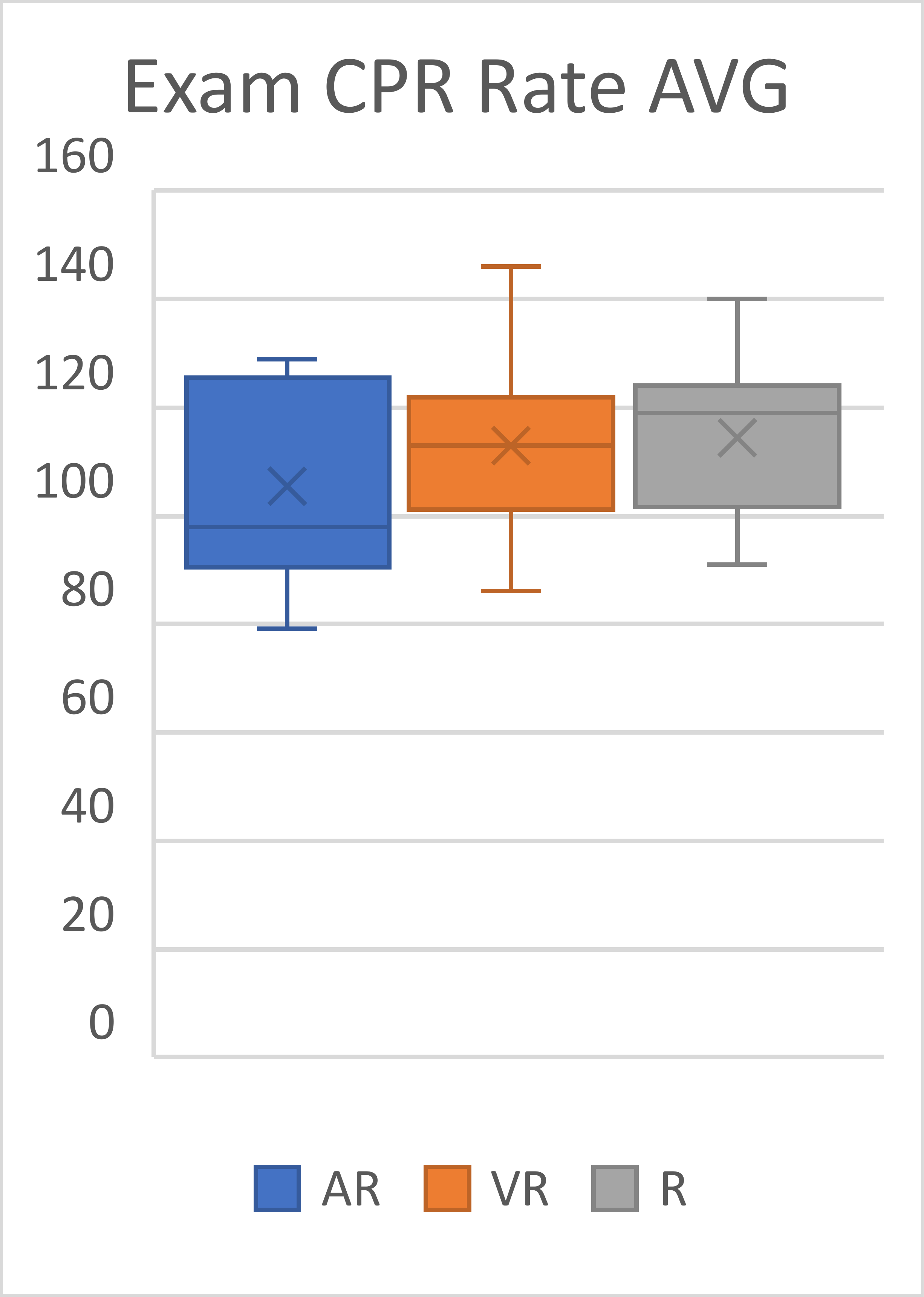}}
    \end{center}
    \caption{CPR Rate Averages}
    \label{fig:eval:rates}
\end{figure}
As chest compressions are an essential part of basic life support, we also measure how well they are executed. To do so, we recorded the compression rate and depth for the trainings and the exams.

Figure \ref{fig:eval:trainingRate} shows the compression rate in the trainings. In our trainings, the users should perform 105 compressions per minute. Here the medians were close to each other (AR:$105$, R:$106$, VR:$101$) whereas the mean values differ more. Users of the AR training tend to be much slower than trained and R training participants tend to be too fast. That can also be observed on the average rate, AR participants had an average rate of $97.29$ and R participants $112.42$. VR participants are on average the closest to the trained rate of $105$ with an average rate of $102.26$.

During the exam (Figure \ref{fig:eval:examRate}), participants of the AR training still tend to be too slow. Their average rate was $105.43$ with a median of $98$. VR and R training participants tend to be a bit too fast with an average and median of $113$ for VR training participants, and an average of $114.23$ and a median of $119$.

\begin{figure}
    \begin{center}  
    \subfloat[Training CPR Depth Average\label{fig:eval:trainingDepth}]{\includegraphics[height=150pt]{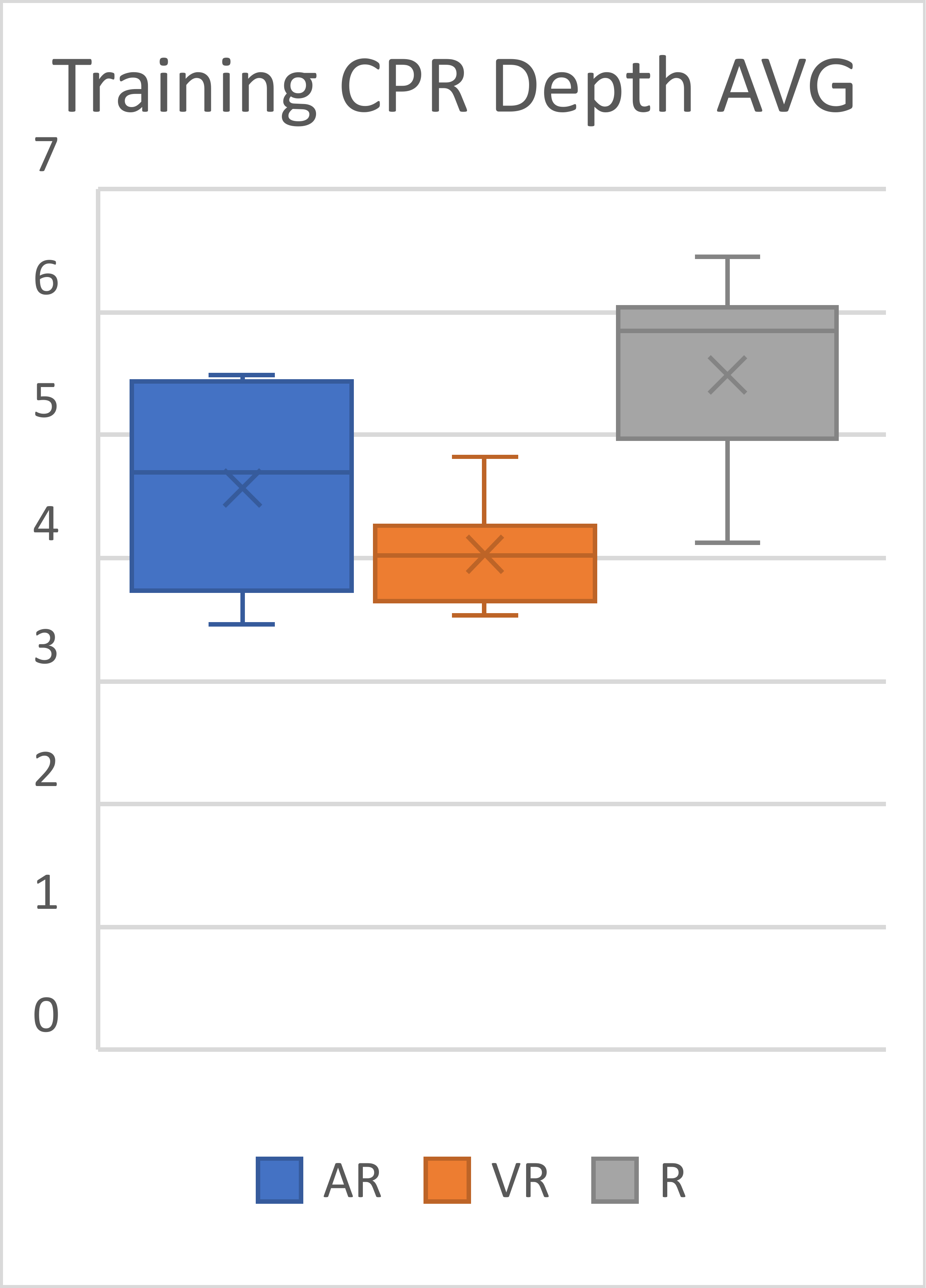}}\hspace*{5pt}
    \subfloat[Exam CPR Depth Average\label{fig:eval:examDepth}]{\includegraphics[height=150pt]{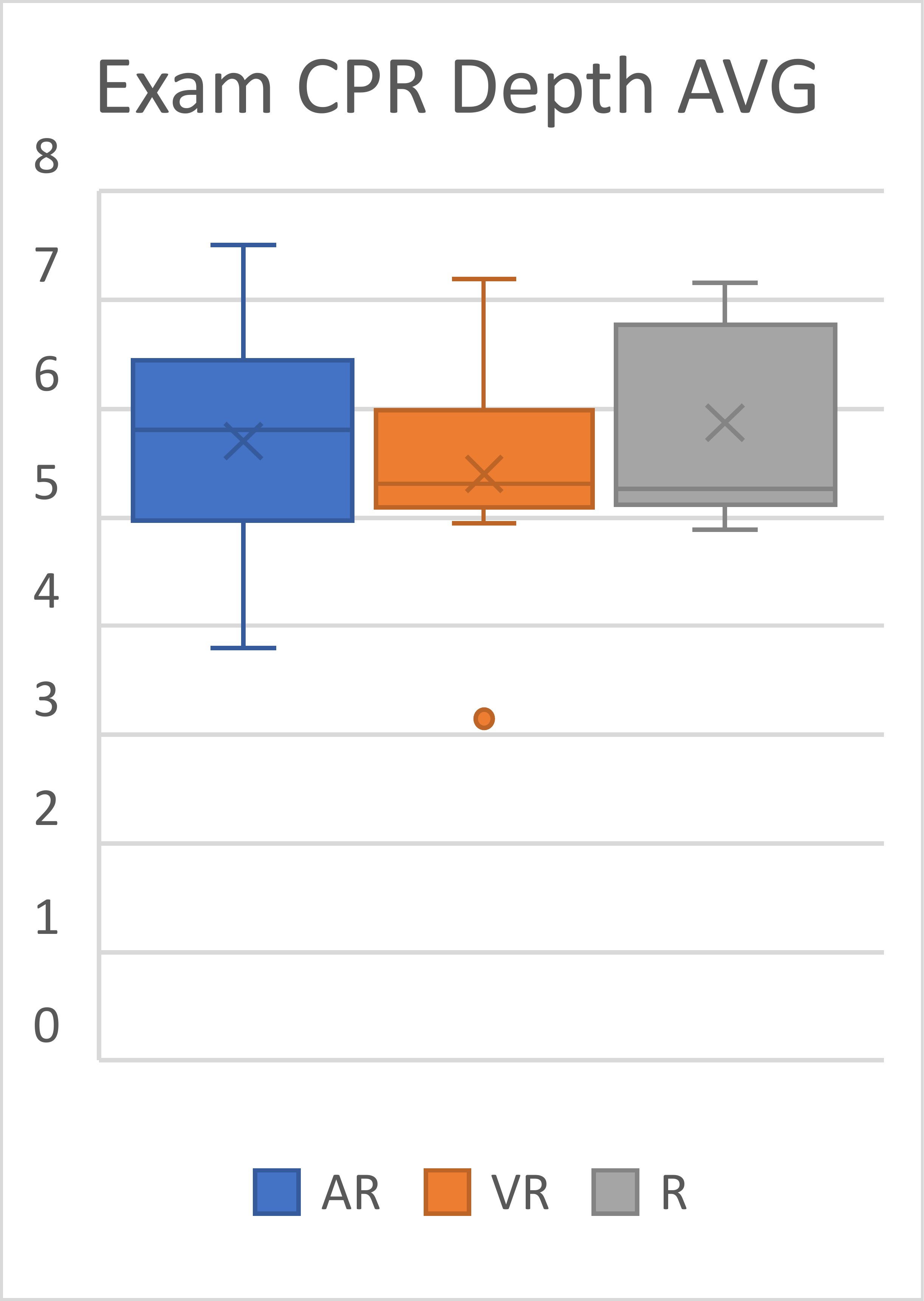}}
    \end{center}
    \caption{CPR Depth Averages}
    \label{fig:eval:depth}
\end{figure}
The compression depth can be seen in Figure \ref{fig:eval:trainingDepth}. In our trainings, we aimed to train the recommended depth of 5-6cm. For the trainings (Figure \ref{fig:eval:trainingDepth}), the R training participants were the best with an average of $5.49$ cm and a median of $5.85$ cm. The participants of the AR training (average: $4.57$ cm, median: $4.7$ cm) as well as the participants of the VR training (average: $4.03$ cm, median: $4.02$ cm did not push far enough.
During the exam, the AR training participants pushed on average $5.7$ cm with a median of $5.8$ cm, but some participants pushed too far or not far enough. VR training participants with an average of $5.4$ cm and a median of $5.31$ cm pushed mostly correct. Some participants pushed a bit too far, and one outlier only pushed $3.15$ cm deep. Some R training participants pushed too far, although the average with $5.87$ cm and the median with $5.26$ cm is still in the desired range.

\subsection{Discussion}\label{section:Evalution:discussion}
A major aspect when comparing the results of the trainings to each other is how well the users could interact with the technology. Considering the SUS score (Figure \ref{fig:eval:susScoreAll}), we can see a wide range of scores. But during the study, we noticed that the participants who gave the application a low SUS score had problems understanding how the technology works and how to use it properly. Whereas the participants who could work with it well gave a high rating. The same aspect can be seen for the TLX score (Figure \ref{fig:eval:TLXScore}). The participants who could work well with the technology felt a smaller workload than users who had problems. Concerning the workload, we can see a great potential of utilizing AR and VR for BLS training. On average, the TLX score for the AR- and VR-based BLS training environment was around 30 and much better than the real-life training that received a TLX score over 40. This shows that independent self-training of the BLS procedure can be eased through an interactive and immersive training environment. A big difference between AR and VR can be seen for the presence score (Figure \ref{fig:eval:PresenceScore}), here the VR application got a way better score. The reason is that using Virtual Reality, we could immerse the users in another place so that it felt more like being at an accident. In AR, they could see the office room they were in the whole time. As the field of view for virtual objects is quite small on the Hololens 2 and we cannot access high computing power, it is hard to immerse the users in another environment. 
When comparing the scores the users reached in the trainings and exams (Figures \ref{fig:eval:trainingScore} and \ref{fig:eval:examScore}), we can see that the scores for AR and VR are always close to each other. For the training, their scores are very high. That is because the application only continues with the next step if the previous step was executed correctly. Only the performance of the CPR can change the results. If a user executes a step wrong in the real-life training, the trainer can correct it afterward, but it was executed wrong in the first place. But considering the exam scores, users with the real-life training had better results than the others. So the 1:1 interaction with the trainer and the corrections still yield a better remembering of the tasks.
The results of the CPR (Figures \ref{fig:eval:rates} and \ref{fig:eval:depth}) are similar for all trainings in the exam. In contrast to the overall score, the execution was learned in an equal quality for all training types.

\subsection{Threats to Validity}\label{section:Evalution:threatsToValidity}
For our study, we had 21 participants and, accordingly, 7 participants in each training type. Especially when comparing the AR and VR versions the results are close to each other and we could not see whether one is significantly better than the other. With that in mind, a bigger user study could yield a more convincing result.

Due to the limited possibility and reliability in recognizing gestures or movement of e.g. the hands, the applications could not verify that all hand and finger movements were 100\% accurate. Thus, it may happen that the application recognizes a step as executed correctly whereas a real trainer could correct even small mistakes. 

In regular first aid training, the trainers are mostly experienced paramedics who not only explain the steps of the BLS sequence but give additional information when needed and enrich the training with personal stories and experiences. We did not integrate such additional content into our trainings, but the user's attention and thus their performance could be increased if they get some context about how the learned techniques really save lives. 

\section{Conclusion and Outlook}
\label{section:conclusion}

In the event of cardiac arrest, early bystander basic life support (BLS) is vital for increased survival chances. Regular training of BLS procedures among people
is required. While AR- and VR-based approaches have been promoted to enhance BLS training in an interactive and practical way, current existing approaches are not fully compliant with the medical BLS guidelines or focus only on specific steps of BLS training such as resuscitation. Furthermore, most of the existing training approaches do not focus on automated assessment to enhance efficiency and effectiveness through fine-grained real-time feedback. To overcome these issues, we have designed and implemented an immersive BLS training environment that supports AR as well as VR training with an interactive and haptic manikin. The main results of our usability evaluation show that AR and VR technology have the potential to increase engagement in BLS training, improve high-quality resuscitation training, and reduce the cognitive workload compared to traditional training with a personal instructor.

In future work, we plan to extend our BLS training environment to also support Advanced Life Support (ALS). While the scope of this work considers independent self-training only, multi-user training becomes more and more important, particularly for Advanced Life Support. As training is not only about learning specific skills such as resuscitation and is also about learning to work together in a team of multiple people, we plan to extend our training environment by collaboration features.


\bibliographystyle{splncs04}
\bibliography{references}

\begin{thebibliography}{10}
\providecommand{\url}[1]{\texttt{#1}}
\providecommand{\urlprefix}{URL }
\providecommand{\doi}[1]{https://doi.org/#1}

\bibitem{almousa2019virtual}
Almousa, O., Prates, J., Yeslam, N., Mac~Gregor, D., Zhang, J., Phan, V.,
  Nielsen, M., Smith, R., Qayumi, K.: Virtual reality simulation technology for
  cardiopulmonary resuscitation training: An innovative hybrid system with
  haptic feedback. Simulation \& Gaming  \textbf{50}(1),  6--22 (2019)

\bibitem{balian2019feasibility}
Balian, S., McGovern, S.K., Abella, B.S., Blewer, A.L., Leary, M.: Feasibility
  of an augmented reality cardiopulmonary resuscitation training system for
  health care providers. Heliyon  \textbf{5}(8),  e02205 (2019)

\bibitem{blome2017vreanimate}
Blome, T., Diefenbach, A., Rudolph, S., Bucher, K., von Mammen, S.:
  Vreanimate—non-verbal guidance and learning in virtual reality. In: 2017
  9th International Conference on Virtual Worlds and Games for Serious
  Applications (VS-Games). pp. 23--30. IEEE (2017)

\bibitem{brooke1996sus}
Brooke, J., et~al.: Sus-a quick and dirty usability scale. Usability evaluation
  in industry  \textbf{189}(194), ~4--7 (1996)

\bibitem{bucher2019vreanimate}
Bucher, K., Blome, T., Rudolph, S., von Mammen, S.: Vreanimate ii: training
  first aid and reanimation in virtual reality. Journal of Computers in
  Education  \textbf{6}(1),  53--78 (2019)

\bibitem{everson2021virtual}
Everson, T., Joordens, M., Forbes, H., Horan, B.: Virtual reality and haptic
  cardiopulmonary resuscitation training approaches: A review. IEEE Systems
  Journal  (2021)

\bibitem{ferracani2015natural}
Ferracani, A., Pezzatini, D., Seidenari, L., Del~Bimbo, A.: Natural and virtual
  environments for the training of emergency medicine personnel. Universal
  Access in the Information Society  \textbf{14}(3),  351--362 (2015)

\bibitem{DBLP:conf/hcse/GottschalkYSE20}
Gottschalk, S., Yigitbas, E., Schmidt, E., Engels, G.: Model-based product
  configuration in augmented reality applications. In: Bernhaupt, R., Ardito,
  C., Sauer, S. (eds.) Human-Centered Software Engineering - 8th {IFIP} {WG}
  13.2 International Working Conference, {HCSE} 2020, Eindhoven, The
  Netherlands, November 30 - December 2, 2020, Proceedings. Lecture Notes in
  Computer Science, vol. 12481, pp. 84--104. Springer (2020).
  \doi{10.1007/978-3-030-64266-2\_5},
  \url{https://doi.org/10.1007/978-3-030-64266-2\_5}

\bibitem{DBLP:conf/hcse/GottschalkYSE20a}
Gottschalk, S., Yigitbas, E., Schmidt, E., Engels, G.: Proconar: {A} tool
  support for model-based {AR} product configuration. In: Bernhaupt, R.,
  Ardito, C., Sauer, S. (eds.) Human-Centered Software Engineering - 8th {IFIP}
  {WG} 13.2 International Working Conference, {HCSE} 2020, Eindhoven, The
  Netherlands, November 30 - December 2, 2020, Proceedings. Lecture Notes in
  Computer Science, vol. 12481, pp. 207--215. Springer (2020).
  \doi{10.1007/978-3-030-64266-2\_14},
  \url{https://doi.org/10.1007/978-3-030-64266-2\_14}

\bibitem{grasner2020survival}
Gr{\"a}sner, J.T., Wnent, J., Herlitz, J., Perkins, G.D., Lefering, R.,
  Tjelmeland, I., Koster, R.W., Masterson, S., Rossell-Ortiz, F., Maurer, H.,
  et~al.: Survival after out-of-hospital cardiac arrest in europe-results of
  the eureca two study. Resuscitation  \textbf{148},  218--226 (2020)

\bibitem{hart1988development}
Hart, S.G., Staveland, L.E.: Development of nasa-tlx (task load index): Results
  of empirical and theoretical research. In: Advances in psychology, vol.~52,
  pp. 139--183. Elsevier (1988)

\bibitem{hsieh2018preliminary}
Hsieh, M.C., Lee, J.: Preliminary study of vr and ar applications in medical
  and healthcare education. J Nurs Health Stud  \textbf{3}(1), ~1 (2018)

\bibitem{ingrassia2020augmented}
Ingrassia, P.L., Mormando, G., Giudici, E., Strada, F., Carfagna, F., Lamberti,
  F., Bottino, A.: Augmented reality learning environment for basic life
  support and defibrillation training: usability study. Journal of medical
  Internet research  \textbf{22}(5),  e14910 (2020)

\bibitem{javaheri2018stayin}
Javaheri, H., Gruenerbl, A., Monger, E., Gobbi, M., Lukowicz, P.: Stayin'alive:
  An interactive augmented: Reality cpr tutorial. In: Proceedings of the 2018
  ACM International Joint Conference and 2018 International Symposium on
  Pervasive and Ubiquitous Computing and Wearable Computers. pp. 365--368
  (2018)

\bibitem{johnson2018holocpr}
Johnson, J.G., Rodrigues, D.G., Gubbala, M., Weibel, N.: Holocpr: Designing and
  evaluating a mixed reality interface for time-critical emergencies. In:
  Proceedings of the 12th EAI International Conference on Pervasive Computing
  Technologies for Healthcare. pp. 67--76 (2018)

\bibitem{DBLP:conf/hcse/JovanovikjY0E20}
Jovanovikj, I., Yigitbas, E., Sauer, S., Engels, G.: Augmented and virtual
  reality object repository for rapid prototyping. In: Bernhaupt, R., Ardito,
  C., Sauer, S. (eds.) Human-Centered Software Engineering - 8th {IFIP} {WG}
  13.2 International Working Conference, {HCSE} 2020, Eindhoven, The
  Netherlands, November 30 - December 2, 2020, Proceedings. Lecture Notes in
  Computer Science, vol. 12481, pp. 216--224. Springer (2020).
  \doi{10.1007/978-3-030-64266-2\_15},
  \url{https://doi.org/10.1007/978-3-030-64266-2\_15}

\bibitem{DBLP:conf/eics/KringsYJ0E20}
Krings, S., Yigitbas, E., Jovanovikj, I., Sauer, S., Engels, G.: Development
  framework for context-aware augmented reality applications. In: Bowen, J.,
  Vanderdonckt, J., Winckler, M. (eds.) {EICS} '20: {ACM} {SIGCHI} Symposium on
  Engineering Interactive Computing Systems, Sophia Antipolis, France, June
  23-26, 2020. pp. 9:1--9:6. {ACM} (2020). \doi{10.1145/3393672.3398640},
  \url{https://doi.org/10.1145/3393672.3398640}

\bibitem{DBLP:conf/eics/KringsYBE22}
Krings, S.C., Yigitbas, E., Biermeier, K., Engels, G.: Design and evaluation of
  ar-assisted end-user robot path planning strategies. In: Winckler, M.,
  Quigley, A. (eds.) {EICS} '22: {ACM} {SIGCHI} Symposium on Engineering
  Interactive Computing Systems, Sophia Antipolis, France, June 21 - 24, 2022,
  Companion Volume. pp. 14--18. {ACM} (2022). \doi{10.1145/3531706.3536452},
  \url{https://doi.org/10.1145/3531706.3536452}

\bibitem{kuyt2021use}
Kuyt, K., Park, S.H., Chang, T.P., Jung, T., MacKinnon, R.: The use of virtual
  reality and augmented reality to enhance cardio-pulmonary resuscitation: a
  scoping review. Advances in Simulation  \textbf{6}(1), ~1--8 (2021)

\bibitem{kwon2014heartisense}
Kwon, Y., Lee, S., Jeong, J., Kim, W.: Heartisense: a novel approach to enable
  effective basic life support training without an instructor. In: CHI'14
  Extended Abstracts on Human Factors in Computing Systems, pp. 1699--1704
  (2014)

\bibitem{olasveengen2021european}
Olasveengen, T.M., Semeraro, F., Ristagno, G., Castren, M., Handley, A.,
  Kuzovlev, A., Monsieurs, K.G., Raffay, V., Smyth, M., Soar, J., et~al.:
  European resuscitation council guidelines 2021: basic life support.
  Resuscitation  \textbf{161},  98--114 (2021)

\bibitem{perkins2015european}
Perkins, G.D., Handley, A.J., Koster, R.W., Castr{\'e}n, M., Smyth, M.A.,
  Olasveengen, T., Monsieurs, K.G., Raffay, V., Gr{\"a}sner, J.T., Wenzel, V.,
  et~al.: European resuscitation council guidelines for resuscitation 2015:
  Section 2. adult basic life support and automated external defibrillation.
  Resuscitation  \textbf{95},  81--99 (2015)

\bibitem{pretto2009augmented}
Pretto, F., Manssour, I.H., Lopes, M.H.I., da~Silva, E.R., Pinho, M.S.:
  Augmented reality environment for life support training. In: Proceedings of
  the 2009 ACM symposium on Applied Computing. pp. 836--841 (2009)

\bibitem{semeraro2009virtual}
Semeraro, F., Frisoli, A., Bergamasco, M., Cerchiari, E.L.: Virtual reality
  enhanced mannequin (vrem) that is well received by resuscitation experts.
  Resuscitation  \textbf{80}(4),  489--492 (2009)

\bibitem{semeraro2019virtual}
Semeraro, F., Ristagno, G., Giulini, G., Gnudi, T., Kayal, J.S., Monesi, A.,
  Tucci, R., Scapigliati, A.: Virtual reality cardiopulmonary resuscitation
  (cpr): Comparison with a standard cpr training mannequin. Resuscitation
  \textbf{135},  234--235 (2019)

\bibitem{strada2019holo}
Strada, F., Bottino, A., Lamberti, F., Mormando, G., Ingrassia, P.L.:
  Holo-blsd-a holographic tool for self-training and self-evaluation of
  emergency response skills. IEEE Transactions on Emerging Topics in Computing
  (2019)

\bibitem{DBLP:journals/presence/UsohCAS00}
Usoh, M., Catena, E., Arman, S., Slater, M.: Using presence questionnaires in
  reality. Presence Teleoperators Virtual Environ.  \textbf{9}(5),  497--503
  (2000). \doi{10.1162/105474600566989},
  \url{https://doi.org/10.1162/105474600566989}

\bibitem{vaughan2019cpr}
Vaughan, N., John, N., Rees, N.: Cpr virtual reality training simulator for
  schools. In: 2019 International Conference on Cyberworlds (CW). pp. 25--28.
  IEEE (2019)

\bibitem{virani2020heart}
Virani, S.S., Alonso, A., Benjamin, E.J., Bittencourt, M.S., Callaway, C.W.,
  Carson, A.P., Chamberlain, A.M., Chang, A.R., Cheng, S., Delling, F.N.,
  et~al.: Heart disease and stroke statistics—2020 update: a report from the
  american heart association. Circulation  \textbf{141}(9),  e139--e596 (2020)

\bibitem{DBLP:conf/models/YigitbasGWE21}
Yigitbas, E., Gorissen, S., Weidmann, N., Engels, G.: Collaborative software
  modeling in virtual reality. In: 24th International Conference on Model
  Driven Engineering Languages and Systems, {MODELS} 2021, Fukuoka, Japan,
  October 10-15, 2021. pp. 261--272. {IEEE} (2021).
  \doi{10.1109/MODELS50736.2021.00034},
  \url{https://doi.org/10.1109/MODELS50736.2021.00034}

\bibitem{DBLP:conf/mc/YigitbasHE19}
Yigitbas, E., Heind{\"{o}}rfer, J., Engels, G.: A context-aware virtual reality
  first aid training application. In: Alt, F., Bulling, A., D{\"{o}}ring, T.
  (eds.) Proc. of Mensch und Computer 2019. pp. 885--888. {GI} / {ACM} (2019)

\bibitem{DBLP:conf/interact/YigitbasJE21}
Yigitbas, E., Jovanovikj, I., Engels, G.: Simplifying robot programming using
  augmented reality and end-user development. In: Ardito, C., Lanzilotti, R.,
  Malizia, A., Petrie, H., Piccinno, A., Desolda, G., Inkpen, K. (eds.)
  Human-Computer Interaction - {INTERACT} 2021 - 18th {IFIP} {TC} 13
  International Conference, Bari, Italy, August 30 - September 3, 2021,
  Proceedings, Part {I}. Lecture Notes in Computer Science, vol. 12932, pp.
  631--651. Springer (2021). \doi{10.1007/978-3-030-85623-6\_36},
  \url{https://doi.org/10.1007/978-3-030-85623-6\_36}

\bibitem{DBLP:conf/interact/YigitbasJ0E19}
Yigitbas, E., Jovanovikj, I., Sauer, S., Engels, G.: On the development of
  context-aware augmented reality applications. In: Abdelnour{-}Nocera, J.L.,
  Parmaxi, A., Winckler, M., Loizides, F., Ardito, C., Bhutkar, G., Dannenmann,
  P. (eds.) Beyond Interactions - {INTERACT} 2019 {IFIP} {TC} 13 Workshops,
  Paphos, Cyprus, September 2-6, 2019, Revised Selected Papers. Lecture Notes
  in Computer Science, vol. 11930, pp. 107--120. Springer (2019).
  \doi{10.1007/978-3-030-46540-7\_11},
  \url{https://doi.org/10.1007/978-3-030-46540-7\_11}

\bibitem{DBLP:conf/vrst/YigitbasJSE20}
Yigitbas, E., Jovanovikj, I., Scholand, J., Engels, G.: {VR} training for
  warehouse management. In: Teather, R.J., Joslin, C., Stuerzlinger, W.,
  Figueroa, P., Hu, Y., Batmaz, A.U., Lee, W., Ortega, F.R. (eds.) {VRST} '20:
  26th {ACM} Symposium on Virtual Reality Software and Technology. pp.
  78:1--78:3. {ACM} (2020)

\bibitem{DBLP:conf/seams/YigitbasKJE21}
Yigitbas, E., Karakaya, K., Jovanovikj, I., Engels, G.: Enhancing
  human-in-the-loop adaptive systems through digital twins and {VR} interfaces.
  In: 16th International Symposium on Software Engineering for Adaptive and
  Self-Managing Systems, SEAMS@ICSE 2021, Madrid, Spain, May 18-24, 2021. pp.
  30--40. {IEEE} (2021). \doi{10.1109/SEAMS51251.2021.00015},
  \url{https://doi.org/10.1109/SEAMS51251.2021.00015}

\bibitem{DBLP:conf/vl/YigitbasKGE21}
Yigitbas, E., Klauke, J., Gottschalk, S., Engels, G.: {VREUD} - an end-user
  development tool to simplify the creation of interactive {VR} scenes. In:
  Harms, K.J., Cunha, J., Oney, S., Kelleher, C. (eds.) {IEEE} Symposium on
  Visual Languages and Human-Centric Computing, {VL/HCC} 2021, St Louis, MO,
  USA, October 10-13, 2021. pp. 1--10. {IEEE} (2021).
  \doi{10.1109/VL/HCC51201.2021.9576372},
  \url{https://doi.org/10.1109/VL/HCC51201.2021.9576372}

\bibitem{DBLP:conf/eics/EnesScaffolding}
Yigitbas, E., Sauer, S., Engels, G.: Using augmented reality for enhancing
  planning and measurements in the scaffolding business. In: {EICS} '21: {ACM}
  {SIGCHI} Symposium on Engineering Interactive Computing Systems, virtual,
  June 8-11, 2021. {ACM} (2021), \url{https://doi.org/10.1145/3459926.3464747}

\bibitem{DBLP:conf/mc/YigitbasTE20}
Yigitbas, E., Tejedor, C.B., Engels, G.: Experiencing and programming the
  {ENIAC} in {VR}. In: Alt, F., Schneegass, S., Hornecker, E. (eds.) Mensch und
  Computer 2020. pp. 505--506. {ACM} (2020)

\end{thebibliography}

\end{document}